 \documentclass[11pt]{article}

 \pdfoutput=1

\usepackage{graphicx} 
\usepackage{amsmath}

\usepackage{enumitem}

\usepackage[francais,english]{babel}

\usepackage{float}
\usepackage{ifthen}
\usepackage{hyperref}

\begin{document}


\title{Definition of a moist-air entropy potential temperature. Application to FIRE-I data flights.}

\author{by Pascal Marquet. {\it M\'et\'eo-France} \hspace*{3mm} (pascal.marquet@meteo.fr)}


\date{15th of January, 2014}

\maketitle


\vspace*{-12mm}
\begin{center}
\noindent
{\em Copy of a paper submitted in May 2010 to the 
 \underline{Quarterly Journal of the Royal Meteorological Society} \\
 (published in Volume 137, Issue 656, pages 768-791, April 2011 Part A).} \\
{\em \underline{V2:} three footnotes (pages 9-19-31); arXiv references to next papers.} \\
{\em \underline{V3:} 1991 $\rightarrow$ 2011 in footnote-p.9; Paper on Enthalpy \underline{accepted}.} \\
{\em \underline{V4:} correct typos, boxed equations (sections 4 and 8).}
\end{center}
\vspace{1mm}


\vspace*{-5mm}
\begin{abstract}
A moist entropy potential temperature -- denoted by ${\theta}_{s}$ -- is defined analytically in terms of the specific entropy for  moist air.
The expression for ${\theta}_{s}$ is valid for a general mixing of dry air, water vapour and possible condensed water species.
It verifies the same conservative properties as the moist entropy, even for varying dry air or total water content.
The moist formulation for ${\theta}_{s}$ is equal to the dry formulation $\theta$ if dry air is considered and it verifies new properties valid for the moist air cases, both saturated or under-saturated ones.
Exact and approximate versions of ${\theta}_{s}$ are evaluated for several Stratocumulus cases, in particular by using the aircraft observations FIRE-I experiment data sets.
It appears that there is no (or small) jump in ${\theta}_{s}$ at the top of the PBL.
The mixing in moist entropy is almost complete in the PBL, with the same values observed in the clear air and the cloudy regions, including the very top of the entrainment region.
The Randall-Deardorff CTEI analysis may be interpreted as a mixing in moist entropy criterion.
The iso-${\theta}_{s}$ lines are plotted on skew $T$-$\ln(p)$ and conserved variable diagrams.
All these properties could suggest some hints on the use of  moist entropy (or ${\theta}_{s}$) in cloud modelling or in mixing processes, with the marine Stratocumulus considered as a paradigm of  moist turbulence.
\end{abstract}


\section{Introduction.} 
\label{section_intro}

One of the conclusions of the IPCC AR4 (2007) is that cloud effects remain the largest sources of uncertainty in GCM based estimates of climate sensitivity, with large cloud radiative feedbacks associated with low-level clouds such as the Marine Stratocumulus.
The increase in the realism of the modelling of clouds is also one of the key features for the improvement of the NWP models (global or LAM ones).

Different projects have already evaluated the quality of the three-dimensional distribution of clouds in the Climate and NWP models (EU\-ROCS: http://www.knmi.nl/\-samenw/\-eurocs/ ; GCSS: http:\-//www.\-gewex.org\-/gcss.html).
The aim of the new European FP5 EUCLIPSE project (http:\-//www.knmi.nl/\-samenw/\-euclipse/) is to promote the comparisons with the new space-borne remote sensing dataset (such as  CloudSat, CALIPSO, TRMM) and by realizing inter-comparisons between GCM, NWP, SCM, CRM and LES outputs.
The goal is to determine what are the main deficiencies in the parameterizations of clouds (either for the stratiform,  shallow or deep convective ones) and to test more accurate updated schemes.

It is also possible to revisit some aspects of the theoretical concepts which form the bases of our understanding of the moist atmospheric processes, such as the definition and the use of  enthalpy,  entropy or  exergy functions.
In particular, the comparison with the existing in-situ datasets could still be of some help in order to assess the different hypotheses presently made to build the turbulent and convective schemes.

In this frame, the PBL region of marine Stratocumuli can be considered as a paradigm of the moist turbulence and it is of common use to realize vertical diffusion of the well-known ``conserved variables'' defined in Betts (1973, hereafter B73).
However, it seems that the in situ observations of the Betts' variables (the liquid potential temperature and the total water content) show that these variables are not constant vertically and that the clear-air and the in-cloud values are different (see for instance the vertical profiles computed with the  FIRE-I data set and published in De Roode and Wang, 2007, hereafter RW07).

The liquid potential temperature ${\theta}_{l}$ is defined in B73 with the aim of being a synonym of moist entropy.
Therefore, ${\theta}_{l}$ may be used in moist turbulent processes as a conserved variable only if the total water content is also a constant and these hypotheses might prevent ${\theta}_{l}$ from being a conservative quantity in case of varying dry-air and total water content, as clearly observed in the vertical profiles of Stratocumulus in-situ measurements.

One of the ways to answer these questions is to remember that, from the general thermodynamics, the moist entropy must be conserved by moist, reversible and adiabatic processes (the ones acting in the moist PBL of Stratocumulus).
Therefore, the aim of this paper will be to compute moist entropy and its associated potential temperature as precisely as possible, and to explain how it is indeed different from, and more interesting than, the Betts' liquid potential temperature.

The use of potential temperatures, instead of  entropy, has a long history in meteorology and the analysis will be made in this paper mainly in terms of a moist potential temperature, denoted by ${\theta}_{s}$ (with ``s'' representing the moist entropy), in order to make the comparisons with all the existing ones easier. 
Nonetheless, the main variable studied in this paper is clearly the moist entropy.

The moist potential temperature ${\theta}_{s}$  is expected to represent all the variations of  moist entropy ``$s$'', whatever the changes in temperature, pressure, specific content of dry air, water vapour or condensed water species (solid and liquid) may be. 
This property would allow us to derive the same conservative properties for ${\theta}_{s}$  as the general ones valid for the moist entropy.

The concept  of  what is nowadays called ``potential temperature'' in atmospheric science was first introduced by von  Helmholtz (1888, 1891), with the use of the name ``waermegehalt'' (warming content) and with the notation $\theta$.
The ``warming content'' of a given mass of air was defined as the absolute temperature $\theta$ which a mass of dry air would assume if it were brought adiabatically to a normal or standard pressure.
This quantity has  been called ``potential temperature'' by von Bezold (1888, 1891) and the link between  $\theta$  and the specific dry air entropy has been discussed later, in Bauer (1908, 1910).

Since these pioneering studies, the concept of potential temperature has been generalized to moist air by using different approaches.
The first method is to compute integrals of different approximate versions of the so-called Gibbs  (1875-76-77-78) differential equation.
With the notations of the Appendix-A, it is written
\begin{align}
  T\: ds & = \; dh \: - \: \alpha\: dp \:  - \: \sum_k \mu_k \: dq_k
  \: . \label{defGibbs} 
\end{align}
The following definitions ensue
\begin{itemize}[label=$\bullet$,leftmargin=3mm,parsep=0cm,itemsep=0.1cm,topsep=0cm,rightmargin=2mm]
\vspace*{-1mm}
\item the liquid  potential temperature ${\theta}_{l}$ of B73, leading to a conservative moist variable, almost constant  within the Stratocumulus regions if the sum of water vapour plus liquid water is a constant;
\item the saturated equivalent potential temperature ${\theta}_{ES}$ obtained in B73 as a companion of ${\theta}_{l}$;
\item the ice-liquid water potential temperature ${\theta}_{il}$, suggested in Deardorff (1976) and derived in Tripoli and Cotton (1981, hereafter TC81), to be applied to the parameterization of the cumulus.
\end{itemize}

\vspace{3mm}
Another set of definitions concerns the impact of the buoyancy force, or other thermodynamic computations,  leading to
\begin{itemize}[label=$\bullet$,leftmargin=3mm,parsep=0cm,itemsep=0.1cm,topsep=0cm,rightmargin=2mm]
\vspace*{-1mm}
\item the equivalent potential temperature  ${\theta}_{E}$, obtained after the condensation level as the dry potential temperature that a parcel will have when all the water is removed from it, via pseudo-adiabatic processes;
\item the virtual potential temperature ${\theta}_{v}$ of Lilly (1968, hereafter L68), used for instance in the thermal production term involved in the turbulent kinetic energy turbulent equations, also in the computation of the CAPE for deep convection;
\item the liquid water virtual potential temperature ${\theta}_{vl}$ described in Grenier and Bretherton (2001, hereafter GB01), suitable for the parameterization of the Stratocumulus top PBL entrainment.
\end{itemize}

\vspace{3mm}
The last method is to start with the analytic formulations for the moist specific entropy $s$, expressed as a sum of the partial specific entropies for  dry air and  water species.
The moist potential (entropic) temperature (let us say ${\theta}_{s}$) is then determined without the use of a Gibbs differential equation, by writing the moist entropy $s$ with some prescribed reference state defined by $s_{r}$, $c_{r}$ and $\theta_{sr}$, leading to
\begin{align}
  s & = \: \sum_k q_k \: s_k
   \; = \: s_{r} \: + \: c_{r} \: \ln({\theta}_{s}/\theta_{sr})
  \: . \label{defThetam0} 
\end{align}
The following definitions ensue
\begin{itemize}[label=$\bullet$,leftmargin=3mm,parsep=0cm,itemsep=0.1cm,topsep=0cm,rightmargin=2mm]
\vspace*{-1mm}
\item different entropy temperatures in Hauf and H\"{o}ller (1987, hereafter HH87), including the one denoted by ${\theta}_{S}^{\ast}$ in what follows (it was denoted by ${\theta}_{S}$ in HH87);
\item a moist potential temperature ${\theta}^{\ast}$ in Marquet (1993, hereafter M93), used in the post-processing of the ARPEGE-IFS models (subroutines PPWETPOINT and PPTHPW) and in the definition of the conservative fluxes and the barycentric equations derived in Catry {\em et al. \/} (2007);
\item the liquid water potential temperature of Emanuel (1994, hereafter E94), denoted by ${\theta}_{l}^{\ast}$ in what follows, including some extra terms when compared to the Betts' formulation ${\theta}_{l}$ (with ${\theta}_{l}^{\ast}$ denoted by ${\theta}_{l}$ in E94).
\end{itemize}

\vspace{3mm}
The paper is organized as follows.
The analytic expression for the moist entropy and for ${\theta}_{s}$ will be obtained starting from the definition (\ref{defThetam0}).
The classical potential temperatures (${\theta}_{v}$, ${\theta}_{ES}$, ${\theta}_{l}$, ${\theta}_{il}$ and ${\theta}_{vl}$) are first recalled in  section \ref{section_mptvl}.
The seldom used moist entropy  potential temperatures ${\theta}_{S}^{\ast}$, ${\theta}^{\ast}$ and ${\theta}_{l}^{\ast}$ are recalled in  section \ref{section_mptthast}.
The new formulation ${\theta}_{s}$ is then derived analytically in  section \ref{section_defTHm} and in the Appendix-B and compared to the previous ones.

A  first-order approximation for ${\theta}_{s}$ is proposed in  section \ref{section_aftha}.
The conservative property verified by ${\theta}_{s}$  is  computed in  section \ref{section_cpftha} and compared to the one verified by ${\theta}_{S}^{\ast}$, ${\theta}^{\ast}$ and ${\theta}_{l}^{\ast}$.

The  moist entropy potential temperature ${\theta}_{s}$ is evaluated in section \ref{section_nefd} by using the FIRE-I experiment, with the PBL aircraft dataset described in RW07.
The impacts of some of the approximations are analysed in  section  \ref{section_sens}.
The vertical fluxes of ${\theta}_{s}$ are computed in section \ref{section_flux_grad}.
Palush conserved variables and skew $T$-$\ln(p)$ diagrams are analysed in sections \ref{section_others_Sc} and \ref{section_diagram} in terms of the new formulation ${\theta}_{s}$.
Some justifications of the constant feature for ${\theta}_{s}$ are suggested in section \ref{section_constths}, including some useful Gibbs-like 3D visions.
Finally, conclusions are presented in section \ref{section_concl}.

\section{Standard moist potential temperatures 
 (B73, TC81, GB01, L68)} 
\label{section_mptvl}

\subsection{The versions of Betts (1973).} 
\label{section_mptvl1}

The potential temperatures ${\theta}_{l}$ and ${\theta}_{ES}$  are defined in B73 (see Eqs.(6), (7) and (9) to (12) in that paper) by approximate Gibbs differential equations and with $r_t$ assumed to be a constant.
The formulation for ${\theta}_{l}$ and ${\theta}_{ES}$ are  obtained with several approximations, such as  $c_p \approx c_{pd}$ and $R \approx R_{d}$, leading to
\vspace{-0.15cm}
\begin{align}
  0 = \frac{ds}{1+r_t}
         \: \approx \:
  c_{pd} \;\frac{d{\theta}_{l}}{{\theta}_{l}}
         & \approx \:
  c_{pd} \;\frac{d{\theta}}{{\theta}}
          \:-
    \frac{L_v(T)}{T} \: d q_l
  , \label{defTHB73a} \\
  0 = \frac{ds}{1+r_t}
         \: \approx \:
  c_{pd} \;\frac{d{\theta}_{ES}}{{\theta}_{ES}}
         & \approx \: 
  c_{pd} \;\frac{d{\theta}}{{\theta}}
          \:+
    \frac{L_v(T)}{T} \: d q_{S}
  . \label{defTHESB73a}
\end{align}
The corresponding values for ${\theta}_{ES}$ and ${\theta}_{l}$  are obtained by integrating (\ref{defTHESB73a}) and (\ref{defTHB73a})  with some further approximations (see also Betts and Dugan, 1973), particularly for the last term and the variations of $L_v(T)/T$ with $T$, giving
\vspace{-0.15cm}
\begin{align}
  {\theta}_{ES}  & = \: {\theta} \; 
         \exp \left( \frac{L_v \:r_{S}}{c_{pd}\: T} \right)
  , \label{defTHESB73b} \\
  {\theta}_{l}  & = \: {\theta} \; 
         \exp \left( - \: \frac{L_v \:q_l}{c_{pd}\: T} \right)
  . \label{defTHB73b}
\end{align}
Eq.(\ref{defTHB73b}) is the equivalent of Eq.(13) in B73, expressed with the notations of the Appendix-A.
The potential temperature (\ref{defTHB73b})  can be further modified by using the Taylor's series approximation $\exp(x) \approx 1 + x$, leading to Eq.(14) in B73 and corresponding to  (\ref{defTHB73c})
\vspace{-0.15cm}
\begin{align}
  {\theta}_{l}  & \approx \: {\theta}  \left( 1 - \frac{L_v \:q_l}{c_{pd} \:T} \right)
  , \label{defTHB73c} \\
  \text{with} \quad  
         q_{t}  & = \:  q_v + q_l
  . \label{defQtB73}
\end{align}
This pair of Betts moist variables $({\theta}_{l}, q_t)$ are nowadays used to compute the moist turbulent fluxes in most of the turbulent schemes (see for example Brinkop and Roeckner (1995) or Cuxart {\em et al.\/} (2000), hereafter BR95 and CBR00)

The variables $({\theta}_{l}, q_t)$ are considered as conservative ones for the hydrostatic and adiabatic motion of a closed parcel of moist air, i.e. if $q_d=1-q_t$ and $q_t=q_v+q_l$ are constant in the clear-air and the in-cloud regions (the precipitating species are not considered).
Accordingly, the equations for the water species correspond to an exchange between the vapour and the liquid phases via evaporation or condensation processes, leading to
\vspace{-0.15cm}
\begin{align}
  d\left( q_v \right)/dt & =\; +(\dot{q})_{eva} \: , \label{def_Dv} \\
  d\left( q_l  \right)/dt & =\; -(\dot{q})_{eva} \: . \label{def_Dl} 
\end{align}
As already mentioned  in Deardorff (1980) the conservative property is verified for ${\theta}_{l}$ only if the change of $L_v(T)/T$ is neglected in the logarithmic derivative of (\ref{defTHB73b}), leading to
\vspace{-0.15cm}
\begin{align}
   \frac{1}{{\theta}_{l}}  \frac{d{\theta}_{l}}{dt}  
& \approx \: 
   \frac{1}{{\theta}}  \frac{d{\theta}}{dt}
   \: - \: \frac{L_v}{c_{pd}\: T}   \frac{dq_l}{dt}
  \:  , \label{defTHB73bapprox} \\
& \approx \: 
   \frac{1}{T}  \frac{dT}{dt}   
   \: - \:  \frac{R_d}{c_{pd}\:p}  \frac{dp}{dt}
   \: - \: \frac{L_v}{c_{pd}\: T}   \frac{dq_l}{dt}
  \:  . \label{defTHB73bapprox2}
\end{align}
The temperature equation must be simplified too, with $c_p$ replaced by $ c_{pd}$ (as in B73) and with $1/\rho \approx R_d\:T/p$, leading to
\vspace{-0.15cm}
\begin{align}
 c_{pd} \: \frac{dT}{dt} & \approx \;\frac{R_d\:T}{p}\: \frac{dp}{dt}
                                  \; - L_v \: (\dot{q})_{eva} 
  \: . \label{def_DTapprox} 
\end{align}
The expected conservative property $d{\theta}_{l}/dt \approx 0$  is obtained with (\ref{def_Dl}) and (\ref{def_DTapprox}) inserted into (\ref{defTHB73bapprox2}).

The even more simple Deardorff's (1976) formula (\ref{defTHB73d}) is sometimes used for ${\theta}_{l}$, as in RW07.
\vspace{-0.15cm}
\begin{align}
  {\theta}_{l}  & \approx \: {\theta}  - \frac{L_v }{c_{pd}} \: q_l
  . \label{defTHB73d}
\end{align}
It is valid if the Exner function $\Pi = T / \: \theta $ is approximated by $1$ in the correction terms including $q_l$ (true for instance within a thin marine PBL, where $\theta \approx T$).

\subsection{The version of Tripoli and Cotton  (1981).} 
\label{section_mptvl2}

The ice-liquid water potential temperature ${\theta}_{il}$ is defined by Eqs.(26) and (28) in the paper TC81, starting from an integral of the Gibbs equation and with the same kind of approximations as in B73, with $L_v$ and $L_s$ considered as constant with $T$ and evaluated at $T_0$.
As suggested in the section 4 of Deardorff (1976), ${\theta}_{il}$ is a three phases generalization of ${\theta}_{l}$ that takes into account the impact of both $r_l$ and $r_i$, in order to be applied to the parameterization of the liquid-ice  cumulus and leading to
\vspace{-0.15cm}
\begin{align}
  {\theta}_{il}  & \equiv \: {\theta} \; 
        \exp \left( - \: \frac{L_v(T_0) \:r_l + L_s(T_0) \:r_i}{c_{pd}\: T} \right)
  , \label{defTHTC81a} \\
  {\theta}_{il}  & \approx \: {\theta} \; 
        \left( 1 - \frac{L_v(T_0) \:r_l + L_s(T_0) \:r_i}{c_{pd}\: T} \right)
  . \label{defTHTC81b}
\end{align}

\subsection{The version of Grenier and Bretherton  (1981).} 
\label{section_mptvl3}

The liquid-water virtual potential temperature ${\theta}_{vl}$ is defined in GB01 (section 3-b ; Appendixes~A and B) in terms of the two Betts variables (\ref{defTHB73c}) and (\ref{defQtB73}) alone.
\vspace{-0.15cm}
\begin{align}
  {\theta}_{vl}  & \equiv \: 
      {\theta}_{l}  \left( 1 + \delta \: q_t \right)
 \; \approx \;
     {\theta}  \left( 1 + \delta \: q_t  - \frac{L_v \:q_l}{c_{pd} \:T}\right)
  . \label{defTHGB01}
\end{align}
It is used in the measure of the buoyancy jump $g\:{\Delta}_i ({\theta}_{v}) /{\theta}_{v} $, with the approximation ${\Delta}_i ({\theta}_{v}) \approx {\Delta}_i ({\theta}_{vl})$ made in GB01 at the top of the PBL of the Stratocumulus.
It is also used in the computation of the top PBL entrainment velocity (see Eqs.(16), (18) and (B7) in the paper GB01).

\subsection{The version of Lilly  (1968).} 
\label{section_mptvl4}

The virtual potential temperature ${\theta}_{v}$ is defined in L68 by a differential equation (see Eq.(22) in that paper) and it is not based on a Gibbs equation.
The aim was to seek for a moist conservative thermodynamic variable in an atmosphere subject to  phase changes which would become a measure of buoyancy.
With the notations of the Appendix-A, it corresponds to (\ref{defTHL68a}) with the use of a mean reference value $\overline{\theta}$, leading to
\vspace{-0.15cm}
\begin{align}
  d {\theta}_{v}  & \approx \: d {\theta} 
                    + \overline{\theta} \: 
                      \left( \delta \:d q_v - d q_l \right)
  . \label{defTHL68a}
\end{align}
The virtual potential temperature ${\theta}_{v}$ is not explicitly computed in Lilly (1968).
It appears in the form of the vertical flux of it, namely $\overline{w'{\theta}_{v}'}$.
Indeed, if $\overline{\theta}$ is a constant term, (\ref{defTHL68a}) corresponds to
\begin{align}
  \overline{w'{\theta}_{v}'}  & \approx 
               \: \overline{w'{\theta}'} 
               \; + \; \overline{\theta} 
               \left( 
                 \delta \: \overline{w'{q}_{v}'} 
                         - \overline{w'{q}_{l}'} 
               \right)
  . \label{defTHL68a2}
\end{align}

The vertical flux of ${\theta}_{v}$ defined by (\ref{defTHL68a2})  is often used as a measure of the buoyancy fluxes, for instance in the moist thermal production $\beta \: \overline{w'{\theta}_{v}'}$, one of the terms acting in the turbulent kinetic energy equations (see  BR95 or CBR00, among others).
This buoyancy potential temperature is also used for the computations of the Bougeault and Lacarr\`ere (1989) non-local mixing length (see CBR00).

It is possible to define the Lilly's virtual potential temperature ${\theta}_{v}$ by integrating (\ref{defTHL68a}) for $\overline{\theta}$ considered as a constant term, then with $\overline{\theta}$ replaced by $\theta$, leading to
\vspace{-0.15cm}
\begin{align}
  {\theta}_{v}  & \equiv \: {\theta} \; \left( 1 +  \delta \:q_v - q_l \right)
  , \label{defTHL68b} \\
  {\theta}_{v}  & = \: {\theta} \; \left( 1 +  \delta \:q_t - \eta \:q_l \right)
  . \label{defTHL68b2}
\end{align}
It can be remarked that the actual temperature associated with ${\theta}_{v}$ corresponds to the ``density temperature'' denoted by  $T_{\rho} \: \approx \: {\theta}_{v} \:(p/p_0)^{\kappa}$ in E94.

\section{The moist entropy potential temperatures (HH87, M93, E94).} 
\label{section_mptthast}

\subsection{The version of Hauf and  H\"{o}ller (1987).} 
\label{section_mptthast1}

In the paper HH87, the specific entropy $s$ is defined by Eqs.(3.23) and (3.25) in terms of an entropy temperature denoted by ${\theta}_{S}^{\ast}$ hereafter.
It can be rewritten, with some algebra and with the notation of the Appendix-A, to give
\vspace{-0.15cm}
\begin{align}
          s        \: & \equiv \: \: 
                q_d \: s_d^0 
              + q_t \: s_l^0
              + q_d \: {c}^{\ast} \: 
               \ln\left({\theta}_{S}^{\ast} / T_0 \right)
  , \label{defTHsHH87}
\end{align}
where
\begin{align}
  {\theta}_{S}^{\ast} \:   & \equiv \: 
    T \:\left( \frac{p_d}{p_0}\right)^{- {R_d}/{{c}^{\ast}}}
        \left( \frac{e}{e_{ws}}\right)^{- {\:(r_v R_v)}/{{c}^{\ast}}}
        \left( \frac{e_{is}}{e_{ws}}\right)^{- {\:(r_i R_v)}/{{c}^{\ast}}}
        \exp \left(
                     \frac{L_v\:r_l - L_f\:r_i}{{c}^{\ast}\:T}
                \right)
  \: . \label{defTHast1HH87}
\end{align}
As noted in HH87, this formulation for ${\theta}_{S}^{\ast}$ supposes the existence of liquid water, at least implicitly, from the use of $s_l^0$ in (\ref{defTHsHH87}) and $c_{l}$ in the definition of ${c}^{\ast}$.
This can be a drawback, since it may not be true for the most general case of an arbitrary parcel or moist air, either saturated or under-saturated, with possibly only liquid water or only solid water.
As for the contribution due to $r_l$ into the exponential of (\ref{defTHast1HH87}), it  is a positive term contrary to what happens in ${\theta}_{l}$ or ${\theta}_{il}$.

\subsection{The Available Enthalpy version (1993).} 
\label{section_mptthast2}

Similarly to the method used in HH87, another moist entropy potential temperature is obtained in M93 as a by-product of the formulation for the moist exergy of an open atmospheric parcel.
It is denoted by ${\theta}^{\ast}$ and, as in HH87, it is directly derived in its analytic form starting from the 
general formulation for $s$,  the specific moist entropy of the system, written in M93 as
\vspace{-0.15cm}
\begin{align}
          s         & \equiv \: 
                q_d \: (s_d)_r 
              + q_t \: (s_v)_r
              + q_d \: {c}^{\ast}_p \: 
               \ln\left({\theta}^{\ast} / {\theta}^{\ast}_r \right)
  . \label{defTHs}
\end{align}
It is suggested in M93 to define the moist entropy potential temperatures ${\theta}^{\ast}$ and ${\theta}^{\ast}_r$ as
\vspace{-0.15cm}
\begin{align}
  {\theta}^{\ast}  \: & \equiv \: 
      T \left( \frac{p_d}{p_0}\right)^{- {R_d}/{{c}^{\ast}_p}}
        \left( \frac{e}{p_0}\right)^{- {\:(r_t R_v)}/{{c}^{\ast}_p}}
        \exp \left( - \:
                     \frac{L_v\:r_l + L_s\:r_i}{{c}^{\ast}_p\:T}
                \right)
  , \label{defTHast1} \\
  {\theta}^{\ast}_r \:  & \equiv \: 
      T_r \left( \frac{(p_d)_r}{p_0}\right)^{- {R_d}/{{c}^{\ast}_p}}
        \left( \frac{e_r}{p_0}\right)^{- {\:(r_t R_v)}/{{c}^{\ast}_p}}
  . \label{defTHast2}
\end{align}
The interest of writing $s$ in M93 by (\ref{defTHs}) as a complement to $q_d \: (s_d)_r + q_t \: (s_v)_r$ was to avoid the problem encountered in HH87, where the definition of ${\theta}_{S}^{\ast}$ by (\ref{defTHsHH87}) supposes the existence of liquid water or ice, with the use of the dry-air ($s_d^0$) and the liquid ($s_l^0$) standard values.
On the contrary, ${\theta}^{\ast}$ defined by (\ref{defTHs}) is valid for both under-saturated conditions ($r_l=r_i=0$) or saturated conditions ($r_l\neq 0$ or $r_i\neq0$), with only the dry-air and water-vapour reference values $(s_d)_r$ and $(s_v)_r$ involved, where $r_d$ and $r_v$ always exist in the atmosphere.

The exponential term in (\ref{defTHast1}) is almost the same as the one of B73, at least for the common liquid water part.
The difference with (\ref{defTHB73b}) is the term ${c}^{\ast}_p$, approximated by $c_{pd}$ in B73.
It is also similar to the exponential term of TC81 recalled in (\ref{defTHTC81a}), for both the liquid and the solid water parts. 
The term ${c}^{\ast}_p$ is approximated by $c_{pd}$ and with  $L_v(T) \approx L_v(T_0)$ and $L_s(T) \approx L_s(T_0)$.

Even if the purpose of HH97 was to show that modified versions of the Gibbs equation verified by ${\theta}_{S}^{\ast}$ could  lead to most of the potential temperature introduced in section \ref{section_mptvl}, the entropy temperatures ${\theta}_{S}^{\ast}$ and ${\theta}^{\ast}$ given by (\ref{defTHast1HH87}) and (\ref{defTHast1}) are not directly expressed with the usual notations, as is done for ${\theta}_{v}$, ${\theta}_{l}$, ${\theta}_{il}$ or ${\theta}_{vl}$.
It could be one of the reasons that have prevented ${\theta}_{S}^{\ast}$ or ${\theta}^{\ast}$ to be applied in most of subsequent meteorological studies.

In order to overcome this drawback, one of the purposes of the present paper is to rewrite ${\theta}^{\ast}$ in a more conventional way.

The equations for the dry air and the water vapour are $p_d = R_d \: {\rho}_d \: T$ and $e = R_v \: {\rho}_v \: T$.
The fraction $p_d/e$ is expressed in terms of $r_v=q_v/q_d={\rho}_v/{\rho}_d$, $\eta=R_v/R_d$ and $e=r_v \: \eta \:p_d$.
With $p=p_d+e$, the result is
\vspace{-0.15cm}
\begin{align}
  p_d   & = 
      \frac{1}{1+\eta\:r_v} \: p
  \: , \label{def0pd} \\
  e   & = 
      \frac{\eta\:r_v}{1+\eta\:r_v} \: p
  \: . \label{def0e}
\end{align}

When (\ref{def0pd}) and (\ref{def0e}) are inserted into (\ref{defTHast1}), the terms rearrange into
\begin{align}
  {\theta}^{\ast}   & \: = \:
      T \; \:\left( \frac{p}{p_0}\right)^{- R^{\ast}/{c}^{\ast}_p}\:\:
     \left[ \:
      \frac{{(1+\eta\:r_v)}^{R^{\ast}/{{c}^{\ast}_p}} }
           {{(  \eta\:r_v)}^{{(r_t R_v)}/{{c}^{\ast}_p}}}
     \: \right] \;
        \exp \left( - \:
                     \frac{L_v\:r_l + L_s\:r_i}{{c}^{\ast}_p\:T}
                \right)
  , \label{defTHast3}
\end{align}
provided that ${R}^{\ast} = R_d  + r_t \: R_{v}$, the companion of ${c}^{\ast}_p = c_{pd} + r_t \: c_{pv}$.

For the dry atmosphere $r_l=r_i=0$ and $r_v$ tends to zero. 
Therefore the exponential term is equal to $1$, ${R}^{\ast}/{c}^{\ast}_p$ has the limit $\kappa$, ${R}_v/{c}^{\ast}_p$ has the limit $\eta$ and the bracketed term has the limit $1$, since $r_v\:\ln(\eta \: r_v)$ has limit $0$ when $r_v$ tends to zero.
As a consequence ${\theta}^{\ast}$ has the correct dry-air limit ${\theta}$.

For the moist clear-air case $r_l=r_i=0$, $r_t=r_v$ and the exponential term is equal to $1$.
Nevertheless the bracketed term is different from $1$ and it can impact on ${\theta}^{\ast}$  not only in cloudy regions but also for the moist clear-air case, with ${\theta}^{\ast}$ different from the dry-air version ${\theta}$.

\subsection{The Emanuel's version (1994).} 
\label{section_mptthast3}

The liquid-water virtual potential temperature is defined in E94 starting from some approximated analytic definition of the entropy of moist air, considered as the sum of dry air, water vapour and liquid water components, with no ice content ($q_i=0$ and $q_t=q_v+q_l$).
It is assumed that
\vspace{-0.15cm}
\begin{align}
  s   & = 
        q_d\:s_d + q_v\:s_v + q_l\:s_l     \: , \label{def_s_E94a} \\
  s   & = 
        q_d\:s_d + q_t\:s_v 
      + q_l \left( \:s_l-s_v\: \right)        , \label{def_s_E94b} \\
  s_d   &  \approx \: s_d^{\diamond}  \: \equiv \:
       {c}_{pd} \: \ln(T) - R_d  \: \ln(p_d) \: , \label{def_sd_E94} \\
  s_v   &  \approx \: s_v^{\diamond}  \: \equiv \:
       {c}_{pv} \: \ln(T) - R_v  \: \ln(e) \: , \label{def_sv_E94} \\
  s_l   &  \approx \: s_l^{\diamond}  \: \equiv \: 
       {c}_{l} \: \ln(T)                   \: , \label{def_sl_E94} \\
  s^{\diamond}   & = 
        q_d\:s_d^{\diamond} + q_t\:s_v^{\diamond} 
      + q_l \left( \:s_l^{\diamond}-s_v^{\diamond}\: \right)    . \label{def_s_E94c}
\end{align}

 if (\ref{def_s_E94a}) and (\ref{def_s_E94b}) are exact definitions, the partial entropies $s_d^{\diamond} $ to $s_l^{\diamond} $ defined by (\ref{def_sd_E94}) to (\ref{def_sl_E94}) are only approximate formulae, because  additional standard values should be considered, leading for instance to the correct formula $ s_d = s_d^0 + {c}_{pd} \: \ln(T/T_0) - R_d  \: \ln(p_d/p_0)$ valid for the dry air component (with similar definitions for the water components).
If $T_0$ and $p_0$ are set to some prescribed values, the associated standard values $s_d^0$, $s_v^0$ and $s_l^0$ are constant terms.
They however impact on $s$ defined by (\ref{def_s_E94a}) or (\ref{def_s_E94b}) not only via the possibly conservative specific contents $q_d$ and $q_t$, but also for the non-conservative one $q_l$.
As a consequence, $s^{\diamond}$ defined by (\ref{def_s_E94c}) is not  equal to the entropy of moist air.

Nonetheless, with the use of the notation of the Appendix-A, when (\ref{def_sd_E94}) to (\ref{def_sl_E94}) are inserted into (\ref{def_s_E94c}), ${\theta}_{l}^{\ast}$ is defined in E94 (see Eq.(4.5.15), page 121) by
\begin{align}
  s^{\diamond}  &\:\equiv \:\: {c}^{\ast}_p \: 
               \ln\left({\theta}_{l}^{\ast} \right)
  , \label{defE94s}
\end{align}
where
\begin{align}
  {\theta}_{l}^{\ast}   & \: = \:
      T \;\: \left( \frac{p}{p_0}\right)^{- R^{\ast}/{c}^{\ast}_p}\:\:
     \left[ \:
      \frac{{(1+\eta\:r_v)}^{R^{\ast}/{{c}^{\ast}_p}} }
           {{(  \eta\:r_v)}^{{(r_t R_v)}/{{c}^{\ast}_p}}}
     \: \right] \;\;
     \left[
      \frac{{(  \eta\:r_t)}^{{(r_t R_v)}/{{c}^{\ast}_p}}}
           {{(1+\eta\:r_t)}^{R^{\ast}/{{c}^{\ast}_p}} }
     \right] \;
        \exp \left( - \:
                     \frac{L_v\:r_l}{{c}^{\ast}_p\:T}
                \right)
  \: . \label{defTHl_E94}
\end{align}
It can be remarked that $1/\eta$ is  denoted by $\varepsilon$ in E94, with $r_t=r_v+r_l$ in ${R}^{\ast}$ and ${c}^{\ast}_p$, also with $\chi=R^{\ast}/{c}^{\ast}_p$.

The definition  (\ref{defE94s}) is different from (\ref{defThetam0}), with no reference term included for the entropy or the potential temperature.
It is a consequence of the approximations  (\ref{def_sd_E94}) to (\ref{def_sl_E94}) where the reference values for the entropy are dropped.

It appears that, except the second bracketed term of (\ref{defTHl_E94}), Emanuel's formulation ${\theta}_{l}^{\ast}$ corresponds to ${\theta}^{\ast}$ given by (\ref{defTHast3}) with $r_i=0$.
This second bracketed term is an additional and arbitrary conservative quantity -- i.e. only constant if $r_t$ is a true constant --  introduced in E94 in order to get the formula (\ref{defTHl_E94b}), expressed with $\eta=1/\varepsilon$.
This additional bracketed term is another reason why Emanuel's potential temperature cannot represent the moist-air entropy.
\begin{align}
  {\theta}_{l}^{\ast}  & \:= \:
      T \;\:\left( \frac{p}{p_0}\right)^{- R^{\ast}/{c}^{\ast}_p}
     \:\:
      {\left(
           1- \frac{\eta\:r_l}{1+\eta\:r_t} 
       \right)}^{R^{\ast}/{{c}^{\ast}_p}} \: \:
           {\left( 1-\frac{r_l}{r_t} 
            \right)}^{{-(r_t R_v)}/{{c}^{\ast}_p}}\:\:
        \exp \left( - \:
                     \frac{L_v\:r_l}{{c}^{\ast}_p\:T}
                \right)
  . \label{defTHl_E94b}
\end{align}

\section{The new moist entropy potential temperatures ${\theta}_{s}$.} 
\label{section_defTHm}

The aim of the paper is the same as in HH87, namely ``to arrive at a definition of a moist potential temperature which could be regarded as a direct measure of the moist entropy'', not only for adiabatic and closed systems, but also for open systems where $q_d$ and $q_t$ are not conservative.

The problem encountered with the previous definitions for the moist entropies $s$, either for (\ref{defTHsHH87}), (\ref{defTHs}) or (\ref{defE94s}), is that $q_d$ or $q_t$ appear outside of the logarithm terms.
They appear explicitly in (\ref{defTHsHH87}) and (\ref{defTHs}).
They are also implicitly present  in (\ref{defE94s}), via ${c}^{\ast}$, ${c}^{\ast}_p$ and the mixing ratio $r_t$.
It results that ${\theta}_{l}^{\ast}$, ${\theta}_{S}^{\ast}$ and ${\theta}^{\ast}$ cannot represent all the variations of the moist air entropy $s$ if $r_t$ or $q_t$ vary.

It is possible to overcome this problem by transferring the varying specific contents $q_d=1-q_t$ and $q_t$ inside the logarithm, and to define ${\theta}_{s}$ as
\vspace{-0.15cm}
\begin{equation}
\boxed{\;\;
  \, s  \, \equiv \, (1-q_r)\:(s_d)_r 
                 + q_r\:(s_v)_r 
                 + {c}_{pd} \: 
               \ln\left(\frac{{\theta}_{s}}
                             {{\theta}_{sr}}
                        \right)
    , \,
 \;}
  \label{defTHmSm1}
\end{equation}
where $ {c}_{pd}$ is known and where $q_r$, $(s_d)_r$, $(s_v)_r$ and ${\theta}_{sr}$ are three constants to be determined.

The computation of the quotient ${\theta}_{s} / {\theta}_{sr}$ is presented in the Appendix-B.
It is suggested to define ${\theta}_{s}$ as
\begin{equation}
\left.
\boxed{\;\;
\begin{aligned}
  {\theta}_{s}   & \equiv 
        \: \theta 
        \; \exp \left( \Lambda\:q_t \right) \;
        \exp \left( - \:
                     \frac{L_v\:q_l + L_s\:q_i}{{c}_{pd}\:T}
                \right)
   \\
                    & \quad \times \;
        \left( \frac{T}{T_r}\right)^{\lambda \:q_t}
        \left( \frac{p}{p_r}\right)^{-\kappa \:\delta \:q_t}
     \left(
      \frac{r_r}{r_v}
     \right)^{\gamma\:q_t}
     \:
      \frac{(1+\eta\:r_v)^{\:\kappa (1+\:\delta \:q_t)}}
           {(1+\eta\:r_r)^{\:\kappa \:\delta \:q_t}}
  ,
\label{defTHm1}
\end{aligned} 
\;\;}
\right.
\end{equation}
where the reference potential temperature is written
\vspace{-0.15cm}
\begin{equation}
  \:\:{\theta}_{sr} \; \equiv \;
      T_r \left( \frac{p_0}{p_r}\right)^{\kappa}
     \:
       \exp \left(\Lambda\:q_r\right) \;
     \:
     \left(
     {1+\eta\:r_r}
     \right)^{\kappa}
    . \;
\label{defTHm2}
\end{equation}

The term $(1+\eta\:r_r)^{\:\kappa}$ is different from $1$ and it is put into (\ref{defTHm2}) -- instead of (\ref{defTHm1}) -- in order to fulfil the demand that $\theta_s$ must be equal to $\theta$ for the dry air case (i.e. for $q_t=q_v=q_l=q_i=0$ and $q_d=1$).
The term $\exp(\Lambda\:q_r)$ appears in (\ref{defTHm2}) in order to verify the expected property $\theta_{sr} = \theta_s(T_r, p_r, q_r ; q_l=q_i=0)$, which results from the choice of $(1-q_r)\:(s_d)_r + q_r\:(s_v)_r$ as a reference entropy in (\ref{defTHmSm1}), balancing the term $\exp (\Lambda\:q_r)$  in (\ref{defTHm2}).

Contrary to the potential temperatures ${\theta}_{il}$, ${\theta}_{S}^{\ast}$, ${\theta}^{\ast}$ and ${\theta}_{l}^{\ast}$ where only the mixing ratios are involved, the formula (\ref{defTHm1}) for $\theta_s$ is written in terms of the specific contents in the exponential term, as for ${\theta}_{v}$, ${\theta}_{l}$ and ${\theta}_{vl}$.

The main difference from all other formulations is the term $\exp(\Lambda\:q_t)$ in (\ref{defTHm1}), with \begin{equation}
\boxed{\;\;
\Lambda= \frac{(s_{v})_r - (s_{d})_r }{c_{pd}} \: .
\;\;}
\nonumber
\end{equation}
It is thus necessary to deal with the difference in the absolute values for the dry air and the water vapour reference partial entropies defined in HH87.

The value for $s$ in  (\ref{defTHmSm1}) is independent on any arbitrary choice for the reference temperature $T_r$,  pressures $p_r = (p_d)_r + e_r$ and specific contents  $(q_d)_r =1 -  q_r$.
However, another choice for $T_r$, $e_r$  and $q_r$ would modify the reference values $ (s_{v})_r$, $ (s_{d})_r$ and $ {\theta}_{sr}$, and also $\theta_s$ in (\ref{defTHm1}).
As a consequence, it will be important to choose accurately the reference values so that the variations of $\theta_s$ with $T$, $p$, $q_v$, $q_l$ and $q_i$ could be similar to the equivalent variations of $s$ (see the sensitivity experiments presented at the end of  section \ref{section_sens}).
\footnote{
\hspace*{1mm}This sentence published in 2011 must be updated.
Indeed, it is shown in Section~\ref{section_sens} that $\theta_s$ is independant on $T_r$ and $p_r$ (and thus on $e_r$, $r_r$ and $q_r$), as expected.
This result is valid if the reference state is defined by a just-saturated parcel of moist-air at $T_r$, with $e_r$ equal to $e_{sw}(T_r)$ or $e_{si}(T_r)$, depending on $T_r>T_0$ or $T_r<T_0$.
The moist-air entropy $s$ is thus fully determined by $\theta_s$ given by (\ref{defTHm1}).
The problem of choosing relevant values for $T_r$ and $p_r$ only concerns the first-order approximation $(\theta_s)_1$ defined in Section~\ref{section_aftha}.
}

The reference entropies $ (s_{v})_r$ and $ (s_{d})_r$ are determined at the temperature $T_r$ and at the partial pressures $(p_d)_r = p_r - e_r$  and $e_r$.
They are computed from the standard values $s_d^0$ and $s_v^0$ (see the Appendix-A),  by the use of  equivalent of  (\ref{defAPPsd}) and (\ref{defAPPsv}), yielding
\vspace{-0.15cm}
\begin{align}
  (s_d)_r  & = s_d^0
    \:+ c_{pd} \,\ln(T_r/T_0)
    \:- R_{d} \,\ln[\:(p_d)_r/p_0\:]
   \: ,
  \label{defSdr} \\
   (s_v)_r  & = s_v^0
    \:+ c_{pv} \,\ln(T_r/T_0)
    \:- R_{v} \,\ln(e_r/p_0)
   \: .
  \label{defSvr}
\end{align}

\section{First-order approximations for ${\theta}_{s}$.} 
\label{section_aftha}

For practical purposes, it would be interesting to write a simple version for (\ref{defTHm1}), with the use of the  first order approximation $\exp(x)\approx 1+x$ valid for the exponential terms and for small values of $x$, as used before to derive  (\ref{defTHB73c}) from (\ref{defTHB73b}) in B73 and (\ref{defTHTC81b}) from (\ref{defTHTC81a}) in TC81.
The other power terms of the form $a^b$ will be rewritten as $\exp[b\ln(a)]$ and they will be approximated by $1+b\ln(a)$ for small values of $b\ln(a)$.
All the products $(1+x)(1+y)$ will be approximated by $1+x+y$ for small values of $x$ and $y$ (for instance equal to $q_v$, $q_t$ or $q_r$), with second order terms like $x\:y$ discarded.

As indicated in (\ref{defTHast4a}), the moist entropy potential temperature ${\theta}_{s}$ can be written as a sum of two terms.
\vspace{-0.15cm}
\begin{align}
  {\theta}_{s}   
  & \approx \; 
    ({\theta}_{s})_1 
        \: + \: 
    ({\theta}_{s})_2 
    \: . \label{defTHast4a}
\end{align}
The first term  $ ({\theta}_{s})_1$ is given by the first line of (\ref{defTHm1}), leading to the expressions (\ref{defTHast4b0}) to (\ref{defTHast4b2}).
It will be shown in the section \ref{section_sens} that $({\theta}_{s})_1$ is indeed the leading order term of  ${\theta}_{s}$ for the Stratocumulus cases in FIRE-I.
\begin{align}
 ({\theta}_{s})_1
  & \: = \: 
    \theta_l
    \;
    \exp\! \left[ \:  
          \Lambda \: q_t
     \: \right]
   , \label{defTHast4b0}
\\
 ({\theta}_{s})_1
  & \: = \: 
    \theta 
    \:
    \exp\! \left[ \:  
          \Lambda \: q_t
     \: \right]
    \: \: 
    \exp\! \left[
       -\: 
          \frac{L_v\:q_l + L_s\:q_i}
               {{c}_{pd}\:T} \: 
     \right]
   , \label{defTHast4b1}
\\
 ({\theta}_{s})_1
  & \: = \: 
     \theta \; 
    \exp\! \left[
       \:    
          \Lambda \: q_t        
       \: - \:
          \frac{L_v\:q_l + L_s\:q_i}
               {{c}_{pd}\:T} \: 
     \right]
   . \label{defTHast4b2}
\end{align}
The second term $ ({\theta}_{s})_2$ is given by (\ref{defTHast4c}). 
It is derived from a leading order approximation of the remaining part of (\ref{defTHm1}), i.e. the second line, valid for small values of $q_t$ and $r_v$.
\begin{align}
 ({\theta}_{s})_2
  & \:\approx \; 
     ({\theta}_{s})_1  \:
     \left[
          \; \eta\:\kappa\:r_v
       \: - \: \gamma\:\ln\!\left(\frac{r_v}{r_r}\right)\:q_t
          + \: \lambda\:\ln\!\left(\frac{T}{T_r}\right)\:q_t
          - \: \kappa \:\delta \:\ln\!\left(\frac{p}{p_r}\right)\:q_t
     \: \right]
  . \label{defTHast4c}
\end{align}
\begin{align}
 ({\theta}_{s})_1
  & \:\approx \;
     {\theta}_{l}\:\:
     \left[ \: 1 
       \: + \:    
          \Lambda \: q_t       
    \:  
     \right]\:
   ,  \label{defTHast4bl} \\
 ({\theta}_{s})_1
  & \:\approx \;
     \theta \:
     \left[ \: 1 
       \: + \:    
          \Lambda \: q_t       
       \: - \:
          \frac{L_v\:q_l + L_s\:q_i}
               {{c}_{pd}\:T} \:  
     \right]\:
   .  \label{defTHast4b}
\end{align}

All the formulae (\ref{defTHast4b0}) to (\ref{defTHast4b2}), (\ref{defTHast4bl}) or (\ref{defTHast4b}) valid for the first term $({\theta}_{s})_1$ contain the term $\Lambda \: q_t $.
It is an extra term in comparison with the liquid-water (B73) and the ice-liquid water potential temperatures (TC01), recalled in (\ref{defTHB73b}), (\ref{defTHB73c}), (\ref{defTHTC81a}) and (\ref{defTHTC81b}).
The ice component $ L_s(T)\:q_i$ is the logical complement to B73's formula,  with the latent heat $L_v$ and $L_s$ expressed for the actual temperature $T$, and not at $T_0$ as in the TC01's formula.
Also, the formulations (\ref{defTHast4b0}), (\ref{defTHast4b2}) or (\ref{defTHast4b}) are similar to the GB01's formulation (\ref{defTHGB01}), with $\delta$ replaced by $\Lambda$.

The formula (\ref{defTHast4c}) for the second term $({\theta}_{s})_2$  can always be computed because both $- \: \gamma\:\ln(r_v/r_r )\:q_t$  and $- \: \kappa\:\delta\:\ln(p/p_r)\:q_t$ has limit $0$ as $q_t$  tends to zero, providing that $q_t$ decreases more rapidly than $\ln(r_v)$ and $\ln(p)$.

\section{The conservative properties verified by ${\theta}_{s}$.} 
\label{section_cpftha}

The three entropy potential temperatures ${\theta}_{S}^{\ast}$, ${\theta}^{\ast}$, and ${\theta}_{s}$ verify conservative properties if $q_d$, $q_t$ and $r_t$ are constant, whatever the possible reversible exchanges existing between the vapour, liquid or solid water species may be.

These properties are not easy to prove starting directly from (\ref{defTHast1HH87}), (\ref{defTHast3}) or (\ref{defTHm1}), where changes in $r_l$ and $r_i$ must be carefully analysed.
It is much easier to analyse the corresponding moist entropy definitions (\ref{defTHsHH87}),  (\ref{defTHs}) or  (\ref{defTHmSm1}), because all the terms except ${\theta}_{S}^{\ast}$, ${\theta}^{\ast}$  and ${\theta}_{s}$ only depend on $q_t$ or $r_t$ which results in a partial conservative feature for the moist potential temperatures, only valid for constant values of $q_t$ and $r_t$ and only if the moist entropy $s$ is a constant for adiabatic and reversible processes occurring within a closed parcel of fluid.

The same partial conservative property is verified by ${\theta}_{l}^{\ast}$ if $q_d$, $q_t$ and $r_t$ are constant.
Even if (\ref{defTHl_E94}) is based in E94 on the approximate moist entropy $s^{\diamond}$ given by (\ref{def_s_E94c}), different from the true moist entropy (\ref{def_s_E94a}), the definition (\ref{defTHl_E94}) for ${\theta}_{l}^{\ast}$ only differs from the one (\ref{defTHast3}) for ${\theta}^{\ast}$ by the aforementioned second bracketed term of (\ref{defTHl_E94}), and this bracketed term only depends on $r_t$, from which the same partial conservative property holds for ${\theta}_{l}^{\ast}$.

A more general conservative property is verified by ${\theta}_{s}$ for a region where the entropy is well-mixed, either by diffusion, turbulent, convective or dynamical processes.
In that case, for constant values of $s$ given by (\ref{defTHmSm1}), ${\theta}_{s}$ defined by (\ref{defTHm1}) is also a constant even if $q_d$, $q_t$ and $r_t$ vary in the vertical or in the horizontal.
A more precise analysis is derived in the Appendix~C.

\section{Numerical evaluations: the FIRE-I dataset.} 
\label{section_nefd}

\subsection{The entropy for the flights RF03B, 02B, 04B, 08B.} 
\label{section_entr03B}

The exact and approximate versions of ${\theta}_{s}$ are analysed with the aircraft observations of the Stratocumulus boundary layer during the First ISCCP Regional Experiment (FIRE I), performed off the coast of Southern California in July 1987.
As in RW07, ``the mean values computed from the aircraft data may be loosely interpreted as typical grid-box mean values in a general circulation model and the standard deviation as a measure of the sub-grid variability''.

The aircraft  measurements of the temperature, the water vapour concentrations and the liquid water content are not local ones.
They are at least averaged during the radial flights with a $100$~m sampling or so.
However, these sampling aircraft observations will be considered as ``local'' measures hereafter, for the temperature and the specific contents (water vapour and liquid water).
The local measures are conditionally averaged in this study following the RW07's method, by separating the in-cloud from the clear-air conditions with the threshold $q_l>0.01$~g/kg.

As in RW07, the average values are computed within fixed height intervals with a depth $\Delta z=25$~m.
Unknown instrumental errors impact on the accuracy of all the data.
It has been decided to correct two of them, with partial removal of the oversaturated or unsaturated in-cloud regions.
The water vapour specific content $q_v$ will be modified if the measured liquid water is above a critical value $(q_l)_c$.
In that case $q_v$ is set to its saturation value $q_{sw}(T)$ (personal communication of J.L. Brenguier).
It is also ensured that $q_v \leq q_{sw}(T)$.
These corrections may have not been done in RW07 and they can explain the small differences from RW07 results.
Another difference with RW07 is the use of the exact definition for the Betts' potential temperature (\ref{defTHB73b}) in the present study, whereas the Deardorff's formulation (\ref{defTHB73d}) is used in RW07.

According to several tests discussed later at the end of section \ref{section_sens},  the reference values have been set to $T_r=T_0=273.15$~K, $p_r = p_0 = 1000$hPa, $\: e_r = e_{ws}(T_0) \approx 6.11$~hPa and $r_{r} \equiv \varepsilon\:e_r / (p_r - e_r) \approx 3.82$~g~kg${}^{-1}$.
The corresponding constant $\Lambda \approx 5.87$  is obtained with $(s_{v})_r$ and $(s_{d})_r$ given by (\ref{defSdr}) and (\ref{defSvr}).

\begin{figure}[hbt]
\centering
\includegraphics[width=0.5\linewidth,angle=0,clip=true]{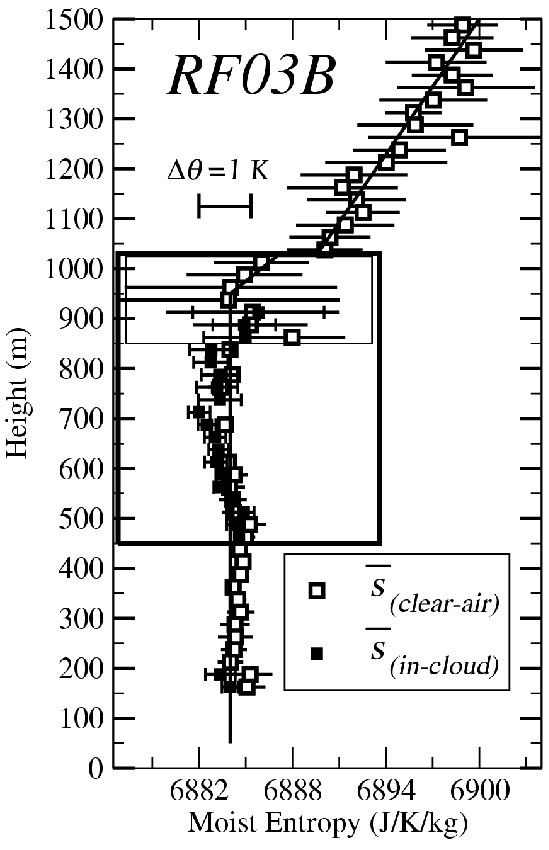}
\caption{\em\small The in-cloud (dark square) and the clear-air (open square) vertical profiles for the average moist entropy $\overline{s}$, depicted for the flight RF03B (2nd of July, 1987).
The large rectangular boxes represent the cloud region (heavy line) and the smaller rectangular boxes represent the top-PBL entrainment zone (thin line), with the same definitions and values for the top-PBL height and the Free-Air base height as the ones published in RW07.
The horizontal bars indicate one standard deviation from the mean values, with small vertical lines at the end of the in-cloud bars.
Other informations are available in the text (for $\Delta \theta = 1$~K and for the sketch profile denoted by a thin solid segment lines).
\label{fig_entropy}}
\end{figure}

\begin{figure}[hbt]
\centering
\includegraphics[width=0.32\linewidth,angle=0,clip=true]{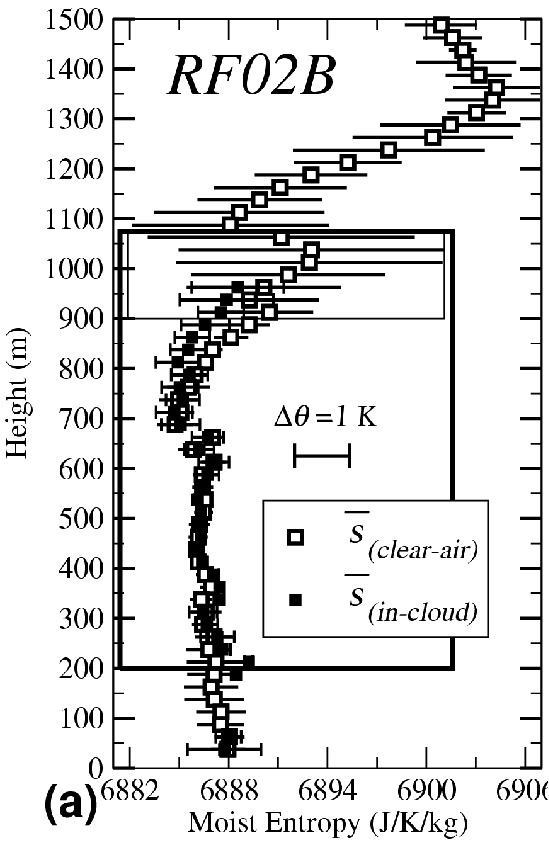}
\includegraphics[width=0.32\linewidth,angle=0,clip=true]{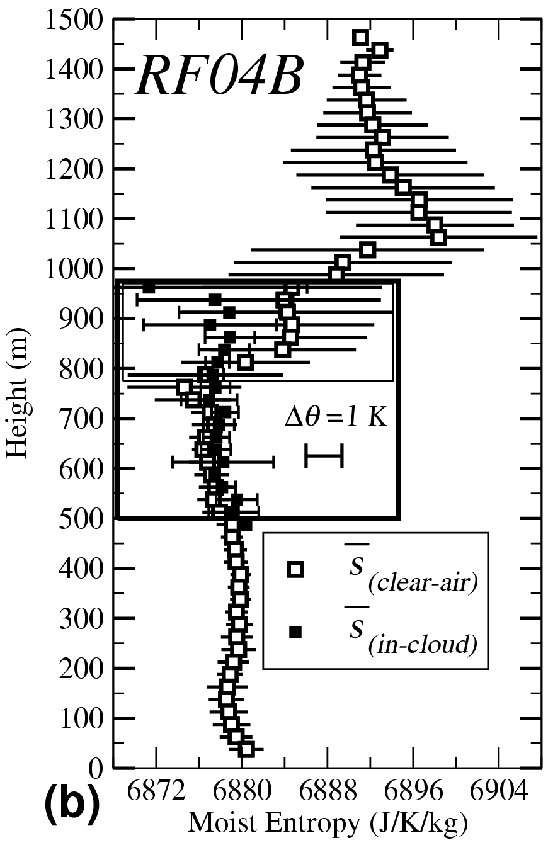}
\includegraphics[width=0.32\linewidth,angle=0,clip=true]{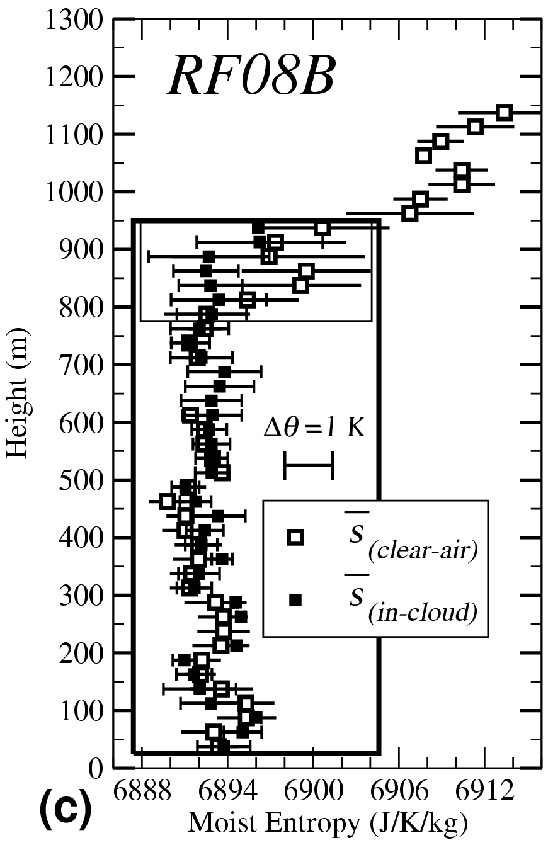}
\caption{\em\small The same as on Fig.(\ref{fig_entropy}), except in  (a) for the flight RF02B (30th of June, 1987), in (b) for the flight  RF04B (5th of July, 1987) and in (c) for the flight RF08B (14th of July, 1987).
\label{fig_entropy2}}
\end{figure}

The $\Delta z=25$~m average values of the moist entropy $\overline{s}$ are depicted  for the flight RF03B in  Fig.(\ref{fig_entropy}).
They are evaluated from (\ref{defTHmSm1}), with ${\theta}_{s}$ and ${\theta}_{sr}$ given by (\ref{defTHm1}) and (\ref{defTHm2}) and with the averaging operator derived in the Appendix~D.

The important result is that, for a given level, the clear-air and the in-cloud values have the same moist entropy, with the standard deviations of the two conditionally averaged subsets crossing over.
Moreover, the moist entropy is almost constant up to $1050$~m or so, including the entrainment region.

In order to make the comparison easier with the usual jumps in ${\theta}_{l}$ of more than $8$ to $10$~K, a width $\Delta \theta = 1$~K is plotted, indicating the small impact in terms of a change in entropy associated with a change in potential temperature from $300$ to $301$~K, leading to $c_{pd}\ln(301/300)\approx 3.34$~J/K/kg.

It appears that the entrainment region is characterized by the largest standard deviations of the PBL, for both the clear-air and the in-cloud conditions.
It could be interpreted as an increase in the sub-grid variability for $\overline{s}$ with a possible partial mixing in moist entropy in the entrainment region, where the moist PBL air and the dry-air above entrain or possibly detrain (see RW07).

A series of (solid) line segments are plotted in  Fig.(\ref{fig_entropy}).
They form a sketch profile for $\overline{s}$, with a constant value of $6884$~J/K/kg plotted up to $950$~m corresponding to a full mixing of $\overline{s}$ within the PBL.
It is observed that the top-PBL mixing is realized with no obvious inversion jump in moist entropy, or corresponding to a possible small jump of less than $1.5$~K in potential temperature.
There is a linear trend above the top-PBL height ($1025$~m), due to the impact of the radiation and to the subsidence processes.

All these results suggest that the moist PBL is homogeneous in $\overline{s}$, with a continuous transition with the dry-air above.
As a consequence, $\overline{s}$ could be an interesting candidate for being a true conservative variable to be used somehow in atmospheric turbulent schemes, where no vertical mixing in $\overline{s}$ may result in zero turbulent tendencies (for all the clear-air, in-cloud or grid-cell average parts).

The properties suggested by the analyses of the moist entropy computed for flight RF03B can be strengthened with the  same analyses applied to the three other flights, as shown in  Fig(\ref{fig_entropy2}).
Even if the differences in the average values for the clear-air and the in-cloud subsets are larger in the top PBL entrainment regions for the flights RF04B and RF08B, the average values of one subset are located within the horizontal bars of the other.
The conclusion is that the clear-air and the in-cloud subsets seems to have almost the same moist entropy for all the  FIRE-I data flights, with a common value for $\overline{s}$ almost constant within the PBL and with a smooth transition occurring with the dry subsiding air located above the PBL regions.

\subsection{Other parameters for the flight RF03B.} 
\label{section_other03B}

\begin{figure}[hbt]
\centering
\includegraphics[width=0.450\linewidth,angle=0,clip=true]{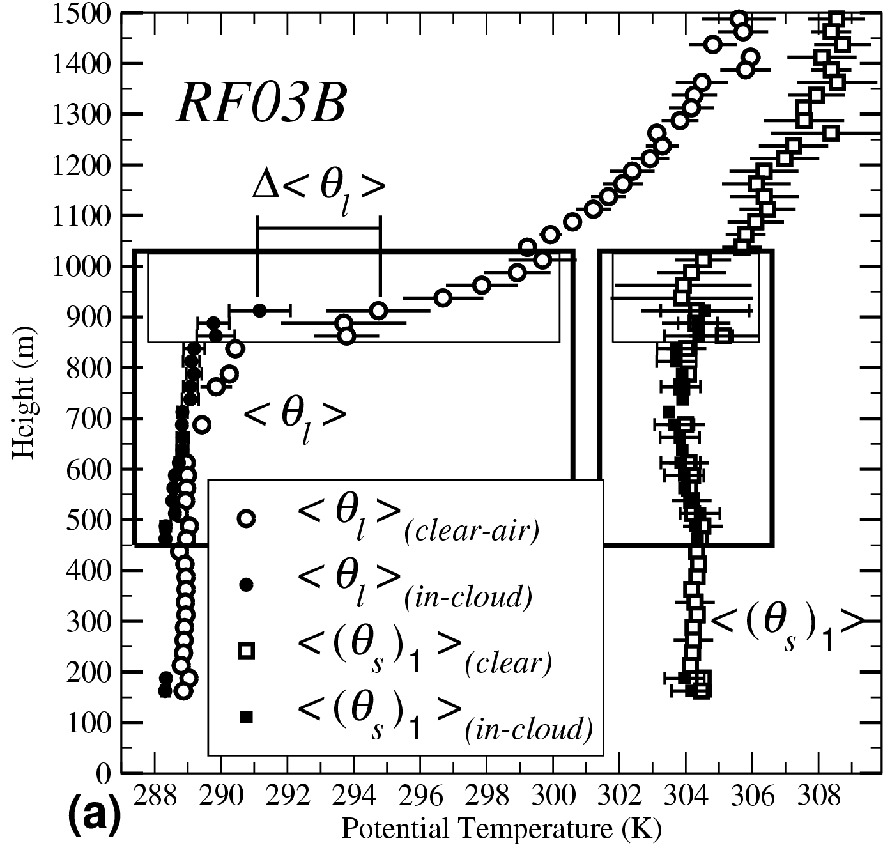}
\includegraphics[width=0.285\linewidth,angle=0,clip=true]{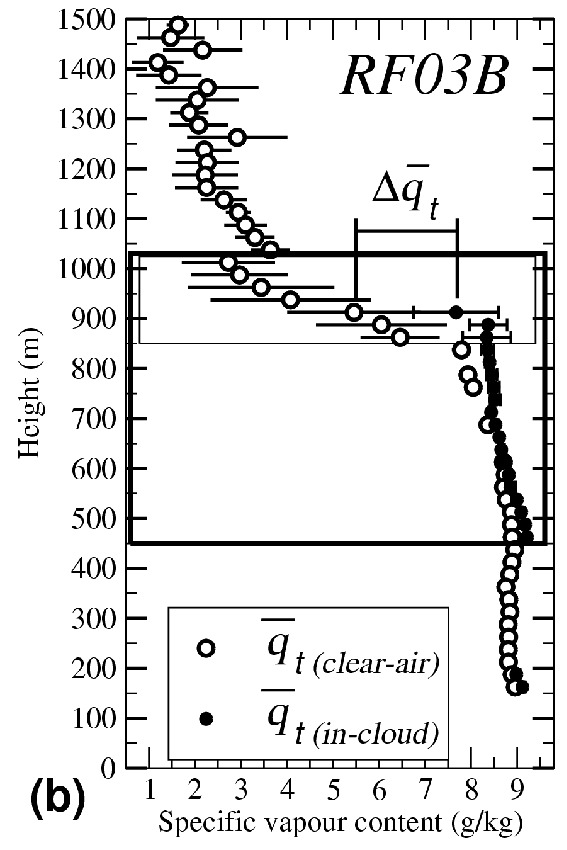}
\includegraphics[width=0.235\linewidth,angle=0,clip=true]{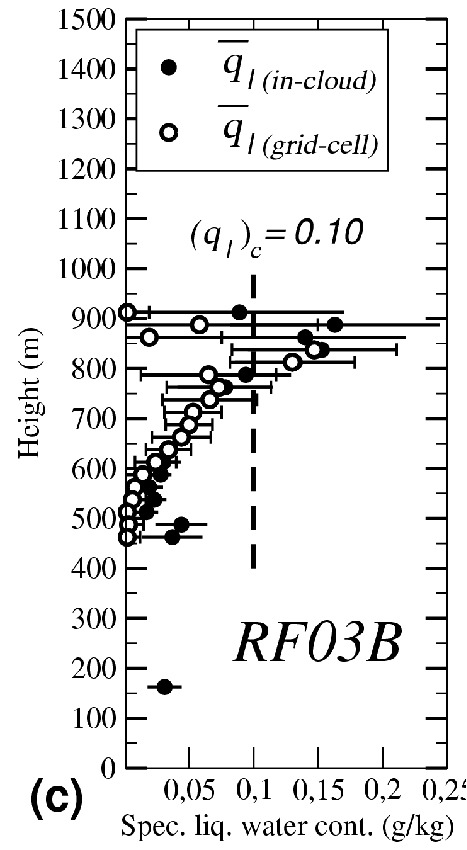}
\caption{\em\small For the flight RF03B (2nd of July, 1987). 
(a) The in-cloud (dark circle or dark square) and the clear-air (open circle or open square) mean values for the moist potential temperatures $<\!\theta_l\!>$ (left) and $<\!(\theta_s)_1\!>$ (right), with $\theta_l$ and $(\theta_s)_1$  computed with (\ref{defTHB73b}) and (\ref{defTHast4b2}).
(b) The in-cloud (dark circle) and the clear-air (open circle) mean values for the total water specific contents $\overline{q_t}$.
(c) The in-cloud (dark circle) and the grid-cell (open circle) mean values for the liquid water specific content $\overline{q_l}$.
The threshold $(q_l)_c$ is represented by a vertical dashed line, above which $q_v$ is set to its saturation value $q_{sw}(T)$.
See the comments in Fig.(\ref{fig_entropy}) concerning the rectangular boxes and the standard deviation horizontal bars.
\label{fig_Ths_Qt_Ql_RF03B}}
\end{figure}

The average values for the two moist potential temperatures $<\!\theta_l\!>$ and $<\!(\theta_s)_1\!>$, the specific total water contents $\overline{q_t}$  and the liquid water content $\overline{q_l}$  are depicted in  Fig.(\ref{fig_Ths_Qt_Ql_RF03B})  for the flight RF03B.
The values of $\theta_l$ and $(\theta_s)_1$ are computed with the exponential expressions  (\ref{defTHB73b}) and (\ref{defTHast4b2}), respectively.
The panel (c) for the liquid water content shows that RF03B corresponds a thin layer and homogeneous Stratocumulus.

The shape of the vertical profiles of $<\!(\theta_s)_1\!>$ in  Fig.(\ref{fig_Ths_Qt_Ql_RF03B})  (a)  is  close to the one observed for $\overline{s}$ in Fig.(\ref{fig_entropy}).
It confirms that, at least for this case and for the aforementioned set of reference values, $<\!(\theta_s)_1\!>$ is indeed a relevant synonym for $\overline{s}$.
It is not true for the B73's mean values $<\!{\theta}_{l}\!>$ in Figs.(\ref{fig_Ths_Qt_Ql_RF03B}) (a) and for $\overline{q_t}$  in  (b), for which linear trends exist in the PBL ($+1$~K and $-1$~g/kg from the surface to $850$~m, even much larger in the cloud and the entrainment region).

Large values are observed for the differences in  $<\!{\theta}_{l}\!>$ between the clear-air and the in-cloud regions,  denoted by $\Delta\!<\!{\theta}_{l}\!>$.
They increase with height, reaching about $4$~K in the entrainment region, as indicated in  Fig.(\ref{fig_Ths_Qt_Ql_RF03B}) (a).
There is an associated decrease with height of $\Delta\:\overline{q_t}$ in the entrainment region, with $\Delta\:\overline{q_t} \approx -2$~g/kg at the top of the entrainment region, as indicated in  Fig.(\ref{fig_Ths_Qt_Ql_RF03B}) (b).

The clear-air values of  $<\!{\theta}_{l}\!>$ are $4$~K  warmer than the in-cloud ones.
They lead to a difference of $1.3$~\% or so.
The term $\exp(\Lambda\:q_t)$ corresponds to an opposite impact of the order of $-1.2$~\%.
Since the  liquid water term $\overline{q_l}$ depicted in Fig.(\ref{fig_Ths_Qt_Ql_RF03B})  (c) gives the same contribution for $<\!{\theta}_{l}\!>$ as for $<\!(\theta_s)_1\!>$, the almost opposite numerical impacts of $\pm1.2$~\% explain how the new term $\exp(\Lambda\:q_t)$ acts in (\ref{defTHast4b0}) to (\ref{defTHast4b2}) in order to make $<\!(\theta_s)_1\!>$ constant with height and to give the same clear-air and in-cloud values.

Large jumps in $<\!\theta_l\!>$ and $\overline{q_t}$ are  observed within the entrainment region in  Fig.(\ref{fig_Ths_Qt_Ql_RF03B})  (a) and (b).
They are in agreement with the values indicated in RW07 for this flight  ($10.1$~K and $-4.9$~g/kg).
As for $\overline{s}$ or $<\!(\theta_s)_1\!>$,  the entrainment region is characterized for $<\!\theta_l\!>$ and $\overline{q_t}$ by larger standard deviations and may be interpreted as an increase in sub-grid variability.

The jump in $<\!(\theta_s)_1\!>$ is much smaller than the one for $<\!{\theta}_{l}\!>$ (i.e. $1$~K to $2$~K  versus $10.1$~K), or possibly does not exist.

For a given level, the standard deviation bars of the clear-air and in-cloud conditionally averaged subsets do not cross over for $<\!\theta_l\!>$ and $\overline{q_t}$.
It seems that the clear-air and the in-cloud values cannot be considered as equal for $<\!\theta_l\!>$ and $\overline{q_t}$, in contrast with the result obtained with $\overline{s}$ and $<\!(\theta_s)_1\!>$.

\subsection{All parameters for the flight RF02B, 04B, 08B.} 
\label{section_paramALL}

\begin{figure}[hbt]
\centering
\includegraphics[width=0.450\linewidth,angle=0,clip=true]{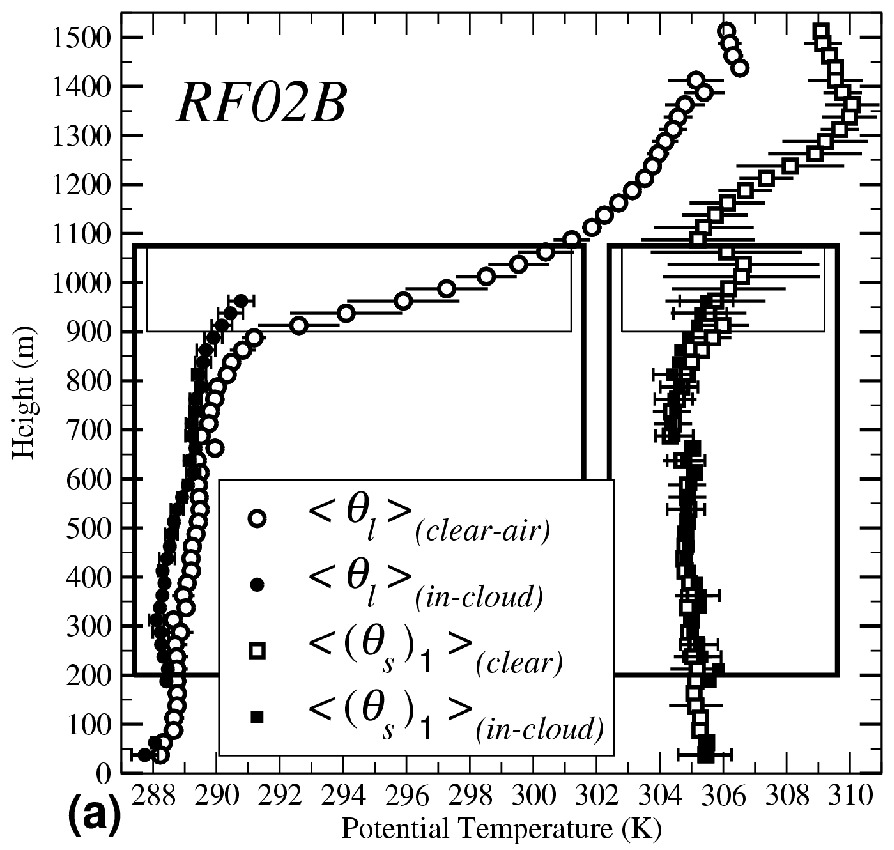}
\includegraphics[width=0.285\linewidth,angle=0,clip=true]{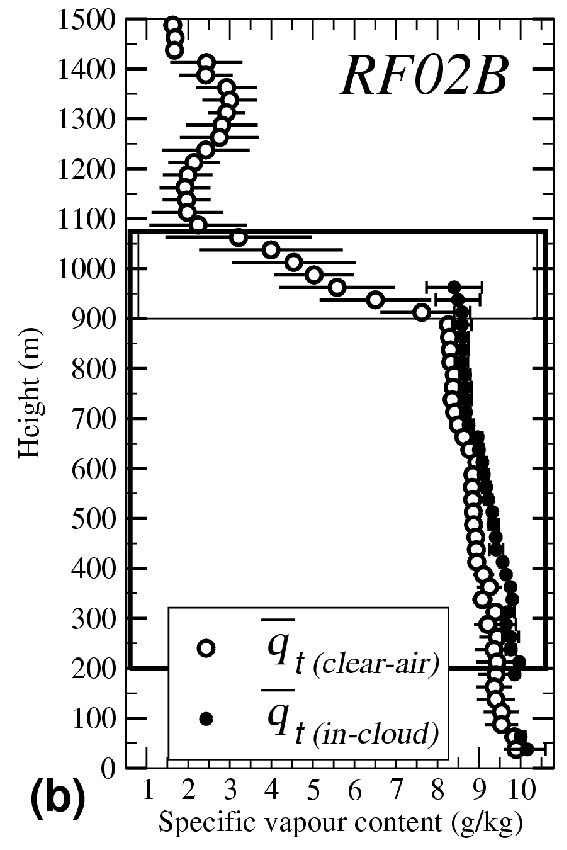}
\includegraphics[width=0.235\linewidth,angle=0,clip=true]{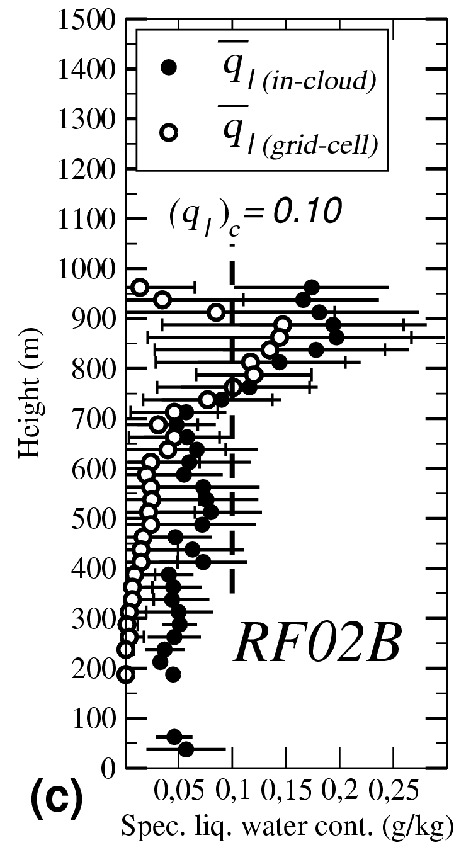}
\caption{\em\small The same as on Fig.(\ref{fig_Ths_Qt_Ql_RF03B}) but for the flight RF02B (30th of June, 1987).
\label{fig_Ths_Qt_Ql_RF02B}}
\end{figure}

\begin{figure}[hbt]
\centering
\includegraphics[width=0.450\linewidth,angle=0,clip=true]{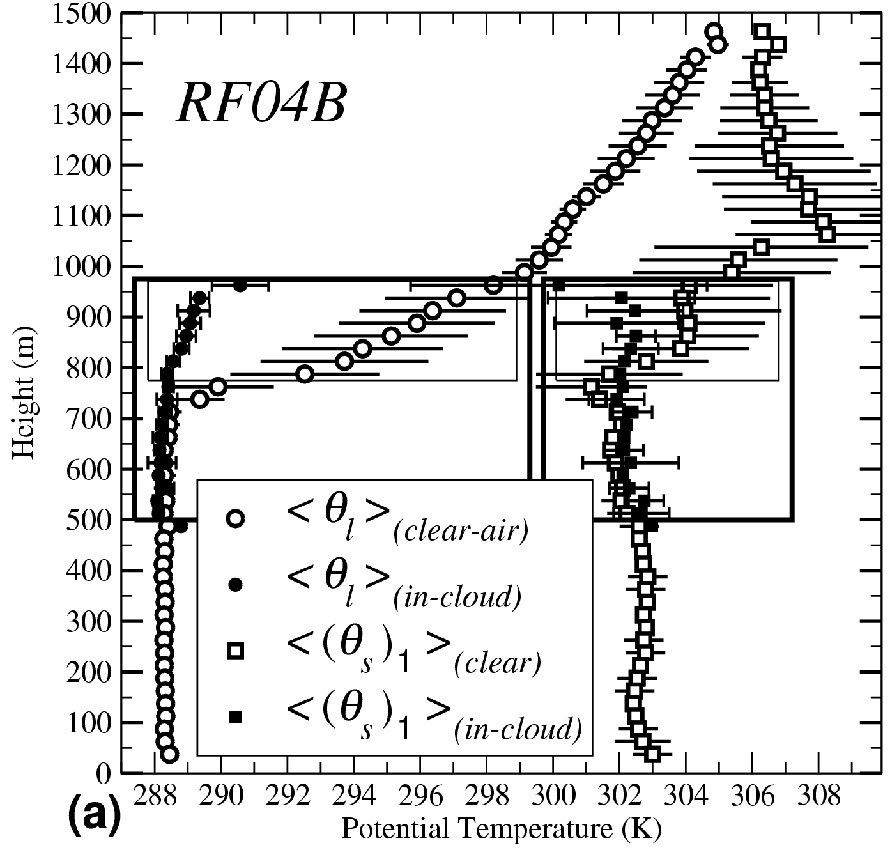}
\includegraphics[width=0.285\linewidth,angle=0,clip=true]{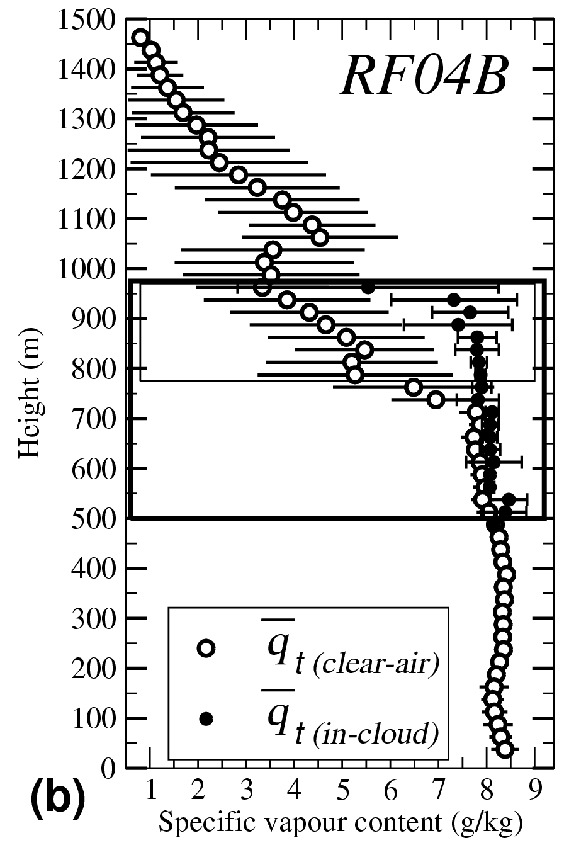}
\includegraphics[width=0.235\linewidth,angle=0,clip=true]{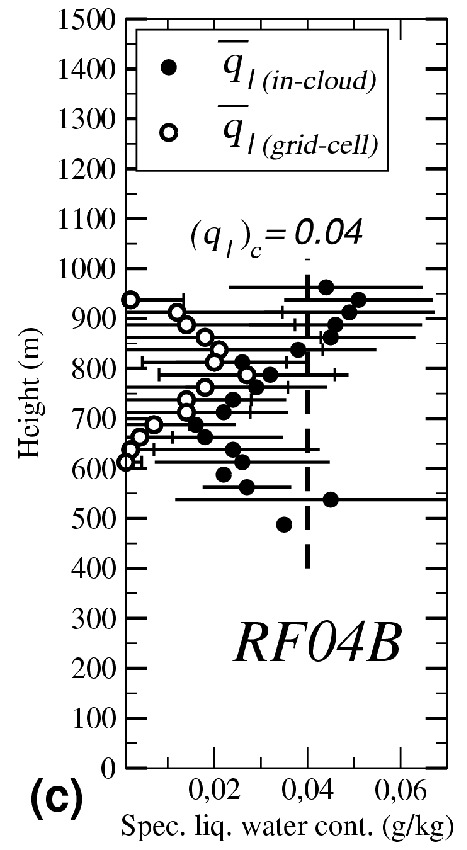}
\caption{\em\small The same as on Fig.(\ref{fig_Ths_Qt_Ql_RF03B}) but for the flight RF04B (5th of July, 1987).
\label{fig_Ths_Qt_Ql_RF04B}}
\end{figure}

\begin{figure}[hbt]
\centering
\includegraphics[width=0.450\linewidth,angle=0,clip=true]{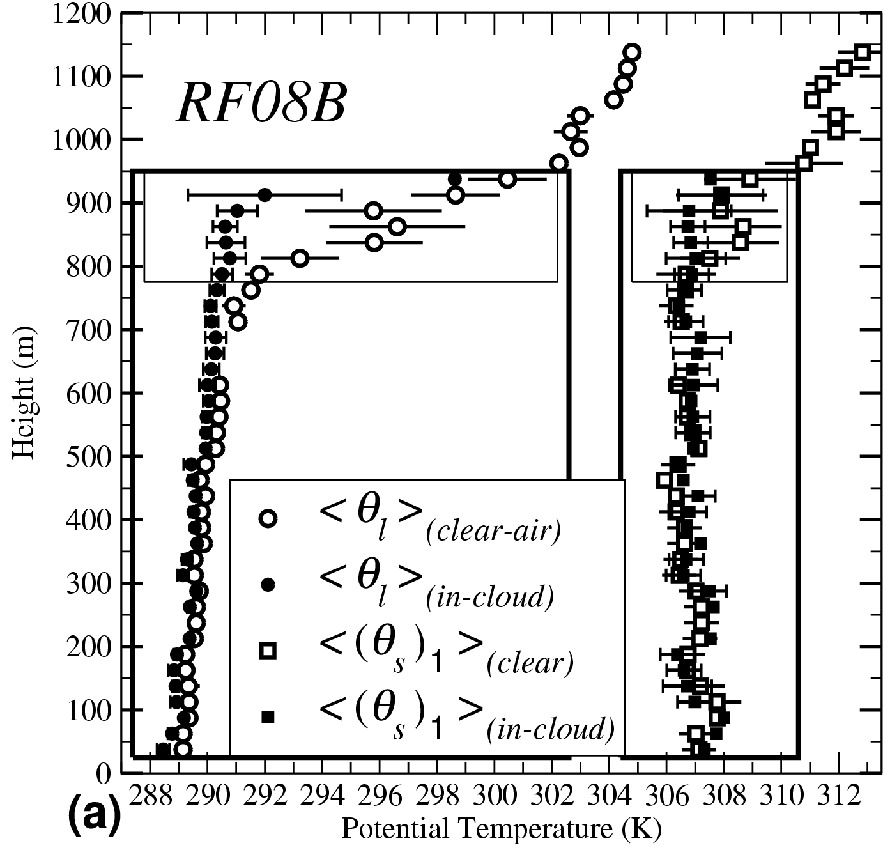}
\includegraphics[width=0.285\linewidth,angle=0,clip=true]{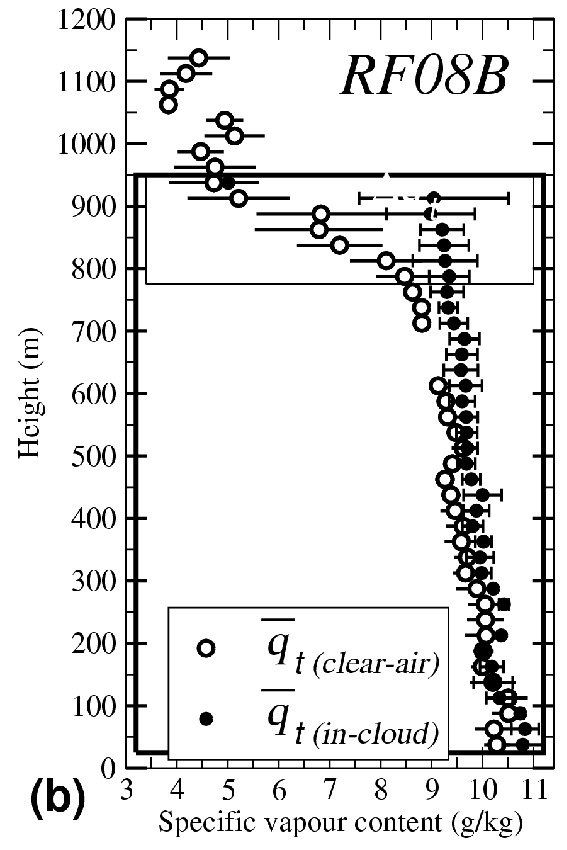}
\includegraphics[width=0.235\linewidth,angle=0,clip=true]{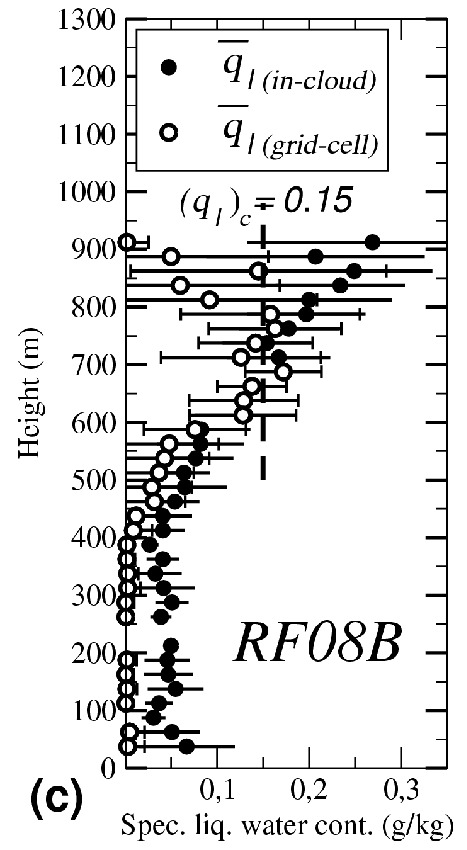}
\caption{\em\small The same as on Fig.(\ref{fig_Ths_Qt_Ql_RF03B}) but for the flight RF08B (14th of July, 1987).
\label{fig_Ths_Qt_Ql_RF08B}}
\end{figure}

Other computations made for the flights RF02B, RF04B and RF08B are presented in  Figs.(\ref{fig_Ths_Qt_Ql_RF02B}) to (\ref{fig_Ths_Qt_Ql_RF08B}).
The clear-air and the in-cloud values of $\overline{{\theta}_{l}}$, $\overline{q_t}$ and $\overline{q_l}$ are  similar to the corresponding results shown in RW07.
The panels (c) for the liquid water content show that RF04B corresponds a thin layer and heterogeneous Stratocumulus, whereas RF02B and RF08B correspond to thick layers and rather heterogeneous clouds (liquid water exists in almost the whole PBL).

The same properties observed for the flight RF03B are verified by the other ones.
In particular, the vertical profiles of $<\!(\theta_s)_1\!>$  are almost constant within the whole PBL, including the entrainment regions, especially for the flight RF08B.
Also, in contrast with the large differences observed with  $\overline{{\theta}_{l}}$ , the values for $<\!(\theta_s)_1\!>$  are almost equal in clear-air and in-cloud conditions, with the same impact found for the term $\exp(\Lambda\:q_t)$ for the three flights.
The impacts are $\pm 1.7$~\% for RF02B, $\pm 2.3$~\% for RF08B, with a partial balance of $+2.7$~\%  and $-2.1$~\%  for the flight RF04B (however, the standard deviations of the two conditionally averaged subsets also cross over for this flight RF04B, indicating that the difference may not be significant).

It can be noted that the standard deviations in the clear-air above the top PBL are much larger for $<\!(\theta_s)_1\!>$ than for $<\!{\theta}_{l}\!>$ for the flight RF04B.
It is an impact of the high level of sub-grid variability existing for $\overline{q_v} = \overline{q_t} - \overline{q_l}$ in this flight, with an influence on $<\!(\theta_s)_1\!>$ only and with no impact on $<\!{\theta}_{l}\!>$.

The variation with height of $<\!(\theta_s)_1\!>$ for the flights RF04B and RF02B and above the top PBL height is more complex than for RF03B.
The vertical gradients of $<\!(\theta_s)_1\!>$ are largely influenced (may be dominated) by the vertical gradients of $\overline{q_v}=\overline{q_t}$.
The almost constant values for $\overline{q_v}$ depicted for the flights RF03B and RF08B above the top PBL height can explain the linear positive trend observed for these flights, where the increase in $<\!(\theta_s)_1\!>$ follows the  increase in $<\!\theta\!>$.

\subsection{The grid-cell mean values.} 
\label{section_gridcellALL}

\begin{figure}[hbt]
\centering
\includegraphics[width=0.24\linewidth,angle=0,clip=true]{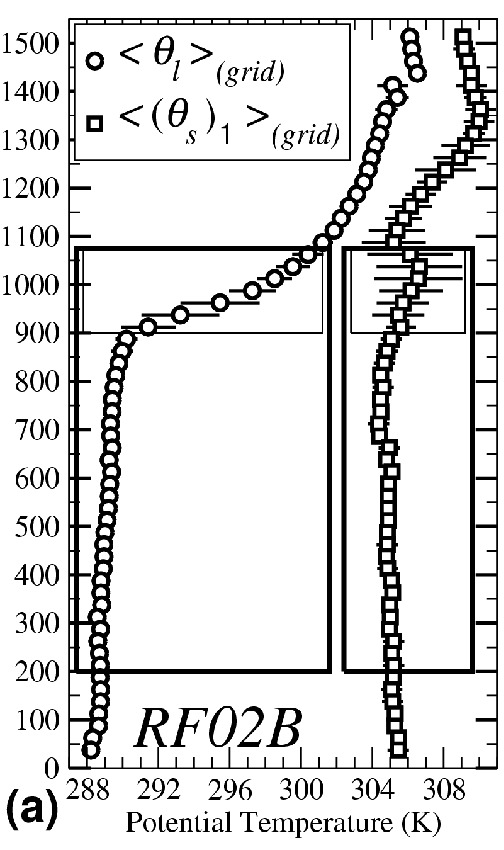}
\includegraphics[width=0.24\linewidth,angle=0,clip=true]{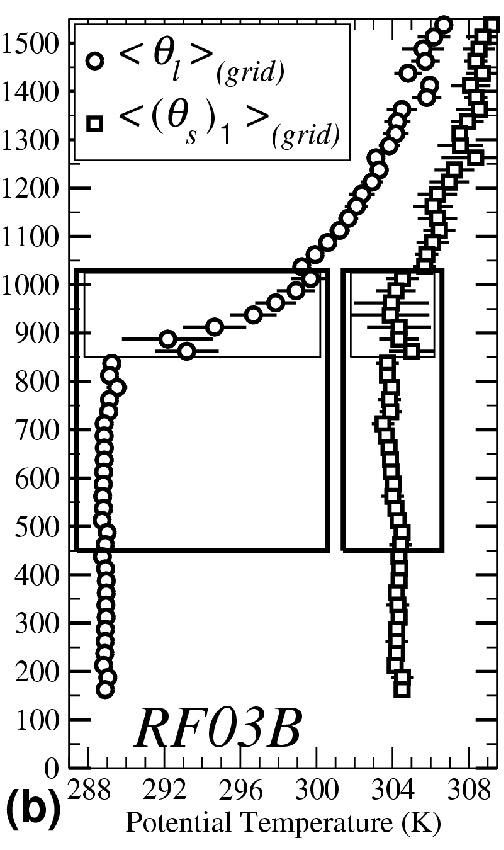}
\includegraphics[width=0.24\linewidth,angle=0,clip=true]{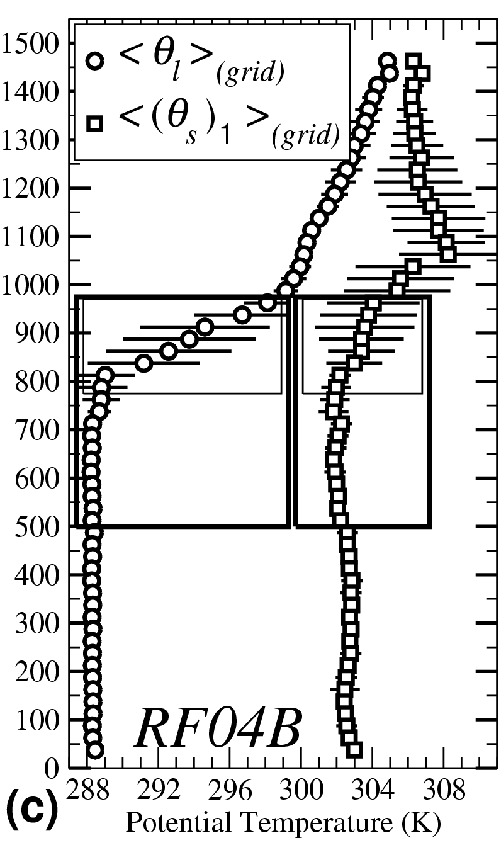}
\includegraphics[width=0.24\linewidth,angle=0,clip=true]{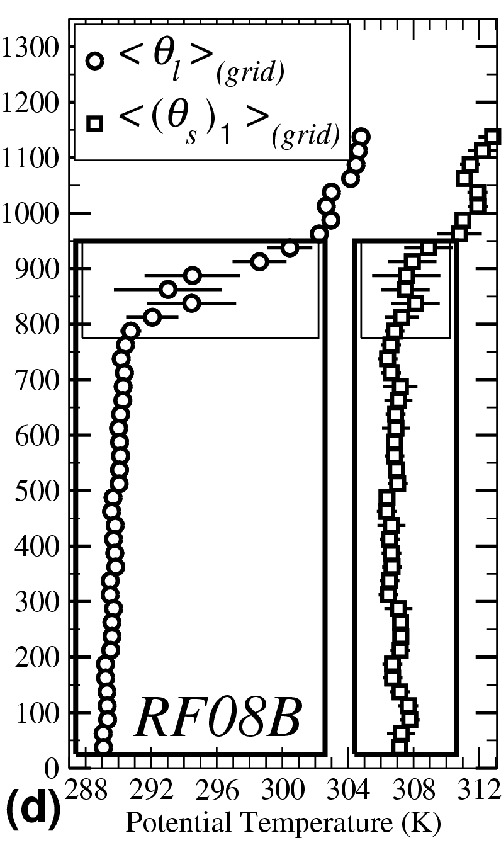}
\caption{\em\small The grid-cell mean values of the moist potential temperatures $<\!\theta_l\!>$ (on the left, open circle) and $<\!(\theta_s)_1\!>$ (on the right, open square) are depicted in (a) to (d) for the flights RF02B to RF08B, respectively.
See the comments in Fig.(\ref{fig_entropy}) concerning the rectangular boxes. 
\label{fig_Ths_grid}}
\end{figure}

The grid-cell mean values for $<\!\theta_l\!>$ and $<\!(\theta_s)_1\!>$ are depicted in  Figs.(\ref{fig_Ths_grid}) (a) to (d), for the four radial flights.
the grid-cell values represent the internal variables available in the NWP models, GCM or SCM.

The computations of the grid-cell average values are more relevant for the moist entropy -- or for $<\!(\theta_s)_1\!>$ -- than for $<\!\theta_l\!>$, because the in-cloud and the clear-air values are equal only for $<\!(\theta_s)_1\!>$, not for $<\!\theta_l\!>$.

The other properties observed for the in-cloud and clear-air averages are also valid for the grid-cell averages.
The jumps in $<\!\theta_l\!>$ within the entrainment region are large and they correspond to the expected results already published for these FIRE-I cases (see for instance RW07).
On the contrary, the jump in $<\!(\theta_s)_1\!>$ does not exist and it is possible to assess the constant value for the grid-cell average of $<\!(\theta_s)_1\!>$ up to the top PBL, with the constant value also valid in the entrainment region since it is located within the horizontal bars, with no more than one standard deviation from the mean values.

There is higher sub-grid variability for $<\!(\theta_s)_1\!>$ in the entrainment region for all flights.
The sub-grid variability is also larger in the dryer air above the top-PBL for the flight RF04B, due to an especially high sub-grid variability for $q_v$ for that flight (see Fig.(\ref{fig_Ths_Qt_Ql_RF04B})(b)).

In order to be more confident in the previous results (i.e. constant PBL values and no jump in  $<\!(\theta_s)_1\!>$), it is interesting to somehow quantify the impact of the instrumental or measurement errors on $<\!(\theta_s)_1\!>$.
It is possible to use a Monte Carlo method by adding a series of perturbations to the original data flight values.
For each of the basic variables ($\theta$, $q_v$, $q_l$), the sets of perturbations are defined by ($\pm 0.1$~\%, $\pm 2$~\%, $\pm 5$~\%) for the weak ones and ($\pm 0.3$~\%, $\pm 5$~\%, $\pm 10$~\%) for the strong ones.
The constraint $q_v<q_{sw}$ is still fulfilled and it can prevent some of the perturbations in $q_v$.
The weighting factors are arbitrarily set to $75$~\% for the original data, $20$~\% for the small perturbations and $5$~\% for the higher ones.

\begin{figure}[hbt]
\centering
\includegraphics[width=0.5\linewidth,angle=0,clip=true]{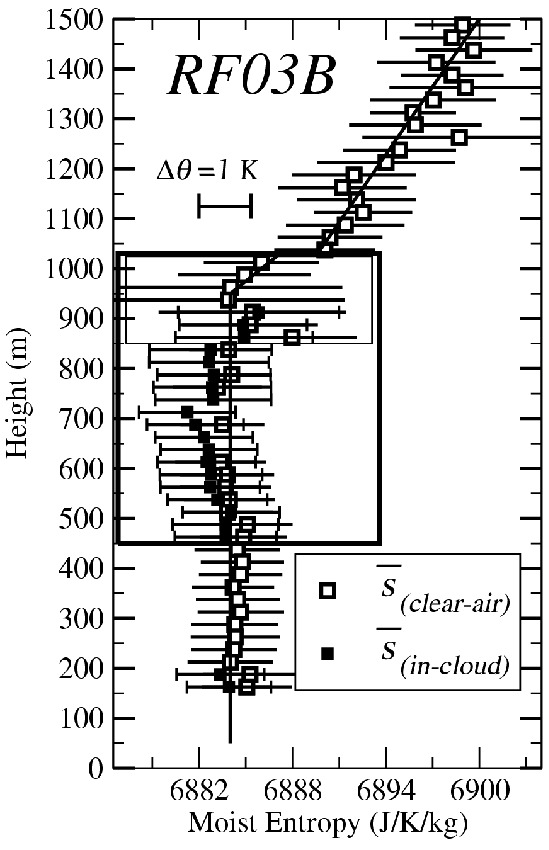}
\caption{\em\small The same as on Fig.(\ref{fig_entropy}) but with the Monte Carlo perturbations added for $\theta$, $q_v$ and $q_l$.
\label{fig_entropy3}}
\end{figure}

The result is depicted in  Fig.(\ref{fig_entropy3}) where the horizontal bars represent the global impact of both the  Monte-Carlo perturbations and the sub-grid variability.
The mean vertical profile of $<\!(\theta_s)_1\!>$ (the sketch thin solid segments lines) is not modified in comparison with  Fig.(\ref{fig_entropy}).
The only differences are the larger horizontal bars, due to the Monte Carlo perturbations perturbations.
The hypothesis of a constant value for the moist entropy ($6884$~J/K/kg) is better supported than in  Fig.(\ref{fig_entropy}), for all levels located within the PBL up to $1025$~m and for both the clear-air and the in-cloud regions.

\subsection{The links between $\Delta\!<\!(\theta_s)_1\!> \: = 0$, CTEI and the ($\Delta \!<\!\theta_l\!>$, $\Delta \,\overline{q_t}$) plane.} 
\label{section_DthlDqt}

\begin{figure}[hbt]
\centering
\includegraphics[width=0.6\linewidth,angle=0,clip=true]{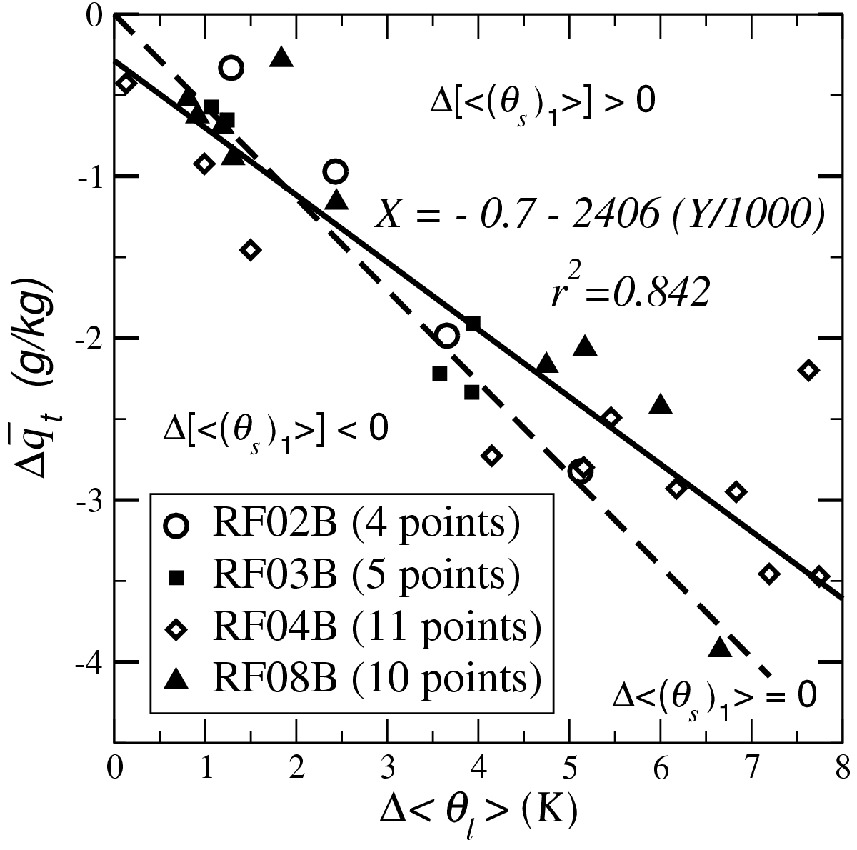}
\caption{\em\small A plot of the differences between ``clear-air'' and ``in-cloud'' values for the mean Betts variables $<\!\theta_l\!>$ and $\overline{q_t}$.
The points of coordinates ($X=\Delta\!<\!\theta_l\!>$, $Y=\Delta\:\overline{q_t}\:$) are plotted for all the $\Delta z=25$~m average layers located within the entrainment regions -- see the thin line boxes in Figs.(\ref{fig_Ths_Qt_Ql_RF03B}) to (\ref{fig_Ths_Qt_Ql_RF08B}) -- for each of the flights RF02B (open circle), RF03B (dark square), RF04B (open diamond) and RF08B (dark triangle).
The solid line represents the least-squared fitted curve.
The dashed line represents the ``moist isentropic'' curve for which $\Delta[<\!({\theta}_{s})_1\!>] = 0$, with positive values above the dashed line and negative values below ($\Lambda = 5.87$  ; ${\theta} \approx 300$~K).
\label{fig_dthldqt}}
\end{figure}

The differences between the clear-air and the in-cloud values for $<\!\theta_l\!>$ and $\overline{q_t}$ are denoted by positive values for $\Delta \!<\!\theta_l\!>$ and negative values for $\Delta \,\overline{q_t}$.
They have been computed for the four FIRE-I flights (02B, 03B, 04B, 08B) and for the few highest in-cloud level located within the entrainment regions (from 4 to 11 points, depending on the flights).
The resulting ($\Delta \!<\!\theta_l\!>$, $\Delta \,\overline{q_t}$) plane is depicted in  Fig.(\ref{fig_dthldqt}).

The reason why the usual jumps in $\theta_l$ and $q_t$ across the cloud-top capping inversion are not used is that these jumps are defined with a poor accuracy, depending on the definition of the free-air base level (see RW07).
On the contrary, the differences between the clear-air and the in-cloud values are unambiguous. 
They are defined for each level and the clear-air values are somehow typical of the clear air located above the inversion, whereas the in-cloud values are typical of the moist PBL values, leading to a difference computed locally at each level that are typical of the ``jump accros the cloud-top capping inversion''.

With $<\!({\theta}_{s})_1\!>$ approximated by (\ref{defTHast4bl}), the differences between clear-air (``$cl$'') and in-cloud (``$in$'') values write
\vspace{-0.15cm}
\begin{align}
 \Delta\!<\!({\theta}_{s})_1\!>
  &\approx \;
 \Delta\!<\!{\theta}_{l}\!>
  + \;
   \Lambda \; ({\theta})_{in}\; 
 \Delta\,\overline{q_t} 
 \; + \;
   \Lambda \; (\overline{q_t})_{cl}\; 
 \Delta\!<\!{\theta}_{l}\!>
   \: ,  \label{defDThsDThlDqt0} \\
 \Delta\!<\!({\theta}_{s})_1\!>
  &\approx \;
 \Delta\!<\!{\theta}_{l}\!>
  \; + \;
   \Lambda \; {\theta}\; 
 \Delta\,\overline{q_t}
   \: .  \label{defDThsDThlDqt}
\end{align}
The last term of (\ref{defDThsDThlDqt0}) is neglected in  (\ref{defDThsDThlDqt}), with  $({\theta})_{in}$ replaced by ${\theta}$.

For ${\theta}_{E}$ approximated by (\ref{defTHE2}) it is possible to express the differences in  $<\!({\theta}_{s})_1\!>$ as (\ref{defDThsDTheDqt}), if the differences in equivalent potential temperature are given by (\ref{defDtheDqtot}).
\vspace{-0.15cm}
\begin{align}
{\theta}_{E} \;
  &  \approx \; {\theta}_{l} \:
   \left( 1 +
   \frac{L_v \: q_t}{c_{pd}\: T}
   \right) 
   \: ,  \label{defTHE2}
\\
 \Delta\!<\!{\theta}_{E}\!>
  & \approx \;
 \Delta\!<\!{\theta}_{l}\!>
  \; + \;
   \frac{L_v}{c_{pd}} \; 
 \Delta\,\overline{q_t} 
   \: ,  \label{defDtheDqtot} \\
 \Delta\!<\!({\theta}_{s})_1\!>
  &\approx \;
 \Delta\!<\!{\theta}_{E}\!>
  \: - \:
   \left(
   \frac{L_v}{c_{pd}}
     -
   \Lambda \:{\theta}  
   \right) 
   \:
 \Delta\,\overline{q_t}
   \: .  \label{defDThsDTheDqt}
\end{align}

The slope of the fitted line in  Fig.(\ref{fig_dthldqt}) is equal to $-2406$~K~(kg/kg)${}^{-1}$.
It corresponds to a value for $\Lambda$ that would make the clear-air and the in-cloud values equal in terms of moist entropy, leading  to $\Delta\!<\!({\theta}_{s})_1\!> = 0$  into (\ref{defDThsDThlDqt}) and to a slope equal to $ \Lambda \; {\theta}$.
For ${\theta} \approx 300$~K, it  corresponds to $\Lambda = 2406 / \theta \approx 8$.
This value if higher than $\Lambda = 5.87$ obtained with $(s_{v})_r$ and $(s_{d})_r$ given by (\ref{defSdr}) and (\ref{defSvr}).
The explanation for this difference is that $\Delta\!<\!({\theta}_{s})_1\!>$ is not exactly equal to zero in  (\ref{defDThsDThlDqt}) and in the entrainment regions of the four FIRE-I flights (even if the mean values are located within the error bars of the others).

The dashed line depicted in  Fig.(\ref{fig_dthldqt}) corresponds to a ``Mixing In Moist Entropy'' (MIME hereafter), where the clear-air and the in-cloud values of $<\!({\theta}_{s})_1\!>$ are equal.
It seems that this dashed line looks like the ``cloud-top instability criterion'' proposed by Randall (1980) and Deardorff (1980), also called ``buoyancy reversal criterion'' or ``Cloud-Top Entrainment Instability'' (CTEI).
The CTEI line is depicted as $\Delta_2 = 0$ in WR07, with a plot of the points corresponding to the jump across the inversion for the four FIRE-I flights (02B, 03B, 04B, 08B).

It is possible to interpret differently the CTEI line, in terms of a MIME (i.e. with the same values for the potential temperature $(\theta_s)_1$ above the cloud and for the in-cloud and the clear-air subparts of the entrainment region).
From (\ref{defDThsDTheDqt}) and (\ref{defDThsDThlDqt}), the hypothesis $\Delta\!<\!(\theta_s)_1\!> \: = 0$ corresponds to the straight lines defined by 
\vspace{-0.15cm}
\begin{align}
 \Delta\!<\!{\theta}_{E}\!>
  &  = \; \;
   \left(
   \frac{L_v}{c_{pd}}
     -
   \Lambda  \: {\theta} 
   \right) 
   \:
 \Delta\,\overline{q_t}
   \: ,  \label{defCTEI2}
\\
 \Delta\!<\!{\theta}_{l}\!>
  & = \;
   - \; \Lambda \; {\theta}\; 
 \Delta\,\overline{q_t}
   \: .  \label{defCTEI} 
\end{align}

According to Yamagushi and Randall (2008), the ``cloud-top instability criterion'' proposed by Randall (1980) and Deardorff (1980) corresponds to (\ref{defCTEI2}).
As suggested by Lilly (2002), the CTEI analysis can also be realized with the help of (\ref{defCTEI}).
Depending on the chosen plane, the CTEI slopes are written either as $\Delta \!<\!\theta_E\!> / \Delta \,\overline{q_t} \:=\: k_{RD}\;L_v\,/c_{pd}\:$ or as $\Delta \!<\!\theta_l\!> / \Delta \,\overline{q_t} \:=\: - \: L_v\,/(k_{L}\;c_{pd})$.
The link between the two parameters $k_{RD}$ and $k_{L}$ and the MIME slope $\Lambda \; {\theta}$ given by (\ref{defCTEI}) is 
\vspace{-0.15cm}
\begin{align}
 k_{RD} 
 & = \; 1 \:-\: \frac{1}{k_{L}} \:=\: 1 \:-\: \frac{ c_{pd}}{L_v} \: \Lambda \: {\theta}
   \: .  \label{defkRDkL}
\end{align}

The CTEI criterion parameter $k_{RD}$ has the standard value of $0.23$ in Kuo and Schubert (1988).
It is mentioned in Yamagushi and Randall (2008) that $k_{RD}$ must vary with the mean potential temperature of the PBL, coming from $0.18$ to $0.48$ for $\theta$ varying from $275$ to $325$~K.
MacVean and Mason (1990) has derived different values, depending on the saturated or unsaturated conditions observed for the above-cloud versus in-cloud conditions: $0.23$ for saturated / saturated (the Randall-Deardorff value) and $0.70$ for unsaturated / saturated (the more relevant one).
Lilly (2002) has derived a real situation value of $k_{RD} = 0.61$ (for $k_{L}=2.55$), with the standard value $k_{RD} = 0.22$ obtained as a limit case for $k_{L}=1.28$.
From the  ($\Delta \!<\!\theta_l\!>$, $\Delta \,\overline{q_t}$) plane published in RW07 and Duynkerke et al. (2004), $k_{RD}$ are set to $0.26$ and $0.18$, respectively.

From the relation (\ref{defkRDkL}), the value $\Lambda \approx 5.87$ retained in this paper and the mean condition ${\theta} \approx 300$~K valid for the FIRE-I data sets lead to $k_{RD} = 0.29$.
\footnote{
\hspace*{1mm} 
This value $k_{RD} \approx 0.29$ corresponds to the use of the first-order approximation $(\theta_s)_1$.
Unpublished results indicate that the use of $\theta_s$ given by (\ref{defTHm1}) leads to a more relevant larger value of about $k_{RD} \approx 0.34$.
}
This value corresponds to a MIME criterion and it compares with the previous values obtained in the studies of the CTEI criterion (coming from $0.18$ to $0.70$).

\section{Sensitivity experiments.} 
\label{section_sens}

\begin{figure}[hbt]
\centering
\includegraphics[width=0.202\linewidth,angle=0,clip=true]{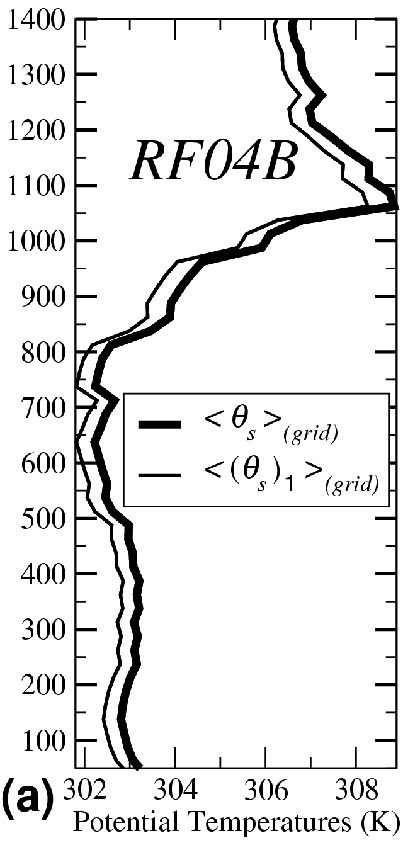}
\includegraphics[width=0.150\linewidth,angle=0,clip=true]{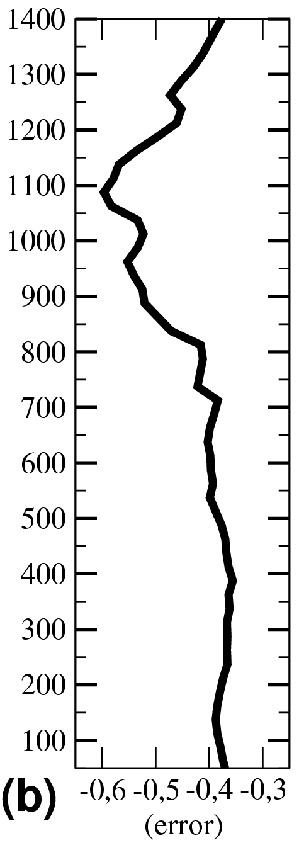}
\includegraphics[width=0.310\linewidth,angle=0,clip=true]{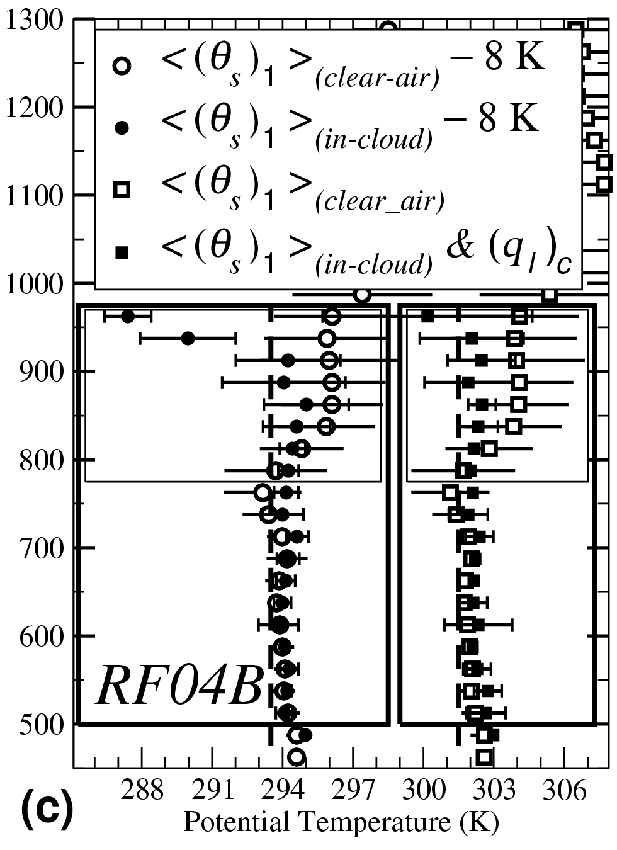}
\includegraphics[width=0.315\linewidth,angle=0,clip=true]{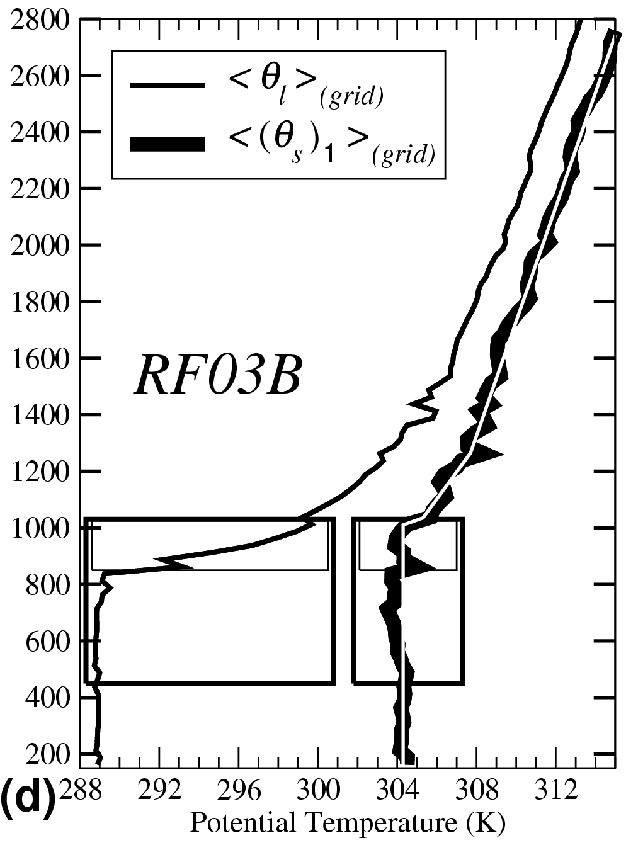}
\caption{\em\small Sensitivity experiments.
(a) The profiles for RF04B and for the exact values $<\!\theta_s\!>$ (heavy line) and the corresponding leading order approximate formulation $<\!(\theta_s)_1\!>$ (thin line).
(b) The profiles for RF04B and for the difference between $<\!(\theta_s)_1\!>$ and $<\!\theta_s\!>$.
(c) The impact on $(\theta_s)_1$ of the threshold value $(q_l)_{c}$ above which $q_v$ is set to its saturation value $q_{sw}(T)$, with the RF04B regular values shifted by the amount $-8$~K on the left and the RF04B modified values located on the right. The two vertical dashed lines are shifted by the same amount of $-8$~K, in order to make easier the comparisons.
(d) The whole RF03B dataset extended above the PBL up to $2800$~m, for both $<\!\theta_l\!>$ (thin black line) and $<\!(\theta_s)_1\!>$ (heavy black line). Thin white line segments are plotted over the vertical profile of $<\!(\theta_s)_1\!>$, indicating a possible linearised description of it.
See the comments in Fig.(\ref{fig_entropy}) concerning the rectangular boxes and the standard deviation bars. 
\label{fig_sens1}}
\end{figure}

The first test depicted in  Fig.(\ref{fig_sens1}) (a) and (b) concerns the evaluation of the error between  the approximate version $<\!(\theta_s)_1\!>$ and the exact one $<\!\theta_s\!>$.
There is a small negative bias of $-0.35$ to $-0.55$~K.
It corresponds to an error of less than $0.2$~\%.
It justifies the use of $<\!(\theta_s)_1\!>$ in the previous analyses.

The second test is shown in  Fig.(\ref{fig_sens1}) (c).
It corresponds to the impact of the threshold value $(q_l)_{c}$ on the clear-air and in-cloud values of $(\theta_s)_1$, as described in the section \ref{section_entr03B}.
According to  Fig.(\ref{fig_Ths_Qt_Ql_RF04B}) (c), $(q_l)_{c}=0.04$~g/kg for the flight RF04B and the possible   impacts could only concern the upper in-cloud levels located between $850$ and $975$~m height, for which $\overline{q_v}>(q_l)_{c}$.
It appears that the modified in-cloud values get closer to the clear-air ones for the layers $925$-$950$~m and $950$-$975$~m, with the horizontal bars crossing over.
It justifies the use of $q_v = q_{sw}(T)$ where $q_v>(q_l)_{c}$ locally.

The third test concerns the analysis of the full vertical range for the flight RF03B, including the extended levels reaching $2800$~m and above.
The aim is to check if the vertical profile of the  approximated new potential temperature $<\!(\theta_s)_1\!>$ exhibits a standard stable layer pattern far above the PBL, or not.
It appears that the ``stable linear regime'' already depicted as  black solid line segments in  Fig.(\ref{fig_entropy}) can be extended above the PBL, as suggested for the grid-cell average depicted in  Fig.(\ref{fig_sens1}) (d) as white solid segments.

As a consequence, it may be more relevant to search for a description by line segments starting with the vertical profiles of  $\overline{s}$ or $<\!(\theta_s)_1\!>$, rather than with the vertical profiles of $<\!\theta_l\!>$.
Applications could be found in the building of idealized initial profiles as used in the SCM, CRM or LES inter-comparison cases.

\begin{figure}[hbt]
\centering
\includegraphics[width=0.31\linewidth,angle=0,clip=true]{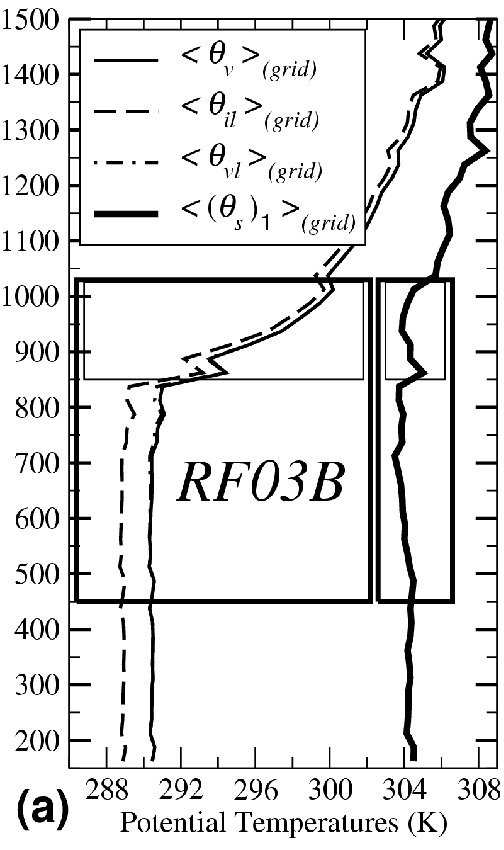}
\includegraphics[width=0.31\linewidth,angle=0,clip=true]{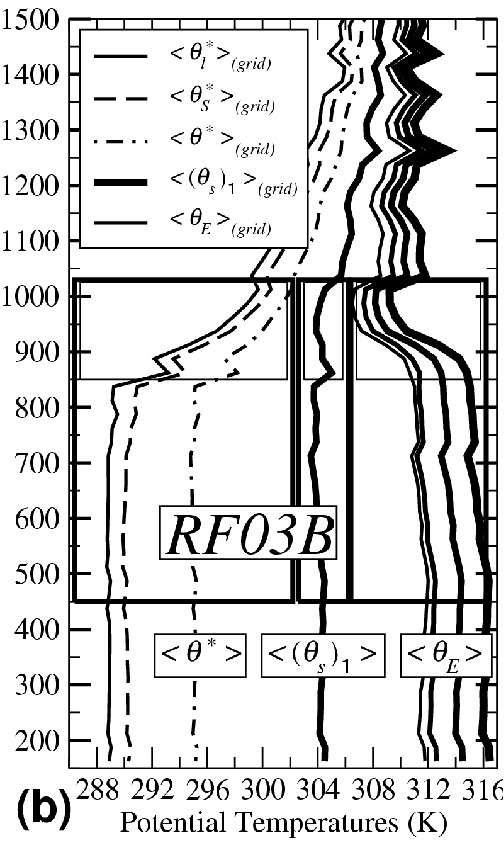}
\includegraphics[width=0.31\linewidth,angle=0,clip=true]{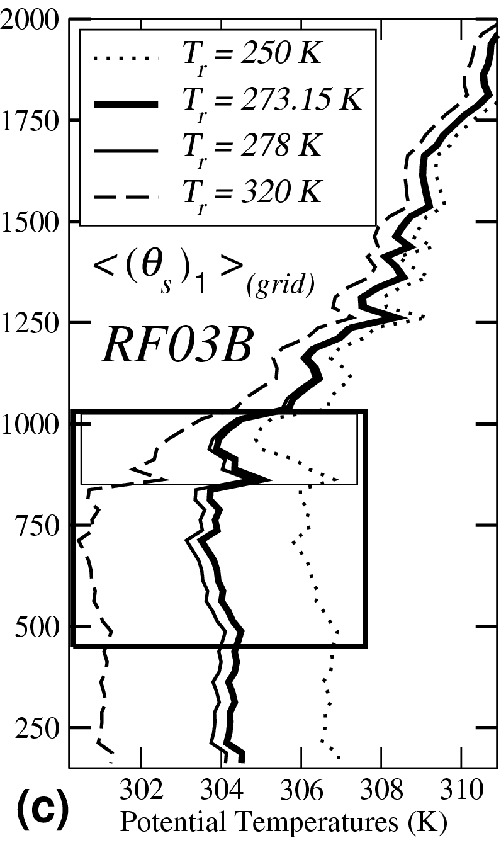}
\caption{\em\small Vertical profiles for the grid-cell averages of several potential temperatures, for the flight RF03B.
(a) Comparison of $({\theta}_{s})_1$ with (from the left to the right): ${\theta}_{il}$ (TC81) , ${\theta}_{vl}$ (GB01) and ${\theta}_{v}$ (L68).
(b) Comparison of $({\theta}_{s})_1$ with (from the left to the right): ${\theta}_{l}^{\ast}$ (E94), ${\theta}_{S}^{\ast}$ (HH87) and ${\theta}^{\ast}$ (M93). The four last profiles located on the right of (b) correspond  (from the left to the right) to the four ${\theta}_{E}$ formulations of B73, E94, Bolton (1980) -- Eqs.(21) and (43) -- and to a numerical computation made with a code developped by J.M. Piriou from the ARPEGE model.
(c) The impact on $({\theta}_{s})_1$ of different choice for $T_r$ equal to $250$~K, $273.15$~K, $278$~K and $320$~K.
See the comments in Fig.(\ref{fig_entropy}) concerning the rectangular boxes. 
\label{fig_sens2}}
\end{figure}

Another set of tests are shown in  Fig.(\ref{fig_sens2}) (a) and (b), where grid-cell average values  have been computed for the flight RF03B and for all the potential temperatures described in the sections \ref{section_mptvl} and \ref{section_mptthast}.
It appears that TC81's and E94's values for ${\theta}_{il}$ and ${\theta}_{l}^{\ast}$ are very close to the Betts one ${\theta}_{l}$.
The (buoyancy) virtual potential temperatures ${\theta}_{v}$ (L68) and ${\theta}_{vl}$ (GB81) are $2$~K warmer than the Betts-like ones.
The same is true for the entropy potential temperature ${\theta}_{S}^{\ast}$ (HH87).

The profile for $<\!(\theta_s)_1\!>$ in (b) is different from all others, with a difference of more than $14$~K from the Betts-like or virtual potential temperatures and with the moist available enthalpy potential temperature  ${\theta}^{\ast}$ leading to in-between values.
Clearly, ${\theta}_{l}$ cannot represent the moist entropy.

The last warmest profiles in the right part of (b) allow a comparison between $<\!(\theta_s)_1\!>$ and four different formulations for the equivalent potential temperature. The coldest profile for $<\!\theta_E\!>$ is based on the simplified formulation (\ref{defTHE2}), with  (\ref{defTHast4bl})  representing the first order expression for (\ref{defTHast4b0}).
The comparison of  (\ref{defTHast4bl})  with (\ref{defTHE2}) explains the reason why the vertical profile of $<\!(\theta_s)_1\!>$ is rougthly in a $2/3$rd position between  ${\theta}_{l}^{\ast} \approx {\theta}_l$ and ${\theta}_{E}$, with $L_v \: / (c_{pd} \:T)$ and $\Lambda$ indeed close to $9$ and $6$, respectively.

As a consequence, it seems that the moist entropy $\overline{s}$ and the associated moist potential temperatures $<\!\theta_s\!>$ or  $<\!(\theta_s)_1\!>$ cannot be represented by any of the other potential temperatures.

The last test concerns the choice of the reference potential temperature $T_r$.
The variations of  $\Lambda$ with $T_r$ and $p_r$ are presented in Table~\ref{Table_LAMBDA}.
\begin{table}
\caption{\em\small The values for $\Lambda = [ (s_{v})_r - (s_{d})_r ] / c_{pd}$ given as a function of $T_r$ (K) and $p_r$ (hPa).
The values $p_r=1000/\exp(1)\approx368$~hPa and $T_r=250$~K have been used in M93.
The bold value $\Lambda=5.87$ corresponds to $p_r=1000$~hPa and $T_r=273.15$~K, as retained in the present study.
\label{Table_LAMBDA}}
\centering
\vspace*{2mm}
\begin{tabular}{|c||c|c|c|c|}
\hline
    $p_r \setminus T_r$  
           & $250$ & $\bf 273.15$ & $300$ & $320$ \\ 
\hline \hline
    $368$  & $6.47$ & $5.58$ & $4.83$ & $4.31$  \\ 
\hline
    $800$  & $6.69$ & $5.80$ & $5.06$ & $4.59$   \\
\hline
$\bf 1000$  & $6.75$ & $\boxed{\bf 5.87}$ & $5.13$ & $4.67$   \\
\hline
\end{tabular}
\end{table}
The sensitivity associated with changes in $T_r$ is more important than with changes in $p_r$.
The value $\Lambda = 5.87$ corresponds to the special choice for $T_r$ and $p_r$ indicated in the Appendix~A.

Four profiles are depicted for the flight RF03B in  Fig.(\ref{fig_sens2}) (c), corresponding to the grid-cell average of $\theta_s$ and for $T_r=250$, $273.15$, $278$ or $320$~K.
One of the rules for choosing a relevant right value for $T_r$ is to search for the ``same vertical profile'' for  $<\!(\theta_s)_1\!>$ in  Fig.(\ref{fig_sens2}) (c) as in  Figs.(\ref{fig_entropy}) for the vertical profile of the moist entropy $\overline{s}$.
It is also useful to compare the vertical profiles for $<\!(\theta_s)_1\!>$ and for  $\overline{s}$ for the three other flights, as described in  Figs.(\ref{fig_entropy2}) for the entropy and  (\ref{fig_Ths_Qt_Ql_RF02B}) to (\ref{fig_Ths_Qt_Ql_RF08B}) for the corresponding potential temperatures.

It seems that the values $T_r = 273.15$~K (chosen in the present study) or $T_r = 278$~K are appropriate ones, at least for these FIRE-I flights.
It can be noted that the change in $<\!(\theta_s)_1\!>$ is less than $\pm 3$~K in the PBL, even for the extreme variations of $T_r$ from $250$ to $320$~K, and it is less than $\pm 1$~K above the PBL.
These changes may be considered as small in comparison with the large differences between $<\!(\theta_s)_1\!>$ and the other potential temperatures, as depicted in  Figs.(\ref{fig_sens2}) (a) and (b).

In spite of these encouraging sensitivity experiments, one may consider that for global applications of $({\theta}_{s})_1$ in GCM or in NWP models with sufficiently large horizontal domains, it may be difficult to find a value for $T_r$ (and for $\Lambda$) which may be relevant for all points, going from equatorial to polar regions?

It is however important to remember that $({\theta}_{s})_1$ is only the first order approximation of the exact formulation (\ref{defTHm1}) and it can be verified that the numerical values for the exact moist entropy $s$  and the moist potential temperature ${\theta}_{s}$ do not depend at all on $T_r$ or $p_r$, as indicated in Table~\ref{Table_TrPr}.
The large changes in the two terms $s_r$ and ${\theta}_{sr}$ balance each other in order to give constant values for the exact potential temperature ${\theta}_{s}$ and for the reference entropy $s_{ref}$, whith $s_{ref}$ defined by 
\begin{equation}
\boxed{\;\;
s  \; = \;  s_{ref} + c_{pd} \,\ln({\theta}_{s})
\;\;}
   \: , \label{defssrTHs}
\end{equation}
where
\vspace*{-3 mm}
\begin{align}
s_{ref} & \: = \: s_r - c_{pd} \,\ln({\theta}_{sr})
    \: , \label{defsref} \\
s_r & \: = \: (1-q_r)\:(s_d)_r + q_r\:(s_v)_r
    \: . \label{defsr}
\end{align}
The quantity  $s_{ref}$ can be evaluated with (\ref{defSdr}), (\ref{defsr}) and (\ref{def0rr}) inserted into (\ref{defsref}), leading to 
\begin{equation}
\boxed{\;\;
s_{ref} \: = \: s_d^0 - c_{pd} \,\ln(T_0)\: \approx \: 1138.56\mbox{~J~K${}^{-1}$~kg${}^{-1}$}
\;\;}
\nonumber
\end{equation}
for the standard values of $s_d^0$ and $T_0$ given in the Appendix-A.

\begin{table}
\caption{\em\small Numerical values computed for the same parcel of cloud ($p=800$~hPa, $T=280$~K, $q_v=7.74$~g/kg, $q_l=1$~g/kg,  $q_i=0$~g/kg) but with different values of $T_r$ (in K) and $p_r$ (in hPa).
From (\ref{defTHmSm1}), the moist entropy is equal to $s = s_r + c_{pd} \,\ln({\theta}_{s}/{\theta}_{sr})$, with $s_r$ given by (\ref{defsr}).
The moist potential temperatures ${\theta}_{s}$, ${\theta}_{sr}$ and $({\theta}_{s})_1$ are given by (\ref{defTHm1}), (\ref{defTHm2}) and (\ref{defTHast4b0}).
The reference entropy $s_{ref}$ is defined by (\ref{defsref}).
\label{Table_TrPr}}
\vspace*{2mm}
\centering
\begin{tabular}{|c|c||c|c||c|c|c||c|}
\hline
    $T_r$ & $p_r$ & $s$ & ${\theta}_{s}$ & $({\theta}_{s})_1$ & $s_r$ & ${\theta}_{sr}$ & $s_{ref}$ \\ 
\hline \hline
    $220$     & $1000$ & $6907.8$ & $311.76$ & $317.8$ & $6557.7$ & $250.9$ & $1138.56$ \\ 
\hline
    $\bf{273.15}$  & $1000$ & $6907.8$ & $311.76$ & $\bf{311.4}$ & $6799.2$ & $279.8$ & $1138.56$ \\
\hline
    $320$     & $1000$ & $6907.8$ & $311.76$ & $308.12$ & $7284.2$ & $340.7$ & $1138.56$ \\
\hline \hline
    $\bf{273.15}$  &  $800$ & $6907.8$ & $311.76$ & $\bf{311.2}$ & $6869.0$ & $300.0$ & $1138.56$ \\
\hline
    $\bf{273.15}$  &  $400$ & $6907.8$ & $311.76$ & $\bf{310.7}$ & $7096.2$ & $376.3$ & $1138.56$ \\
\hline
\end{tabular}
\end{table}

The formula (\ref{defssrTHs}), where $c_{pd}$ and $s_{ref}$ are equal to two thermodynamic constants, demonstrates that ${\theta}_{s}$ is a true synonym of the moist entropy.
The consequence is that the analysis of the vertical profiles of $s$ can be realized with no approximation in terms of ${\theta}_{s}$, whatever the choices for $T_r$ and $p_r$ may be!

If an approximate version of ${\theta}_{s}$ is needed, the bold values of $({\theta}_{s})_1$ presented in Table~\ref{Table_TrPr} show that $273.15$~K is a relevant value for $T_r$, with a negative bias in the computation of $({\theta}_{s})_1$ less than $1$~K and corresponding to the values depicted in Fig.(\ref{fig_sens1})(a) and (b).

\section{Vertical fluxes of ${\theta}_{s}$.} 
\label{section_flux_grad}

According to the formulation (\ref{defTHmSm1}), the moist entropy depends on the logarithm of ${\theta}_{s}$.
It is approximated by the logarithm of  $({\theta}_{s})_1$ given by (\ref{defTHast4b0}), leading to
\begin{align}
 \ln[({\theta}_{s})_1]  & \: = \: 
 \ln({\theta}_{l})\:+\:\Lambda\,q_t
    \: . \label{defTHs2a}
\end{align}
The differential of (\ref{defTHs2a}) is written
\vspace{-0.15cm}
\begin{align}
\frac{ds}{{c}_{pd}}  & \: \approx \: 
\frac{d({\theta}_{s})_1}{({\theta}_{s})_1}
\; = \;
\frac{d{\theta}_{l}}{{\theta}_{l}}
\: + \:
\Lambda\; dq_t
    \:  , \label{defdTHs1}
\end{align}
and the flux of moist entropy is then approximated by
\begin{align}
 \!\!\overline{w'{s}'} 
& \: \equiv \: 
 \frac{{c}_{pd}}{\;\overline{({\theta}_{s})}\;} \: \overline{w'{\theta}_{s}'}
  \; \approx \; 
 \frac{{c}_{pd}}{\;\overline{({\theta}_{s})_1}\;} \: \overline{w'({\theta}_{s})_1'}
    \: , \label{defFluxTHs1} 
\\
\!\!
& \: \approx \:
\frac{{c}_{pd}}{\;\overline{({\theta}_{l})}\;} \:
   \overline{w'{\theta}_{l}'}
      \; + \; 
 {c}_{pd} \;  \Lambda \; \overline{w'q_t'}
    \: . \label{defFluxTHs2a}
\end{align}
The flux of $({\theta}_{s})_1$  is written
\vspace{-0.15cm}
\begin{align}
 \overline{w'({\theta}_{s})_1'} & \: \approx \:
 (1+\Lambda\,\overline{q_t}) \; \overline{w'{\theta}_{l}'}
      \; + \; 
  \Lambda \; \overline{({\theta}_{s})_1}  \;\; \overline{w'q_t'}
    \: . 
 \label{defFluxTHs2}
\end{align}

If the moist entropy is a constant  within the PBL -- as observed for the FIRE-I flights -- then $\overline{w'{s}'}\equiv 0$ and, from (\ref{defFluxTHs1}), $\overline{w'({\theta}_{s})_1'}\approx 0$.
When this assumption  is introduced into (\ref{defFluxTHs2}), it leads to a moist isentropic balance of the Betts' variables fluxes  and, according to  (\ref{defTHast4b0}),  it is written
\vspace{-0.15cm}
\begin{equation}
 \overline{w'{\theta}_{l}'} 
\; \approx 
\; - \; \Lambda \;\; \overline{{\theta}_{l}} \;\; \overline{w'q_t'} \: .
\label{defFluxTHs2b}
\end{equation}
This relation between the Betts' variables fluxes correspond the CTEI criterion and to  (\ref{defCTEI}).

In some parameterizations of the turbulence, the internal variables used in the numerical schemes are based on a modified static stability function defined by ${c}_{pd}\:T + g\:z - L_v\:q_l$.
It replaces the use of ${\theta}_{l}$.
The trick is to take into account the hydrostatic (exact) differential and (approximate) flux equations
\vspace{-0.15cm}
\begin{align}
 {c}_{pd}\: \frac{d\theta}{\theta} \:
 & = \:
  \frac{1}{T} \: d\:( \,{c}_{pd}\:{T}+ g\:z \,)
    \: , \label{defderiv} \\
 \frac{{c}_{pd}}{\;\overline{(\theta)}\;} \: \overline{w'{\theta}'}
 & \approx \:
  \frac{1}{\;\overline{(T)}\;} \: \overline{w'({c}_{pd}\:{T}^{\prime} + g\:z')}
    \: , \label{defFluxTHs3}
\end{align}
and to use the original Betts formula (\ref{defTHB73b}) with the variations of $L_v(T)/T$ with $T$ neglected with respect to the changes in $q_l$, to arrive at
\vspace{-0.15cm}
\begin{align}
 \frac{{c}_{pd}}{\;\overline{(\theta_l)}\;} \: \overline{w'{\theta_l}'}
 \: \approx \:
  \frac{1}{\;\overline{(T)}\;} \: \overline{w' S'_l }
    \: , \label{defFluxTHs4}
\end{align}
where the liquid water static energy $S_l$ is defined in Stevens et al. (2003) by
\vspace{-0.15cm}
\begin{align}
 S_l & = \;  {c}_{pd}\:{T} + g\:z - \overline{L_v}\:q_l
    \: . \label{defSl}
\end{align}
The flux of moist entropy is then obtained with (\ref{defFluxTHs4}) inserted into (\ref{defFluxTHs2a}) and  (\ref{defFluxTHs1}), leading to
\vspace{-0.15cm}
\begin{equation}
 \overline{w'{s}'} \;\approx \;\;
   \frac{1}{\;\overline{(T)}\;} \;\; \overline{w'{S}'_m}  \: ,
\label{defFluxTHs7}
\end{equation}
where $S'_m$ is the perturbation of a kind of  ``moist entropy static energy'' function $S_m$ defined by
\vspace{-0.15cm}
\begin{equation}
 {S}_m \; = \;
   {c}_{pd}\:\left(\:{T}+\Lambda\:\overline{T}\:q_t\: \right)\: + g\:z - \overline{L_v}\:q_l \:
    ,
\label{defFluxTHs8}
\end{equation}
or equivalently by
\vspace{-0.15cm}
\begin{align}
 {S}_m & = \;
   {c}_{pd}\: T \: + \:  g\:z + \overline{L_v} \: q_v
  \nonumber \\
       & \quad - \: \left(\: \overline{L_v} \: - \:  {c}_{pd}\:\overline{T} \: \Lambda\: \right) \:q_t
    \: .
\label{defFluxTHs9}
\end{align}

In comparison with the liquid water static energy (\ref{defSl}), $S$ given by (\ref{defFluxTHs8}) contains the additional part ${c}_{pd}\:\Lambda\:\overline{T}\:q_t$.
This term is not constant with height if $\overline{T}$ varies with $z$, even if $q_t$ is a constant (as an invariant of the moist system).
Only the moist entropy flux (\ref{defFluxTHs7}) is a constant, including the division by $\overline{T}$.
It is the reason why the quantity $S_l/\overline{T}$ is plotted in Stevens et al. (2003) in place of $\theta_l$, corresponding to the flux (\ref{defFluxTHs4}). 

The additional part between $S_m/\overline{T}$ given by (\ref{defFluxTHs8})  and $S_l/\overline{T}$ given by (\ref{defSl})  is ${c}_{pd}\:\Lambda\:q_t$.
It can only be discarded  if $q_t$ is a constant, a property not verified  in the entrainment region where possible large differences could exist between the flux of $S_m/\overline{T}$ and the flux of $S_l/\overline{T}$.

\section{Other Stratocumulus cases ; Conserved variable diagram.} 
\label{section_others_Sc}

\begin{figure}[hbt]
\centering
\includegraphics[width=0.32\linewidth,angle=0,clip=true]{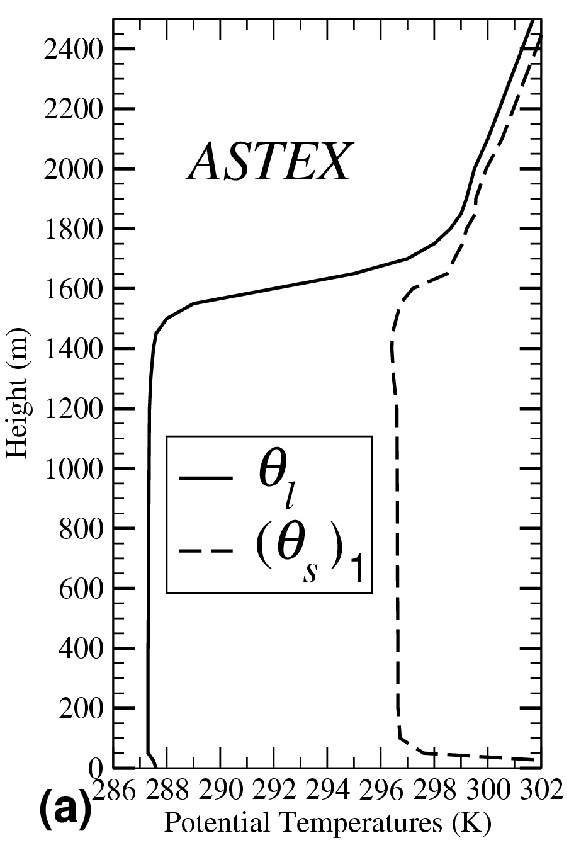}
\includegraphics[width=0.32\linewidth,angle=0,clip=true]{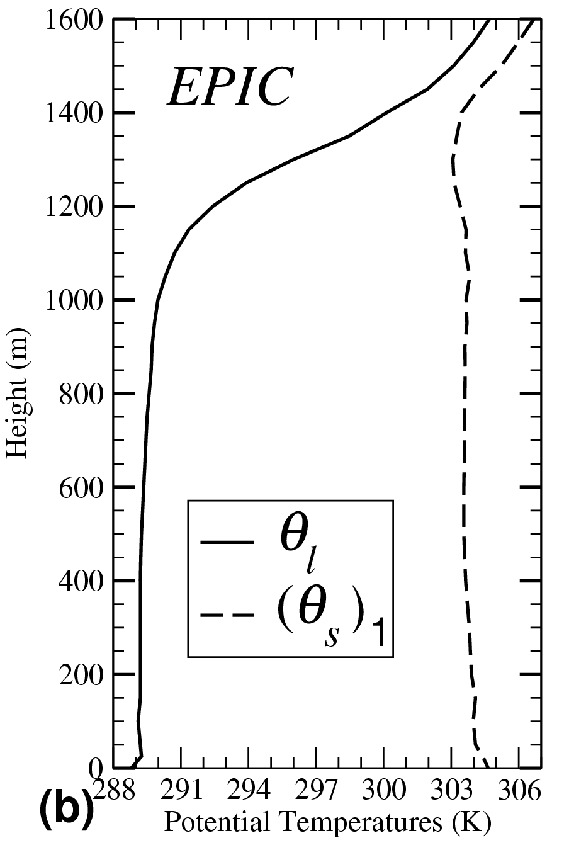}
\includegraphics[width=0.32\linewidth,angle=0,clip=true]{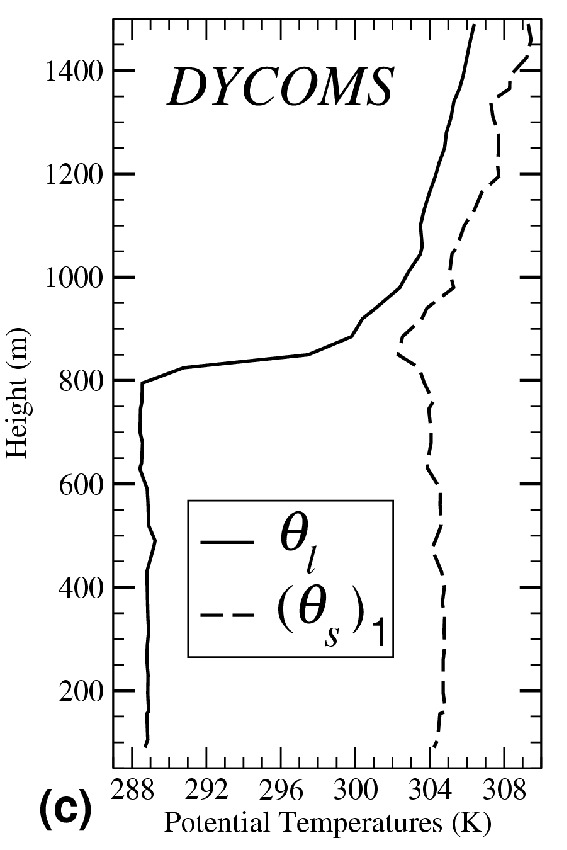}
\caption{\em\small The potential temperatures and specific water content for (a) the ASTEX, (b) the EPIC and (c) the DYCOMS-II (RF01) Stratocumulus cases.
The liquid-water potential temperature $\theta_l$ is depicted as solid lines, $({\theta}_{s})_1$ as dashed lines.
\label{fig_others_Sc}}
\end{figure}

To obtain a more general appreciation of the interest to use the moist entropy -- or $({\theta}_{s})_1$ -- in atmospheric science, three well-known Stratocumulus cases have been numerized from different published papers, corresponding to different regions and time.

The north-eastern Atlantic ocean ``ASTEX'' profiles (June 1992) are plotted for ($\theta_l$, $q_t$) in Cuijpers and Bechtold (1995).
The south-eastern Pacific ocean ``EPIC'' profiles (6-day mean values, October 2001) are plotted for ($\theta$, $q_v$, $\rho\,q_l$) in Bretherton et al. (2004).
The north-eastern Pacific ocean  ``DYCOMS-II'' profiles (RF01 data set, July 2001) are plotted for ($\theta$, $q_t$, $q_l$) in Zhu et al. (2005).

The vertical profiles of $\theta_l$ and $({\theta}_{s})_1$ are plotted for the three cases in  Figure (\ref{fig_others_Sc}).
As for the grid-cell values of the FIRE-I cases depicted in Figure (\ref{fig_Ths_grid}), there is no (EPIC, DYCOMS-II) or small (ASTEX) jump in moist entropy potential temperature at the top of the PBL, with  $({\theta}_{s})_1$ a constant throughout the PBL of the three cases.

\begin{figure}[hbt]
\centering
\includegraphics[width=0.7\linewidth,angle=0,clip=true]{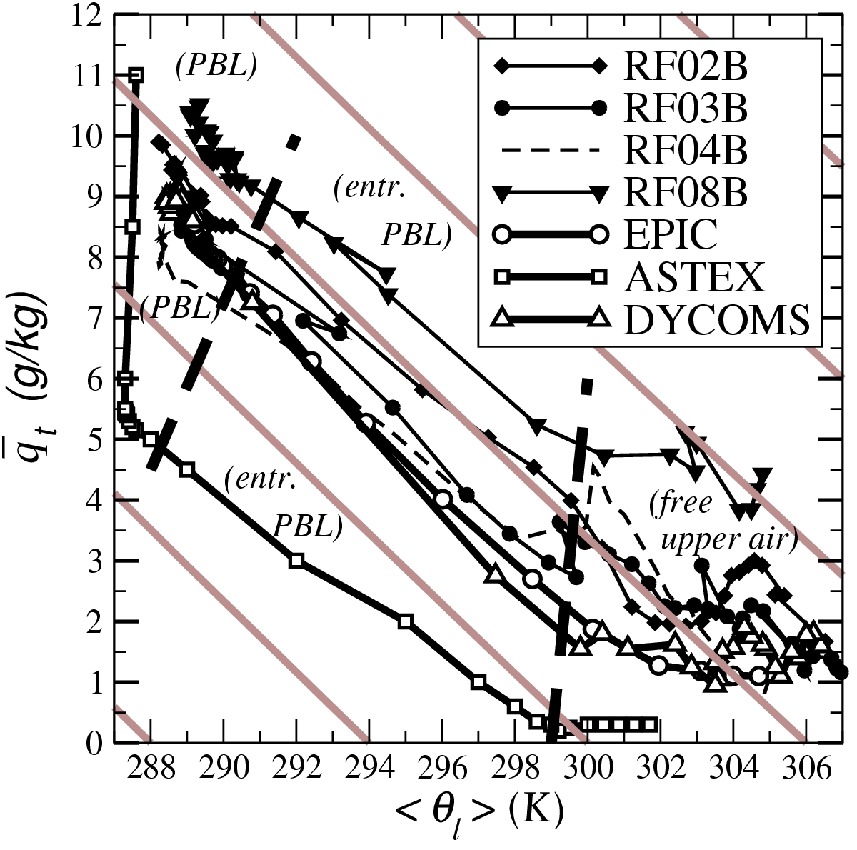}
\caption{\em\small A conserved variable diagram with the total specific content $\overline{q_t}$ plotted against the liquid water potential temperature  $<\!\theta_l\!>$.
The four FIRE-I data flights (02B, 03B, 04B and 08B) are represented together with the three EPIC, ASTEX and DYCOMS-II (RF01) data sets.
The entrainment regions are depicted between the heavy dashed lines, with the free upper air points located above $300$~K on the bottom right of the diagram and the moist PBL points grouped on the other side.
The slantwise greyish solid lines correspond to constant values for $({\theta}_{s})_1$ (they can be labelled every $6$~K with the values of ${\theta}_{l}$ at $q_t=0$).
\label{fig_thlqt}}
\end{figure}

In the conserved variable diagrams, the total specific content of water vapour is plotted against the equivalent potential temperature (Palush, 1979) or the liquid-water potential temperature (Neggers et al., 2002).
Figure (\ref{fig_thlqt}) is the $(q_t, \theta_l)$ diagram for the four FIRE-I data flights and for the three other Stratocumulus cases ASTEX, EPIC and DYCOMS-II (RF01).
This diagram can be used as a graphical method to demonstrate (or to appreciate) the constant moist entropy regime and the MIME processes occurring within the PBL of these Stratocumulus cases.

The moist PBL values are assembled on the left side of the diagram, with small increases in $\theta_l$ with height and associated decreases in $q_t$.
The upward variations of the points in the PBL and then in the entrainment regions correspond to changes along slantwise patterns following approximately the constant $({\theta}_{s})_1$ lines, defined by $\Lambda \: q_t = \ln[({\theta}_{s})_1/{\theta}_{l}]$.
Clearly, from the left to the right there are constant regime or smooth transitions for all flights in terms of $({\theta}_{s})_1$ between the moist PBL, the entrainment region and the free upper air, where $({\theta}_{s})_1$ starts to increase due to the diabatic heating processes and to the subsidence of the dry air located above.

The ASTEX curve depicted in the conserved variable diagram of  Figure (\ref{fig_thlqt}) is different from the others, with values of $({\theta}_{s})_1$ varying rapidly close to the surface and in the entrainment region.
Indeed, the ASTEX vertical profiles presented in Cuijpers and Bechtold (1995) correspond to a moist surface layer with a dryer and colder PBL than the other FIRE-I, EPIC or DYCOMS-II observed vertical profiles.
This kind of diagram can illustrate the method to appreciate to which extent a vertical profile may be typical of a Stratocumulus distinctive pattern.

\section{Thermodynamic diagrams.} 
\label{section_diagram}

As stated by Emanuel (1994, chapter 5), ``The stability characteristics and thermodynamic properties of convective clouds and of convecting atmospheres are most easily seen by making plots of the thermodynamic variables.
Various thermodynamic transformations can also be easily calculated using thermodynamic diagrams, avoiding the often tedious calculations necessary in moist thermodynamics''.

Accordingly, it is possible to add a new set of moist entropy curves (based on $({\theta}_{s})_1$) on the so-called skew $T$-$\ln(p)$ diagram, as a companion set of the dry entropy curves (dry convection / $\theta$) and of the  pseudo-potential temperature curves (deep convection / $\theta'_w$).

\begin{figure}[hbt]
\centering
\includegraphics[width=0.7\linewidth,angle=0,clip=true]{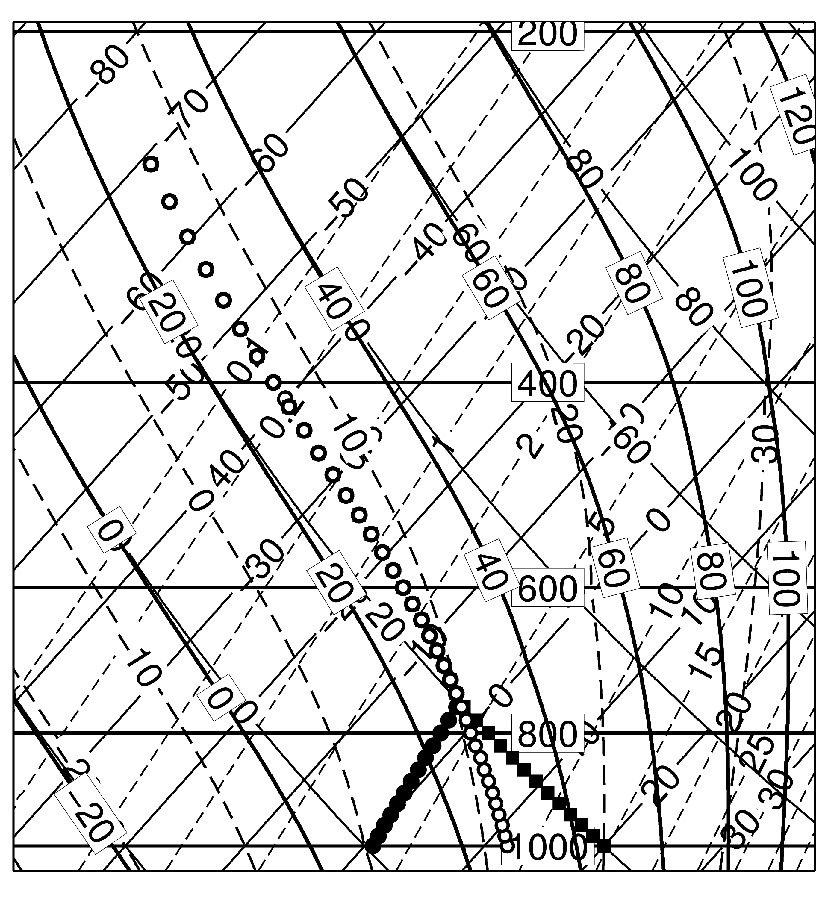}
\caption{\em\small The skew $T$-$\ln(p)$ diagram.
The classic isolines of $T$, $q_v$, $\theta$ and $\theta'_w$ are depicted in the usual way.
The moist entropy solid  lines are defined by constant values of ${\theta}_{s}$ and they are labelled by the boxed values going from $-20$ to $120$~C.
They tend toward the corresponding dry adiabatic values $\theta$ for small values of $q_v$ (upper left) and the differences increase more and more for larger values of $q_v$ (bottom right).
An ideal and adiabatic ascent of a parcel is depicted with dark circles for $T_d$, open circles for $T_s$ and dark squares for $T$ (see further explanations in the text).
\label{fig_diagram1}}
\end{figure}

Figure (\ref{fig_diagram1}) is an example of a skew $T$-$\ln(p)$ diagram where an initial parcel defined by $p=1000$~hPa, $T=20$~C and $q_v=4$~g/kg is shifted upwards adiabatically up to $250$~hPa, with the assumption of a constant value for the moist entropy (surface value of ${\theta}_{s}=27$~C).
The moist entropy temperature $T_s$ (open circle) is defined for each level as the value of ${\theta}_{s}$ measured at the corresponding condensation level, in a way similar to the graphical process used to evaluate $\theta'_w$.

For this ideal case study and above the condensation level, the ${\theta}_{s}=27$~C line is located in between the unsaturated dry adiabatic line ($\theta=20$~C) and the saturated pseudo-adiabatic one ($\theta'_w=10$~C).
It can be noted that, above the condensation level, liquid or ice cloud water exist and are taken into account in the computations of ${\theta}_{s}$. 

For real non-precipitating ascents (such as  shallow convection), diabatic processes exist (horizontal or vertical advections, radiation, lateral mixing with the environment).
They all modify the ascent in a way to be determined for each case.

\begin{figure}[hbt]
\centering
\includegraphics[width=0.7\linewidth,angle=0,clip=true]{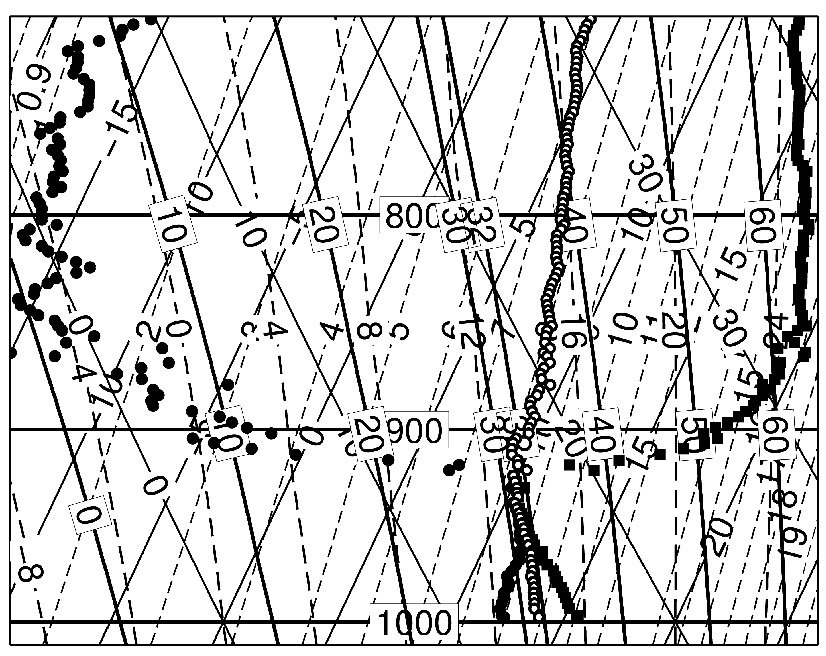}
\caption{\em\small A zoom of the skew $T-\ln(p)$ diagram for the FIRE-I RF03B Stratocumulus.
The dark circles (on the left) represent the dew point temperatures $T_d$, the open circle (middle position) the moist entropy temperatures $T_s$ and the dark squares (on the right) the actual temperatures $T$.
The moist PBL is characterized by constant values for $T_s$ up to the level $904$~hPa, including the entrainment region which extends up to the first level where $q_v < 3$~g/kg (see  Fig.(\ref{fig_entropy})(a)).
See further explanations in the text.
\label{fig_diagram2}}
\end{figure}
A zoom of the skew $T-\ln(p)$ diagram is presented in  Figure (\ref{fig_diagram2}), where the vertical profile of the FIRE-I RF03B data set is plotted up to $700$~hPa.
The top PBL height is $904$~hPa for that flight.
The moist entropy temperature $T_s$ (open circle) corresponds to the value of $({\theta}_{s})_1$ computed at each level from the data flight and taking into account the cloud liquid water.

The PBL is characterized by an almost constant value of $T_s$, remaining close to $304.5$~K (or $31.5$~C) for both the saturated and the unsaturated layers, as already suggested in  Figure (\ref{fig_Ths_Qt_Ql_RF03B}).
The two lines ${\theta}_{s}=30$ and $32$~C are depicted, in order to make easier the analysis.

It can be noted that, up to the surface condensation level (about $960$~hPa), the RF03B ascent looks like the ideal ascent depicted in  Figure (\ref{fig_diagram1}), with a saturated constant $T_s$ path up to the top PBL $904$~hPa level.
Above the top PBL height, the jump in $T_s$ is small (less than $1$~C?) and the moist entropy temperature $T_s$ increases linearly with $z$ or $\ln(p)$ in the dry and warm subsiding air, due to the diabatic processes (radiation and subsidence).

\section{The budget equation for the moist entropy.} 
\label{section_constths}

Although it is a central question in this paper, it may be difficult to understand or to explain why moist entropy seems to be almost a constant throughout the PBL region of marine Stratocumulus, as observed in the section (\ref{section_entr03B}).
The difficulty lies in the second principle of thermodynamics, which is uneasy to apply to real atmospheric circulations, particularly if stationary fluxes of heat and water species exist at the surface, transmitted by conduction, turbulent or convective processes to higher atmospheric levels.

One of the ways to understand ``by hand'' why the profile of moist entropy may be a constant within the PBL of marine Stratocumulus clouds is to analyze the properties verified by these clouds in the atmosphere, and by entropy in general thermodynamics.
\begin{itemize}[label=$\bullet$,leftmargin=3mm,parsep=0cm,itemsep=0.1cm,topsep=0cm,rightmargin=2mm]
\vspace*{-1mm}
  \item (Atmosphere)
  In marine Stratocumuli it is assumed that the cloud and the sub-cloud regions are in quasi-equilibrium with the surface temperature and the thermal radiations.
This kind of cloud acts as a  ``black body radiator''.
Even if sources and sinks of energy and species exist at the surface and at the top of the cloud, it is an open system in a quasi-equilibrium and in a quasi-stationary state.
  \item (Thermodynamics)
  In contrast to a closed system, steady states with constant entropy production are possible for open systems.
If the system is sufficiently close to equilibrium, the local equilibrium hypothesis can be made and, from the Prigogine theorem, the entropy production is extremal, with a constant entropy production balanced by removal from the system, so that the entropy may be locally held constant.
  \item (Turbulence)
  Since the moist turbulent processes act in order to mix-up the steady-state properties with no sources or sinks, and since  moist entropy has indeed no (or small) sources or sinks within the PBL of marine Stratocumuli, moist entropy must be well-mixed throughout the PBL (the MIME process), contrary to the Betts' variables which must vary with height in order to be in equilibrium with the steady-state vertical fluxes of energy and water species, respectively.
\end{itemize}

Another way to try to understand why moist entropy may be a constant is the analysis of the material change for  moist entropy.
From (\ref{defAPPC_Qi}) and (\ref{defAPPeqS}), the following statements are verified
\begin{align}
\rho \: T \: \frac{d\,s}{dt}  
  & =\: 
\rho \: \left( 
      \dot{Q}_i \:+\: \dot{D} 
       \right)
  \: -  \: \rho \: 
      \left[ \:
       \mu_k\: \frac{d_i\,q_k}{dt}
      \: \right]
  \nonumber \\
  & \quad
  \: - \:\mbox{\boldmath $J\!$}_k \: 
  .    \:\mbox{\boldmath $\!\nabla$} (h_k)
  \: - \: T \:s_k\:
      \left(\:
      \mbox{\boldmath $\!\nabla$} . \: \mbox{\boldmath $J\!$}_k
       \:\right)
  \: .\label{defAPPeqS0}
\end{align}

\begin{figure}[hbt]
\centering
\includegraphics[width=0.7\linewidth,angle=0,clip=true]{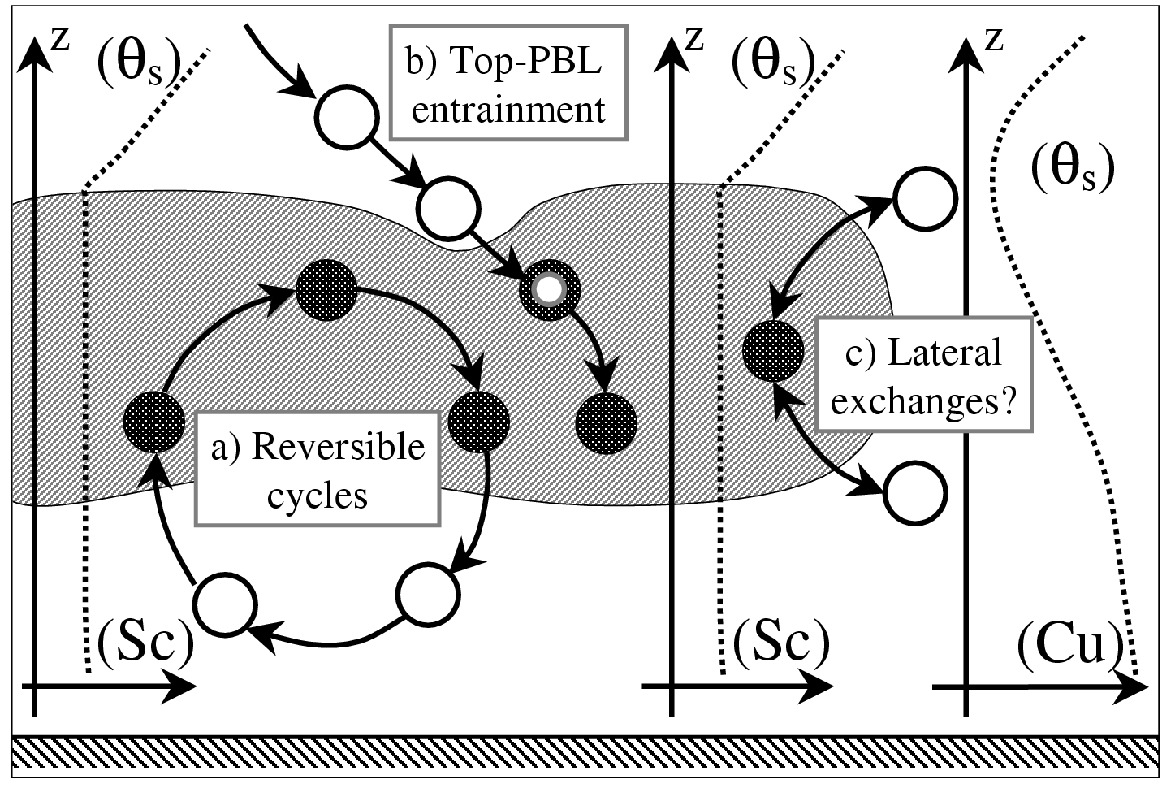}
\caption{\em\small Schematic representations of three Lagrangian motions occurring inside (or close to) a Stratocumulus region. 
(a) A reversible and moist adiabatic cycle.
(b) A top-PBL entrainment of a clear-air parcel within the Stratocumulus.
(c) Lateral mixing or exchanges between the Stratocumulus air (moist or dry) and the warmer cloud free environment (or for the transition to a Cumulus case).
\label{fig_StratoCu}}
\end{figure}

If the marine Stratocumulus clouds are in quasi-equilibrium and quasi-stationary state, with a net energy flux due to radiation almost equal to zero inside the cloud, or somehow balanced with other sources/sinks, the net value $\dot{Q}_i$ may be considered as a small term in (\ref{defAPPeqS0}).
It is also assumed that, except close to the surface, the dissipation term $\dot{D}$ is  a small term.
As demonstrated in the Appendix~C, the bracketed term in (\ref{defAPPeqS0}) represents the condensation and evaporation processes and it is canceled out for a set of reversible changes of phases.
As a consequence, the first line on the RHS of (\ref{defAPPeqS0}) is almost equal to zero for a marine Stratocumulus and for the reversible and moist adiabatic cycle represented in  Fig.(\ref{fig_StratoCu})(a).
In that case, the budget equation for moist entropy is controlled by the two last terms of (\ref{defAPPeqS0}), which both depend  on the diffusion fluxes $\mbox{\boldmath $J\!$}_k$ for dry air and water species.
If no precipitation exists and if no external mixing occurs between the different species of the moist air, then the diffusion fluxes are small or equal to zero, leading to $ds/dt = 0$ and to a possible explanation for the conservative property verified by the moist entropy within the PBL region of marine Stratocumulus.

The  process represented in  Fig.(\ref{fig_StratoCu})(b) corresponds to an entrainment of a  warm and dry clear-air parcel through the top of the Stratocumulus.
When the parcel enters  the cloud, the solar radiation is gradually switched off and  $\dot{Q}_i$ becomes a small term in (\ref{defAPPeqS0}).
The entrainment is then associated with a cooling of the parcel, a saturation toward $e_{ws}$ and a condensation of liquid water.
The cooling occurring after the entrainment may be explained by a thermal equilibrium process between the warm parcel and the colder surrounding cloud air.
The reason why the temperature is colder inside the cloud cannot be explained by the entropy budget.
It corresponds to the first principle and the internal energy of the enthalpy budgets.
The saturation and the condensation processes undergone by the parcel are associated with almost reversible changes of phases, leading to a cancellation of the bracketed term.
Therefore the three terms in the first line on the RHS of (\ref{defAPPeqS0}) are small.
If the diffusion fluxes $\mbox{\boldmath $J\!$}_k$ are assumed to be small, then the entropy and  $({\theta}_{s})_1$ must be conservative quantities, with the top-PBL values retained within the cloud, after the entrainment stage.

At the edges of the cloud (or outside the clouds, for Cumulus cases), the net heating rate due to radiation ($\dot{Q}_i$)  is not equal to zero, leading to higher values close to the surface for the moist entropy and with $({\theta}_{s})_1$ decreasing with height, as depicted in  Fig.(\ref{fig_StratoCu}) (c).
The exchanges between the Stratocumulus and the lateral cloud-free air may gradually modify the moist entropy of the Stratocumulus (and vice versa).
The lateral cloud-free vertical profile for $({\theta}_{s})_1$ corresponds to a composite analysis (not shown), realized by the author for several shallow Cumulus cases (BOMEX, ARM-Cu, RICO-composite, ATEX, GATE, SCMS-RF12).

\begin{figure}[hbt]
\centering
\includegraphics[width=0.7\linewidth,angle=0,clip=true]{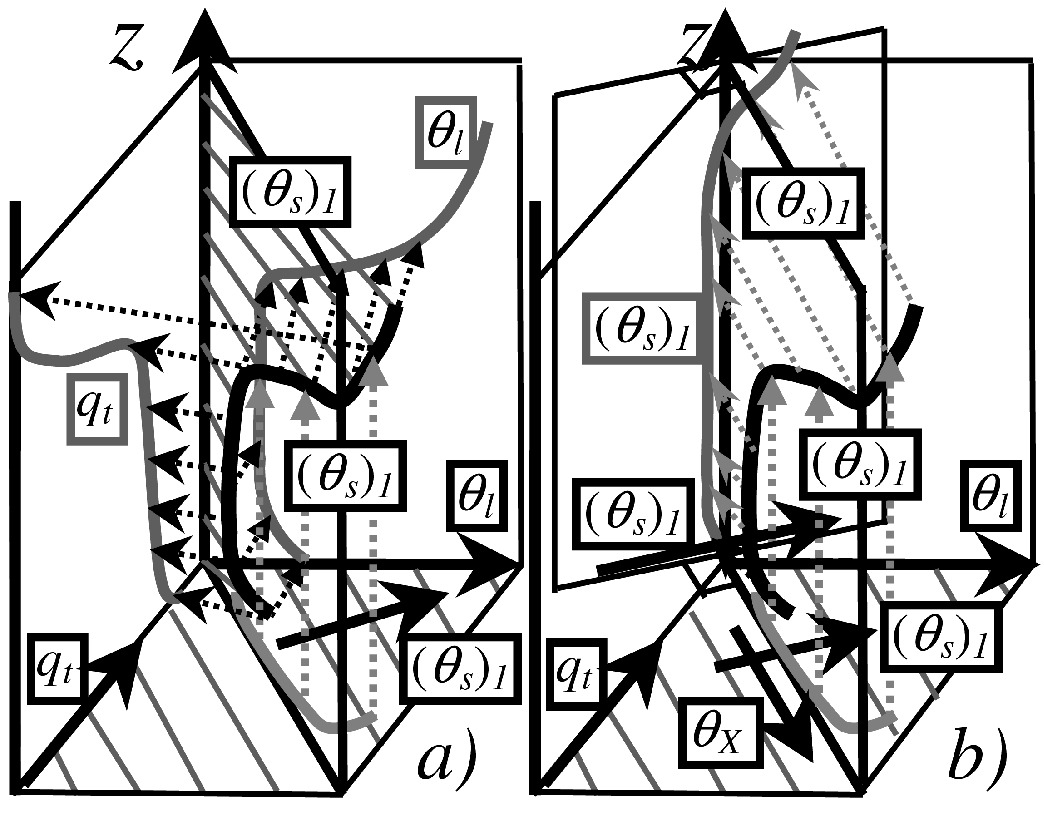}
\caption{\em\small A 3D-representation of the curve $({\theta}_{s})_1 (q_t, {\theta}_{l}, z)$ (heavy black line).
The conserved variable diagram is placed at the bottom, with a schematic curve representing a typical behaviour of the curves depicted in Fig.(\ref{fig_thlqt}), with the same slantwise greyish solid lines corresponding to constant values for $({\theta}_{s})_1$ (moist isentropes).
(a) The 3D-curve is obtained by plotting for each height $(z)$ the point of coordinates $(q_t, {\theta}_{l})$, with the vertical light grey arrows connecting the light grey conserved variable curve to the heavy black 3D-curve.
The curves $q_t(z)$ and ${\theta}_{l}(z)$ are obtained from the 3D curve by projections onto the left and the rear planes, respectively, with large jumps observed not only for $q_t(z)$, ${\theta}_{l}(z)$ but also for the 3D-curve.
(b) The new curve $({\theta}_{s})_1 (z)$ is obtained by a projection onto the slantwise vertical plane normal to the moist isentropic vertical plane.
Even if the jumps in $q_t(z)$ and ${\theta}_{l}(z)$ are large, the jump in $({\theta}_{s})_1 (z)$ almost disappears because the 3D-curve is almost parallel to the iso-$({\theta}_{s})_1 (z)$ vertical plane, leading to a straight line (up to the top of the inversion) created by the projection onto the plane normal to it.
The 3D-curve stars to diverge from the mean iso-$({\theta}_{s})_1 (z)$ plane above the top of the inversion and, accordingly, the curve $({\theta}_{s})_1 (z)$ starts to increase in the clear-air above the Strato-Cumulus.
The direction of increasing potential temperature ${\theta}_{X}({\theta}_{l}, q_t)$ is depicted in the conserved variable diagram (at the bottom) as a normal to the gradient in $({\theta}_{s})_1 ({\theta}_{l}, q_t)$.
\label{fig_3DThetas1}}
\end{figure}
Another way to understand how the existing jumps in ${\theta}_{l}$ and $q_t$ can be in agreement with a continuous profile of the moist entropy and of ${\theta}_{s}$ at the top of the PBL is presented in Fig.(\ref{fig_3DThetas1}). Following the graphical approach of Gibbs (1873), a 3D-curve ${\theta}_{s}({\theta}_{l}, q_t, z)$ is plotted in the panel (a), with the Betts' variables as horizontal  coordinates and with the usual Betts' vertical profiles obtained by projections onto the left and the rear vertical planes, where  large jumps exist for ${\theta}_{l}$ and $q_t$.
The ``mystery'' of the disappearing of the jump in ${\theta}_{s}$ is explained in the panel (b), by a vision ``in profile'' of the 3D-curve of ${\theta}_{s}(z)$ when it is projected onto the slantwise plane normal to the vertical isentropic planes. 

The jumps in ${\theta}_{l}(z)$ and $q_t(z)$  are thus minimized  in the direction normal to the isentropic plane,  labelled by $({\theta}_{s})_1$, whereas they are maximized in the direction parallel to the isentropic plane,  labelled in Fig.(\ref{fig_3DThetas1})(b) by the normal coordinate denoted by ${\theta}_{X}$ and defined from (\ref{defdTHs1}) by
\vspace{-0.15cm}
\begin{align}
\frac{d{\theta}_{X}}{{\theta}_{X}}
\: & = \;
\frac{d{\theta}_{l}}{{\theta}_{l}}
\: - \: 
\frac{dq_t}{\Lambda}
    \:  , \label{defdTHxa} \\
{\theta}_{X}
\: & = \;
{\theta}_{l} \:
\exp \left( - \:
\frac{q_t}{\Lambda}
\right)
     . \label{defdTHxb}
\end{align}

A possible application of these normal variables $({\theta}_{s})_1$ and ${\theta}_{X}$ are the vertical flux of them, approximated by (\ref{defFluxTHs1}) and (\ref{defFluxTHs2a}) for $({\theta}_{s})_1$ and by
\vspace{-0.15cm}
\begin{align}
\frac{\overline{w'{\theta}_{X}'}}{\overline{{\theta}_{X} }}
 & \: \approx \:
\frac{\overline{w'{\theta}_{l}'}}{\overline{{\theta}_{l} }}
      \; - \;
\frac{1}{\Lambda}
    \;\; \overline{w'q_t'} 
 \label{defFluxTHx2}
\end{align}
for the vertical flux of ${\theta}_{X}$.
It is possible to invert (\ref{defFluxTHs1}), (\ref{defFluxTHs2a}) and (\ref{defFluxTHx2}) to express the fluxes of the Betts' variables as
\vspace{-0.15cm}
\begin{align}
\frac{\overline{w'{\theta}_{l}'}}{\overline{ {\theta}_{l} }}
  & \: \approx \:
\frac{1}{1+\Lambda^2}
 \; \frac{\overline{w'({\theta}_{s})_1'}}{\overline{ ({\theta}_{s})_1 }}
      \; + \; 
\frac{\Lambda^2}{1+\Lambda^2}
 \; \frac{\overline{w'{\theta}_{X}'}}{\overline{{\theta}_{X} }}
 \: ,
 \label{defFluxTHl3} \\
 \overline{w'{q}_{t}'} & \: \approx \:
\frac{\Lambda}{1+\Lambda^2}
 \; \frac{\overline{w'({\theta}_{s})_1'}}{\overline{ ({\theta}_{s})_1 }}
      \; - \; 
\frac{\Lambda}{1+\Lambda^2}
 \; \frac{\overline{w'{\theta}_{X}'}}{\overline{{\theta}_{X} }}
    \: . 
 \label{defFluxQt3}
\end{align}
The system (\ref{defFluxTHl3}) and (\ref{defFluxQt3}) corresponds to the local relations
\vspace{-0.15cm}
\begin{align}
(1+\Lambda^2) \: \ln \left( {\theta}_{l} \right)
& \: = \:
 \; \ln \left[ ({\theta}_{s})_1 \: ({\theta}_{X})^{\Lambda^2} \right]
 \label{defTHl4a} \\
(1+\Lambda^2) \: q_t
& \: = \: \Lambda \:
 \; \ln \left[ ({\theta}_{s})_1 / {\theta}_{X} \right]
 \label{defQt4a}
\end{align}
The aim of the flux of $({\theta}_{s})_1$ is to reduce the departures from an isentropic profile, whatever the flux of ${\theta}_{X}$ may be.
The aim of the flux of ${\theta}_{X}$ is to jointly reduce the vertical gradients in ${\theta}_{l}$ and $q_t$, under the constraint of a conserved moist entropy.
This system (\ref{defFluxTHl3}) and (\ref{defFluxQt3}) may lead to new analyses or modelling of the moist turbulent processes.

\section{Conclusions.} 
\label{section_concl}

It is demonstrated in this paper that the  moist potential temperature $\theta_s$ is a true synonym of moist entropy, whatever the standard and reference values $T_0$, $T_r$ or $p_r$ may be.
It is suggested that $\theta_s$ could be an answer to the questions raised in the introduction of HH87:  it ``can be regarded as a direct  measure of (moist) entropy'', it ``stresses the importance of (moist) entropy in atmospheric dynamics'', and it could suggest some hints on ``how entropy can be used in cloud modelling''.

The analysis of the FIRE-I data flights shows that the Stratocumulus exhibits an almost constant moist entropy regime within the  whole PBL (from the surface to the  top of the cloud).
Moreover, it seems that there is no (or small) jump in moist entropy at the top of the Stratocumulus, with a soft and continuous transition  between the moist PBL and the warm and subsiding dry air above.
The explanations for these observed features are still partly unclear, although it has been explained via 3D-visions why it is possible to have at the same time large jumps in $\theta_l$ and $q_t$ and a smooth profile for moist entropy.

It is shown that  moist entropy can be approximated by a simple expression denoted by $({\theta}_{s})_1$ and given by any of (\ref{defTHast4b0}),  (\ref{defTHast4b1}),  (\ref{defTHast4b2}), (\ref{defTHast4bl}) or (\ref{defTHast4b}), with a good accuracy and with the common values $\Lambda=5.87$ valid for all flights.
It can be noted that all these formulae can be applied to either liquid water or ice cloud drops.
Therefore, they can be applied in GCM or LAM, including over Polar Regions.

The comparison of ${\theta}_{s}$ and  $({\theta}_{s})_1$ with the well-known Betts (1973) liquid-water potential temperature $\theta_l$ shows that an extra term $\Lambda\:q_t$ appears, with the coefficient $\Lambda$ corresponding to the difference between the dry-air and the water vapour partial entropies.
It is a way to take into account the impact of the change in entropy when some dry air enters a parcel of fluid and when it is replaced by water vapour, and vice versa. 
These kinds of processes were not fully represented in any of the previous potential temperature computations.

The mixing in moist entropy process (MIME)  appears to correspond to the CTEI criterion curves suggested by Randall (1980) and Deardorff (1980). 
The slantwise lines representing constant values for $({\theta}_{s})_1$  can be used in conserved variable diagrams to represent the Stratocumulus curves.
It is also possible to represent the moist entropic lines -- or iso-$({\theta}_{s})_1$ curves --  in the skew $T$-$\ln(p)$ diagrams, with clear distinctive patterns valid for marine Stratocumulus clouds, as observed in many real soundings (not shown).

Since moist entropy and the corresponding moist potential temperature $({\theta}_{s})_1$  are constants within the moist PBL in all FIRE-I data flights, also for the ASTEX, EPIC and DYCOMS-II (RF01) cases, it may be interesting to use  $({\theta}_{s})_1$ to study the non-precipitating Stratocumulus.
The applications may also concern the more general case of non-adiabatic turbulent fluxes, with the Betts' variables fluxes expressed in (\ref{defFluxTHl3}) and (\ref{defFluxQt3}) in terms of two weighted sums of the turbulent fluxes of $({\theta}_{s})_1$ and ${\theta}_{X}$.
This formulation offers new perspectives, with the flux of $({\theta}_{s})_1$ acting as a relaxation term toward a constant vertical profile of entropy, whereas the flux of ${\theta}_{X}$ may act as an isentropic and joint mixing of ${\theta}_{l}$ and $q_t$.
It can be noted that the problem of re-projection onto the non-conservative variables is not approached in this study.

It may be interesting to express the flux of ${\theta}_{v}$ in the thermal production (involved in the prognostic tke-equations) in terms of the fluxes of $({\theta}_{s})_1$ and may be ${\theta}_{X}$, with possible large impacts for both saturated or unsaturated moist air.

Other applications are can be expected,
\footnote{
\hspace*{1mm} 
Moist-air {\it Brunt-V\"{a}is\"{a}l\"{a} frequency} and {\it Potential Vorticity} are defined in  
     \url{http://arxiv.org/abs/1401.2379} {\tt arXiv:1401.2379 [ao-ph]};
     \url{http://arxiv.org/abs/1401.2383} {\tt arXiv:1401.2383 [ao-ph]};
 and \url{http://arxiv.org/abs/1401.2006} {\tt arXiv:1401.2006 [ao-ph]}.
The same Third Law used to defined $s$ in terms of $\theta_s$ is used to define the {\it moist-air enthalpy} (\url{http://arxiv.org/abs/1401.3125} {\tt arXiv:1401.3125 [ao-ph]}, paper submitted in 2012 to the Q. J. R. Meteorol. Soc, last revision in January 2014).
}
in particular for regions where $q_t$ is not constant and where gradients of the Third-Law quantity $\exp(\Lambda \: q_t)$ may become a new entry for interpreting atmospheric features.



\vspace{5mm}
\noindent{\large\bf Acknowledgements}
\vspace{2mm}

The author is most grateful to J.F. Geleyn, J.L. Brenguier, P. Santurette, I. Sandu and J.M. Piriou for helpful suggestions and encouraging discussions. 
The author would like to thank the anonymous referees for the constructive comments, which help to improve the manuscript.

The validation data from the NASA Flights during the FIRE I experiment have been kindly provided by S. R. de Roode and Q. Wang.


\vspace{4mm}
\noindent
{\bf Appendix A. List of symbols and acronyms.}
             \label{appendixSymbol}
\renewcommand{\theequation}{A.\arabic{equation}}
  \renewcommand{\thefigure}{A.\arabic{figure}}
   \renewcommand{\thetable}{A.\arabic{table}}
      \setcounter{equation}{0}
        \setcounter{figure}{0}
         \setcounter{table}{0}
\vspace{1mm}
\hrule

\begin{tabbing}
 ------------------\=  ------------------------------------ --\= \kill
 ASTEX    \> Atlantic Stratocumulus Transition Experiment \\
 CRM    \> Cloud Resolving Model\\
 CTEI \>  Cloud Top Entrainment Instability\\
 DYCOMS    \> DYnamics and Chemistry Of Marine Strat. \\
 EPIC \> East Pacific Investigation of Climate\\
 EUCLIPSE   \> European-Union CLoud Intercomparison,  \\
            \>  Process Study and Evaluation project \\
 EUROCS \> EUROpean Cloud Systems \\
 FIRE   \> First ISCCP Regional Experiment\\
 GCM    \> General Circulation Model\\
 GCSS   \> Gewex Cloud System Study\\
 IPCC    \>  Intergovernmental Panel on Climate Change \\
 ISCCP   \> International Satellite Cloud Climatology Project \\
 LAM    \> Limited Area Model\\
 LES    \> Large Eddy Simulation\\
 MIME \>  Mixing In Moist Entropy\\
 NWP    \> Numerical Weather Prediction \\
 PBL      \> Planetary Boundary Layer \\
 SCM    \> Single Column Model\\
 $\alpha$ \> $=1/\rho$ the specific volume \\
 $c_{pd}$ \> specific heat for dry air   \>($1004.7$~J~K${}^{-1}$~kg${}^{-1}$) \\
 $c_{pv}$ \> spec. heat for water vapour \>($1846.1$~J~K${}^{-1}$~kg${}^{-1}$) \\
 $c_{l}$  \> spec. heat for liquid water \>($4218$~J~K${}^{-1}$~kg${}^{-1}$) \\
 $c_{i}$  \> spec. heat for ice          \>($2106$~J~K${}^{-1}$~kg${}^{-1}$) \\
 $c_p$ \> specific heat at constant pressure for moist air, \\
       \> $ = \: q_d \: c_{pd} + q_v \: c_{pv} + q_l \: c_l + q_i \: c_i $ \\
       \> $ = \: q_d \: ( \: c_{pd} + r_v \: c_{pv} + r_l \: c_l + r_i \: c_i)$ \\
 ${c}^{\ast}$   \> $=c_{pd} + r_t \: c_{l}$ \\
 ${c}^{\ast}_p$ \> $=c_{pd} + r_t \: c_{pv}$ \\
 $d/dt$         \> the material (Lagrangian) barycentric derivative \\
 $\overline{(\dots)}$ \> horizontal and linear averaging operator  \\
 $<\!\ldots\!>$ \> horizontal and logarithmic averaging operator \\
 $\delta_{kj}$ \> equal to $1$ if $k=j$ ; equal to $0$ otherwise \\
 $\delta$ \> $=R_v/R_d-1 \approx 0.608$ \\
 $\eta$   \> $=1+\delta =R_v/R_d \approx 1.608$ \\
 $\varepsilon$ \> $=1/\eta=R_d/R_v \approx 0.622$ \\
 $\kappa$ \> $=R_d/c_{pd}\approx 0.2857$ \\
 $\gamma$ \> $= \eta \: \kappa \ = R_v/c_{pd} \approx 0.46$ \\
 $\lambda$ \> $= c_{pv}/c_{pd}-1 \approx 0.8375$ \\
 $e$       \> the water vapour partial pressure \\
 $e_r$      \> the water vapour reference partial pressure,\\
            \> with $\: e_r = e_{ws}(T_0) \approx 6.11$~hPa \\
 $e_{ws}(T)$ \> partial saturating pressure over liquid water \\
 $e_{is}(T)$ \> partial saturating pressure over ice \\
 $h$       \> specific enthalpy \\
 ${h}_d$   \> specific enthalpy for the dry air  \\
 ${h}_v$   \> specific enthalpy for the water vapour \\
 ${h}_l$   \> specific enthalpy for the liquid water \\
 ${h}_i$   \> specific enthalpy for the ice water \\
 $\Lambda$ \> $= [ (s_{v})_r - (s_{d})_r ] / c_{pd} \approx 5.87$ \\
 $L_v (T)$ \> $=h_v-h_l$: Latent heat of vaporisation \\
 $L_s (T)$ \> $=h_v-h_i$: Latent heat of sublimation \\
 $L_f (T)$ \> $=h_l-h_i$: Latent heat of fusion \\
 $L_v (T_0)$ \> $= 2.501$~$10^{6}$~J~kg${}^{-1}$ \\
 $L_s (T_0)$ \> $= 2.835$~$10^{6}$~J~kg${}^{-1}$ \\
 $L_f (T_0)$ \> $= 0.334$~$10^{6}$~J~kg${}^{-1}$ \\
 $\mu_k$  \> $=h_k-T\:s_k$ the specific chemical potential for the species $k=(d, v, l, i)$ \\
 $\mu_d$     \> specific chemical potential for dry air\\
 $\mu_v$     \> spec. chemical potential for water vapour\\
 $\mu_l$     \> spec. chemical potential for liquid water\\
 $\mu_i$     \> spec. chemical potential for solid water\\
 $\omega$ \> $=dp/dt$: vertical wind in isobaric coordinate \\
 $p$      \> $=p_d + e$: local value for the pressure \\
 $p_r$  \> $=(p_d)_r + e_r$: reference pressure ($p_r=p_0$)\\
 $p_d$     \> local dry air partial pressure \\
 $(p_d)_r$ \> reference dry air partial pressure ($\equiv p_r-e_r$)\\
 $p_0$     \> $=1000$~hPa: conventional pressure \\
 $\Pi$     \> $= T / \: \theta $: the Exner function \\
 $q_{d}$   \> $={\rho}_d / {\rho}$: specific content for dry air \\
 $q_{v}$   \> $={\rho}_v / {\rho}$: specific content for water vapour \\
 $q_{l}$   \> $={\rho}_l / {\rho}$: specific content for liquid water \\
 $q_{i}$   \> $={\rho}_i / {\rho}$: specific content for ice water \\
 $q_t  $   \> $= q_v+q_l+q_i$: total specific content of water \\
 $q_r  $   \> reference specific content of water, with
              $=r_r/(1+r_r)\approx 3.84$~g~kg${}^{-1}$ (see $r_r$)\\
 $(\dot{q})_{eva}$  \> rate of change of $q_l$ into $q_v$ (evaporation) \\
 $(\dot{q})_{sub}$  \> rate of change of $q_i$ into $q_v$ (sublimation) \\
 $(\dot{q})_{fus}$  \>  rate of change of $q_i$ into $q_l$ (fusion) \\
 $q_{S}$   \> saturation specific content for water vapour  \\
 $r_{v}$   \> $=q_{v}/q_{d}$: mixing ratio for water vapour \\
 $r_{l}$   \> $=q_{l}/q_{d}$: mixing ratio for liquid water \\
 $r_{i}$   \> $=q_{i}/q_{d}$: mixing ratio for ice water \\
 $r_{t}$   \> $=q_{t}/q_{d}$: mixing ratio for total water \\
 $r_{r}$   \> saturation reference mixing ratio of water:
              $\eta\:r_{r} \equiv e_r / (p_d)_r$ and $r_{r} \approx 3.82$~g~kg${}^{-1}$ \\
 $r_{S}$   \> saturation mixing  for water vapour \\
 ${\rho}_d$   \> specific mass for the dry air  \\
 ${\rho}_v$   \> specific mass for the water vapour \\
 ${\rho}_l$   \> specific mass for the liquid water \\
 ${\rho}_i$   \> specific mass for the ice water \\
 ${\rho}$   \> specific mass for the moist air $={\rho}_d+{\rho}_v+{\rho}_l+{\rho}_i$  \\
 $R_v$   \> water vapour gas constant --\= ($461.52$~J~K${}^{-1}$~kg${}^{-1}$) \kill
 $R_d$   \> dry air gas constant     \> ($287.06$~J~K${}^{-1}$~kg${}^{-1}$) \\
 $R_v$   \> water vapour gas constant \> ($461.53$~J~K${}^{-1}$~kg${}^{-1}$) \\
 $R$     \> $ = q_d \: R_d + q_v \: R_v$: gas constant for moist air $ = q_d \:(\: R_d + r_v \: R_v)$\\
 ${R}^{\ast}$ \> $ = R_{d}  + r_t \: R_{v} $ \\
 $S_m$   \> moist entropy static energy \\
 $S_l$   \> liquid-water static energy \\
 $s$       \> specific entropy \\
 ${s}_d$   \> specific entropy for the dry air  \\
 ${s}_v$   \> specific entropy for the water vapour \\
 ${s}_l$   \> specific entropy for the liquid water \\
 ${s}_i$   \> specific entropy for the ice water \\
 ${s}_d^{\diamond}$   \> approximate specific entropy for the dry air\\
 ${s}_v^{\diamond}$   \> approx. spec. entropy for the water vapour\\
 ${s}_l^{\diamond}$   \> approx. spec. entropy for the liquid water\\
 $ s_r $   \> reference entropy \\
 $(s_{d})_r$  \> reference values for the entropy of dry air at $T_0$ and $(p_d)_r$ \\
 $(s_{v})_r$  \> reference values for the entropy of water vapour at $T_0$ and $e_r$\\
 $s^0_d$   \> standard specific entropy for the dry air at $T_0$ and $p_0$: $6775$~J~K${}^{-1}$~kg${}^{-1}$ \\
 $s^0_v$   \> standard specific entropy for the water vapour at $T_0$ and $p_0$: $10320$~J~K${}^{-1}$~kg${}^{-1}$ \\
 $s^0_l$   \> standard specific entropy for the liquid water at $T_0$ and $p_0$: $3517$~J~K${}^{-1}$~kg${}^{-1}$ \\
 $s^0_i$   \> standard specific entropy for the solid water at $T_0$ and $p_0$: $2296$~J~K${}^{-1}$~kg${}^{-1}$ \\
 $T$       \> local temperature \\
 $T_d$     \> dew point temperature \\
 $T_{r}$   \> the reference temperature ($T_r\equiv T_0$) \\
 $T_s$     \> moist entropy temperature corresponding to ${\theta}_{s}$\\
 $T_{0}$   \> zero Celsius temperature ($=273.15$~K) \\
 $\theta$         \> $ = T\:(p_0/p)^{\kappa}$: potential temperature\\
 ${\theta}'_w$   \> wet-bulb pseudo-adiabatic potential temperature  \\
 ${\theta}_E$   \> equivalent potential temperature \\
 ${\theta}_{ES}$   \> saturation equivalent potential temperature \\
 ${\theta}_{v}$   \> virtual potential temperature (L68)  \\
 ${\theta}_{l}$   \> liquid-water potential temperature (B73)  \\
 ${\theta}_{il}$  \> ice-liquid water potential temperature (TC81)  \\
 ${\theta}_{vl}$  \> liquid-water virtual potential temperature (GB01)  \\
 ${\theta}_{l}^{\ast}$ \> liquid-water virtual potential temperature (E94)  \\
 ${\theta}_{S}^{\ast}$      \> entropy temperature (HH87)  \\
 ${\theta}^{\ast}$   \> moist entropy potential temperature (M93)  \\
 ${\theta}^{\ast}_r$ \> reference value for ${\theta}^{\ast}$ (M93) \\
 ${\theta}_{s}$   \> the new moist entropy potential temperature \\
 $({\theta}_{s})_1$   \> approximate version of ${\theta}_{s}$ (1st part)  \\
 $({\theta}_{s})_2$   \> approximate version of ${\theta}_{s}$ (2nd part)  \\
 ${\theta}_{sr}$ \> the reference value for ${\theta}_{s}$ \\
 ${\theta}_{X}$ \> the coordinate normal to ${\theta}_{s}$.
\end{tabbing}

\vspace{1mm}
\hrule

\vspace{4mm}
\noindent
{\bf Appendix B. The moist potential temperature $\theta_s$.}
             \label{appendixThetam}
\renewcommand{\theequation}{B.\arabic{equation}}
  \renewcommand{\thefigure}{B.\arabic{figure}}
   \renewcommand{\thetable}{B.\arabic{table}}
      \setcounter{equation}{0}
        \setcounter{figure}{0}
         \setcounter{table}{0}
\vspace{1mm}

\vspace{2mm}
The specific moist entropy is defined by (\ref{defAPPs1}) as a weighted sum of the specific partial entropies and, following HH87, it can be expressed as (\ref{defAPPs2}), where $q_t=q_v+q_l+q_i$.
\vspace{-0.15cm}
\begin{align}
   s  & = \: q_d\:s_d \: + \: q_v\:s_v
           \: + \: q_l\:s_l \: + \: q_i\:s_i
   \: ,
  \label{defAPPs1} \\
   s  & = \: q_d\:s_d \: + \: q_t\:s_v
     \: + \: q_l\:(s_l-s_v)
     \: + \: q_i\:(s_i-s_v)
   \: .
  \label{defAPPs2}
\end{align}

The differences of the partial entropies express in terms of the differences of the enthalpies and the chemical potentials, leading to
\vspace{-0.15cm}
\begin{align}
   s_l \: - \: s_v  & 
     = \:-\: \frac{h_v\:-\:h_l}{T} 
       \:-\: \frac{\mu_l\:-\:\mu_v}{T} 
   \: ,
  \label{defAPPslv} \\
   s_i \: - \: s_v  & 
     = \:-\: \frac{h_v\:-\:h_i}{T} 
       \:-\: \frac{\mu_i\:-\:\mu_v}{T} 
   \: .
  \label{defAPPsiv}
\end{align}

The differences of the enthalpies are equal to the latent heats $L_v = h_v\:-\:h_l$ and $L_s = h_v\:-\:h_i$.
If metastable states such as supercooled water are ignored, the difference of the chemical potentials are equal to the affinities and they are related to the saturation partial pressures by
\vspace{-0.15cm}
\begin{align}
   \mu_l \: - \: \mu_v  & = 
    \: R_v \: T 
    \: \ln \left(e_{ws}/e\right)
   \: ,
  \label{defAPPmulv} \\
   \mu_i \: - \: \mu_v  & = 
    \: R_v \: T 
    \: \ln \left(e_{is}/e\right)
   \: .
  \label{defAPPmuiv}
\end{align}

When (\ref{defAPPslv}) to (\ref{defAPPmuiv}) are inserted into (\ref{defAPPs2}), it yields
\vspace{-0.15cm}
\begin{align}
   s  & = \: q_d\:s_d \: + \: q_t\:s_v
     \: - \: \left( \frac{q_l\:L_v\:+\:q_i\:L_s}{T} \right)
     \: - \: R_v
     \left[\:
          q_l \: \ln \left(e_{ws}/e\right)
          + 
          q_i \: \ln \left(e_{is}/e\right) \:
     \right]
   \: .
  \label{defAPPs3}
\end{align}
The bracketed terms of (\ref{defAPPs3}) cancels out for clear air regions, where $q_l=q_i=0$.
It is also equal to zero for cloudy air if the partial pressure of the water vapour is equal to $e_{ws}$ if $q_l \neq 0$, or is equal to $e_{is}$ if $q_i \neq 0$ (i.e. with no under or supersaturation).

For the atmospheric conditions where the specific heat and the gas constants do not vary with $T$ or $p$, the dry air and water vapour specific partial entropies $s_d$ and $s_v$ can be expressed analytically as a relative change from a given reference state, defined by $T_r$, $(p_d)_r$, $e_r$ and $(q_v)_r=q_r$.
\vspace{-0.15cm}
\begin{align}
   s_d  & = (s_d)_r
    \:+ c_{pd} \,\ln(T/T_r)
    \:- R_{d} \,\ln[\:p_d/(p_d)_r\:]
   \: ,
  \label{defAPPsd} \\
   s_v  & = (s_v)_r
    \:+ c_{pv} \,\ln(T/T_r)
    \:- R_{v} \,\ln(e/e_r)
   \: .
  \label{defAPPsv}
\end{align}
When (\ref{defAPPsd}) and (\ref{defAPPsv}) are inserted into (\ref{defAPPs3}), with the bracketed terms of (\ref{defAPPs3}) cancelled, it yields
\vspace{-0.15cm}
\begin{align}
   s  & = \: q_d\:(s_d)_r \: + \: q_t\:(s_v)_r
        \: + \: (q_d\:c_{pd}\: + \: q_t\:c_{pv}) \,\ln(T/T_r)
        \: - \: \left( \frac{q_l\:L_v\:+\:q_i\:L_s}{T} \right)
  \nonumber \\
      & \quad
    \:- q_d\:R_{d} \,\ln[\:p_d/(p_d)_r\:]
    \:- q_t\:R_{v} \,\ln[\:e/e_r\:]
   \: .
  \label{defAPPs4}
\end{align}
The M93's formulation of the quotient ${\theta}^{\ast} / {\theta}^{\ast}_r$ follows from (\ref{defAPPs4}) and from the definition (\ref{defTHs}) in  section \ref{section_mptthast}, with a rearrangement of the terms expressed as $ q_d \: {c}^{\ast} \; \ln(\dots)$,

The computation of the quotient ${\theta}_{s} / {\theta}_{sr}$ defined by (\ref{defTHmSm1}) in the section \ref{section_defTHm} is obtained by transforming  $q_d\:(s_d)_r \: + \: q_t\:(s_v)_r$ in (\ref{defAPPs4}) with the property $q_d=1-q_t$, leading to
\vspace{-0.15cm}
\begin{align}
   q_d\:(s_d)_r \: + \: q_t\:(s_v)_r  
   & = 
   (s_d)_r \: + \: c_{pd} \; \Lambda \; q_t
   \: ,
  \label{defAPPs5}
\end{align}
where $\Lambda = [\:(s_v)_r-(s_d)_r\:]/c_{pd}$.
Similarly, 
\vspace{-0.15cm}
\begin{align}
   q_d\:c_{pd} \: + \: q_t\:c_{pv}  
   & = 
   c_{pd} \left(  1 \: + \: \lambda \: q_t \right)
   \: ,
  \label{defAPPs6}
\end{align}
where $\lambda = (\:c_{pv}-c_{pd}\:)/c_{pd}$.

A reference value $q_r$ is introduced in (\ref{defAPPs5}), with the use of a logarithm, to give
\vspace{-0.15cm}
\begin{align}
   q_d\:(s_d)_r \: + \: q_t\:(s_v)_r
   & \: = \:
   (1-q_r)\:(s_d)_r \: + \:q_r\:(s_v)_r 
  \: +\: c_{pd} \;
     \ln\left[
        \frac{\exp(\Lambda \; q_t)}
             {\exp(\Lambda \; q_r)}
     \right]
   \: .
  \label{defAPPs7}
\end{align}

The next step is to insert (\ref{def0pd}), (\ref{def0e}) and  (\ref{defAPPs7}) into (\ref{defAPPs4}), together with the following relations defined for the reference state
\vspace{-0.15cm}
\begin{align}
  (p_d)_r   & = 
      \frac{1}{1+\eta\:r_r} \: p_r
  \: , \label{def0pdr} \\
  e_r   & = 
      \frac{\eta\:r_r}{1+\eta\:r_r} \: p_r
     \; = \;
     \eta\:r_r\:(p_d)_r 
  \: , \label{def0er} \\
  r_r   & = 
      \frac{e_r}{\eta\:(p_d)_r }
  \: , \label{def0rr} \\
  q_r   & = 
      \frac{r_r}{1+r_r}
  \: . \label{def0qr}
\end{align}
After some rearrangement of the terms, the result is written with all the varying terms expressed as $c_{pd} \; \ln(\dots)$, leading to
\vspace{-0.15cm}
\begin{align}
   s  & = 
   \: (1-q_r)\:(s_d)_r \: + \:q_r\:(s_v)_r 
   +\: c_{pd} \;
     \ln\!\left[
        \frac{\exp(\Lambda \; q_t)}
             {\exp(\Lambda \; q_r)}
     \right]
  \nonumber \\
      & 
        + c_{pd}\:
             \ln\!\left[
             (T/T_r)^{\,1+\lambda\:q_t}
             \right]
        \:+ c_{pd}\:
             \ln\!\left[
             (p_r/p)^{\,\kappa\,(1+\delta\:q_t)}
             \right]
  \nonumber \\
      &
        + c_{pd}\:
             \ln\!\left[\:
             \left(
             \frac{1+\eta \:r_v}{1+\eta \:r_r}
             \right)^{\,\kappa\,(1+\delta\:q_t)}
              \:
             \left(
             \frac{r_r}{r_v}
             \right)^{\,\gamma\:q_t}
             \:\right]
  \nonumber \\
      &
     + c_{pd}\:
             \ln\!\left[
           \exp\left(
          - \:\frac{q_l\:L_v\:+\:q_i\:L_s}
                  {c_{pd}\:T} 
             \right)
             \right]
   \: .
  \label{defAPPs8}
\end{align}
The quotient ${\theta}_{s} / {\theta}_{sr}$ and the formulations (\ref{defTHm1}) and (\ref{defTHm2}) for ${\theta}_{s}$ and ${\theta}_{sr}$ follow directly from the identification of all the logarithm terms in  (\ref{defAPPs8}) with the one in (\ref{defTHmSm1}).

\vspace{4mm}
\noindent
{\bf Appendix C. The conservative equation for $({\theta}_{s})_1$.}
             \label{appendixConservThs1}
\renewcommand{\theequation}{C.\arabic{equation}}
  \renewcommand{\thefigure}{C.\arabic{figure}}
   \renewcommand{\thetable}{C.\arabic{table}}
      \setcounter{equation}{0}
        \setcounter{figure}{0}
         \setcounter{table}{0}
\vspace{1mm}

The formalism used in this Appendix is adapted from the approaches of De Groot and Mazur (1962), M93 or Zdunkowski and Bott (2004).
The implicit Einstein's summation rules prevail with $k=0$ representing the dry air, $k=1$ the water vapour and $k=(2,3)$ the condensed liquid water and ice, respectively.
The material derivative $d/dt$ for any variable can be separated into a sum of external and internal changes $d_e/dt + d_i/dt$.

The external changes $d_e/dt$ are generated by the diffusion fluxes of matter $\mbox{\boldmath $J\!$}_k$, with the differential velocity computed for each component with respect to the barycentric mean velocity $\mbox{\boldmath $v$}$, leading to $\mbox{\boldmath $J\!$}_k = \delta_{kj}\: \rho_j\:(\mbox{\boldmath $v$}_j - \mbox{\boldmath $v$} )$.
The external changes of matter $d_e(q_k)/dt$ are equal to $-\:(\rho)^{-1}\:\mbox{\boldmath $\!\nabla$} . \: \mbox{\boldmath $J\!$}_k$.
The internal changes $d_i/dt$ are generated by the physical processes such as the absorption of radiation or the phase changes, regarded as chemical reactions.

The effective diabatic heating rate $\dot{Q}_e$ will be defined as the sum of the true internal diabatic heating rate ($\dot{Q}_i = -\:(\rho)^{-1}\:\mbox{\boldmath $\!\nabla$} . \:\mbox{\boldmath $J\!$}_q$) plus the kinetic energy dissipation ($\dot{D}$) plus the differential diffusion of the partial enthalpy $h_k$, leading to
\vspace{-0.15cm}
\begin{align}
\dot{Q}_e & 
\:=\:
 \dot{Q}_i \:+\: \dot{D} 
\:-\:
 \frac{1}{\rho}
\:\mbox{\boldmath $J\!$}_k\: 
 . 
\:\mbox{\boldmath $\!\nabla$} (h_k)
  \: .\label{defAPPC_Qi}
\end{align}
It can be noted that the latent heat release processes are not included in $\dot{Q}_i$ (nor in $\dot{Q}_e$).
They are represented by the internal changes $d_i(q_k)/dt$.

With the use of (\ref{defAPPC_Qi}), the enthalpy and the entropy equations are given by
\vspace{-0.15cm}
\begin{align}
\frac{d\,h}{dt} & =\: 
       \frac{1}{\rho} \: \frac{d\,p}{dt}
  \: + \dot{Q}_e
  \: + \:   
       h_k\: \frac{d_e\,q_k}{dt}
  \:  , \label{defAPPeqh} \\
T\: \frac{d\,s}{dt}  & =\: 
     \frac{d\,h}{dt} 
  \: - \frac{1}{\rho} \: \frac{d\,p}{dt}
  \: -   
       \mu_k\: \frac{d\,q_k}{dt}
  \: . \label{defAPPeqGibbs}
\end{align}
The entropy equation (\ref{defAPPeqGibbs}) is equivalent to the Gibbs equation (\ref{defGibbs}), with the material derivatives replacing the differentials.

The derivative of $h=q_k\,h_k$ is equal to $q_k\,dh_k/dt + h_k\,dq_k/dt$.
The two terms are equal to $q_k\,c_{pk}\:d\,T/dt = c_p\:d\,T/dt$ and $h_k\,d_eq_k/dt + h_k\,d_iq_k/dt$, respectively.
For an hydrostatic equilibrium $(\rho)^{-1} = \,R \:T / p$.
It results that the temperature and the entropy equations can be written
\vspace{-0.15cm}
\begin{align}
c_p\: \frac{d\,T}{dt} & =\: 
       \frac{R \: T}{p} \: \frac{d\,p}{dt}
  \: + \dot{Q}_e
  \: - \:   
       h_k\: \frac{d_i\,q_k}{dt}
  \:  , \label{defAPPeqT} \\
T\: \frac{d\,s}{dt}  & =\: 
       \dot{Q}_e
  \: + \:   
       T \:s_k\: \frac{d_e\,q_k}{dt}
  \: -   
      \left[ \:
       \mu_k\: \frac{d_i\,q_k}{dt}
      \: \right]
  \: .\label{defAPPeqS}
\end{align}

The bracketed term in (\ref{defAPPeqS}) can be evaluated for a set of adiabatic internal changes given by
\vspace{-0.15cm}
\begin{align}
  d_i\left( q_v \right)/dt & =\; + \,  (\dot{q})_{eva} \; + \, (\dot{q})_{sub} 
\: , \label{def1_Dv2} \\
  d_i\left( q_l  \right)/dt & =\; - \, (\dot{q})_{eva}   \; + \, (\dot{q})_{fus} 
\: , \label{def1_Dl2} \\
  d_i\left( q_i  \right)/dt & = \; - \, (\dot{q})_{sub}   \; - \, (\dot{q})_{fus} 
\: . \label{def1_Di2} 
\end{align}
They represent the conversions between the water species, as in  section  \ref{section_mptvl1} for the Betts approach, except for all the conversion terms included, i.e with evaporation (or condensation), sublimation (or solid condensation) and fusion (or solidification) processes.
The latent heat release processes are represented by (\ref{def1_Dv2}) to (\ref{def1_Di2}), with the corresponding impacts $-\,h_k\: d_i\,q_k / dt$ and $-\,\mu_k\: d_i\,q_k/dt$ in the enthalpy and entropy equations, respectively.

From (\ref{def1_Dv2}) to (\ref{def1_Di2}), the bracketed term in (\ref{defAPPeqS}) is written
\vspace{-0.15cm}
\begin{align}
 &
\: - \:\left( \: \mu_{v} -  \mu_{l} \: \right) \:  (\dot{q})_{eva} 
\: - \:\left( \: \mu_{v} -  \mu_{i} \: \right) \: (\dot{q})_{sub} 
\nonumber \\
 & 
\: - \:\left( \: \mu_{l} -  \mu_{i} \: \right) \: (\dot{q})_{fus}
  \: .\label{defAPPeqbr} 
\end{align}
These terms vanish if changes of phase are assumed to be reversible and to occur with zero affinities, i.e. with the same chemical potentials $\mu_k$.
It is true if no over-saturation nor metastable phases exist (such as liquid water with $T<T_0$).

The aim of this section is to verify that (\ref{defAPPeqS}) is almost valid for the moist entropy $s$ defined by (\ref{defTHmSm1}) and with ${\theta}_{s}$ approximated by $({\theta}_{s})_1$ given by (\ref{defTHast4b1}).
Also, it would be important to understand how the approximate entropy equation defined with $({\theta}_{s})_1$ works with open systems and variable values for $q_d$ and $q_t$.
The resulting equation, valid for $c_{pd}\ln[({\theta}_{s})_1]$ can be written
\vspace{-0.15cm}
\begin{align}
\frac{d\,s}{dt}  & \approx\: 
\frac{c_{pd}}{({\theta}_{s})_1} \: \frac{d\,({\theta}_{s})_1}{dt} 
 \; = \;
\frac{c_{pd}}{{\theta}} \: \frac{d\,{\theta}}{dt} 
  \: + \:
 c_{pd} \:\Lambda \: \frac{d\,q_t}{dt}
\nonumber \\
 & \quad\quad\quad
  \: - \:
\frac{d}{dt}\left[\frac{L_v\: q_l}{T} \right]
  \: - \:
\frac{d}{dt}\left[\frac{L_s\: q_i}{T} \right]
  \: ,\label{defAPPeqS2} 
\end{align}
where
\vspace{-0.15cm}
\begin{align}
T \: \frac{c_{pd}}{{\theta}} \: \frac{d\,{\theta}}{dt} 
 & =\: 
c_{pd} \: \frac{d\,T}{dt} 
\: -\:
\frac{R_{d}\: T}{p} \: \frac{d\,p}{dt} 
  \: .\label{defAPPeqS2bis}
\end{align}

Let us assume the following hypotheses.
\vspace{-0.15cm}
\begin{align}
  q_l \;\frac{d}{dt}\left[\frac{L_v(T)}{T} \right]
& \: \ll \:
 \frac{L_v}{T} \: \frac{d \, q_l}{dt}
  \: ,\label{defHC1a} 
\\
  q_i \;\frac{d}{dt}\left[\frac{L_s(T)}{T} \right]
& \: \ll \:
 \frac{L_s}{T} \: \frac{d \, q_i}{dt}
  \: ,\label{defHC1b} 
\\
c_{pd}\: \frac{d\,T}{dt}
  \, - \,   
   \frac{R_d \: T}{p} \frac{d\,p}{dt} 
& \: \approx \: 
c_p\: \frac{d\,T}{dt}
  \, - \,   
   \frac{R \: T}{p} \frac{d\,p}{dt} 
  \, .\label{defHC2}
\end{align}

When  (\ref{defAPPeqT})  is put into (\ref{defHC2}), and  then into (\ref{defAPPeqS2}) via (\ref{defAPPeqS2bis}), the approximate equation results
\vspace{-0.15cm}
\begin{align}
T\: \frac{d\,s}{dt}  & \approx \: 
       \dot{Q}_e
  \: + \:   
       T \:s_k\: \frac{d_e\,q_k}{dt}
  \: -   
      \left[ \:
       \mu_k\: \frac{d_i\,q_k}{dt}
      \: \right]
 \nonumber \\
 & \quad
  \: - \: L_v \: \frac{d\,q_l}{dt}
  \: - \: L_s \: \frac{d\,q_i}{dt}
  \: - \:  T \:s_k\: \frac{d\,q_k}{dt}
 \nonumber \\
 & \quad
  \: + \:  T \: 
          \left[ \: (s_{v})_r - (s_{d})_r \: \right]
       \: \frac{d\,q_t}{dt}
  \: .\label{defAPPeqS3}
\end{align}

The approximate formula (\ref{defAPPeqS3}) has been obtained with the last term in  (\ref{defAPPeqT}) transformed into
\vspace{-0.15cm}
\begin{align}
- \:   h_k\: \frac{d_i\,q_k}{dt} 
& =\:   
       T \:s_k\: \frac{d_e\,q_k}{dt}
  \: - \:   
       T \:s_k\: \frac{d\,q_k}{dt}
  \: -   
      \left[ \:
       \mu_k\: \frac{d_i\,q_k}{dt}
      \: \right]
   , \nonumber
\end{align}
and with $ c_{pd} \:\Lambda =  (s_{v})_r - (s_{d})_r$ and (\ref{defHC1a}) plus (\ref{defHC1b}) introduced into (\ref{defAPPeqS2}).

All the terms in the second and third lines of (\ref{defAPPeqS3}) do not exist in (\ref{defAPPeqS}).
Therefore, the challenge is to understand in which conditions these terms can vanish in open systems, where not only reversible exchanges can exist between the water species $q_v$, $q_l$ and $q_i$, but  where $q_d$ and $q_t$ can also vary, with however the conservative  constraint $d\,q_d/dt = - \,d\,q_t/dt $.

The next step is to write the following identities
\vspace{-0.15cm}
\begin{align}
- \: L_v \: \frac{d\,q_l}{dt}
& =\: 
- \:  \left( \: h_{v} - h_{l} \: \right)\: \frac{d\,q_l}{dt}
  \: ,\label{defAPPeq4}\\
- \: L_s \: \frac{d\,q_i}{dt}
& =\: 
- \:  \left( \: h_{v} - h_{i} \: \right)\: \frac{d\,q_i}{dt}
  \: ,\label{defAPPeq5}
\end{align}
and
\vspace{-0.15cm}
\begin{align}
  - \:  T \:s_k\: \frac{d\,q_k}{dt}
& =\: 
  - \:  T \: 
          \left( \: s_{v} - s_{d} \: \right)
       \: \frac{d\,q_t}{dt}
\: + \:  T \:\left( \: s_{v} - s_{l} \: \right)\: \frac{d\,q_l}{dt}
\: + \:  T \:\left( \: s_{v} - s_{i} \: \right)\: \frac{d\,q_i}{dt}
  \: .\label{defAPPeq6} 
\end{align}

With  (\ref{defAPPeq4}) to (\ref{defAPPeq6}), the second and third lines of (\ref{defAPPeqS3}) are changed  into
\vspace{-0.15cm}
\begin{align}
& 
  - \:  T \: \left\{\:
          \left( \: s_{v} - s_{d} \: \right)
           \: - \:
     \left[ \: (s_{v})_r - (s_{d})_r \: \right]
            \: \right\}
       \: \frac{d\,q_t}{dt}
\: - \:\left( \: \mu_{v} -  \mu_{l} \: \right)\: \frac{d\,q_l}{dt}
\: - \:\left( \: \mu_{v} -  \mu_{i} \: \right)\: \frac{d\,q_i}{dt}
  \: .\label{defAPPeq7} 
\end{align}
The last two terms of (\ref{defAPPeq7}) depend on differences in chemical potentials.
They must be evaluated for both external and internal changes in $q_k$.
For the external changes the chemical potentials are written with (\ref{defAPPmulv}) and (\ref{defAPPmuiv}) and for the internal changes the set of internal conversions (\ref{def1_Dv2}) to (\ref{def1_Di2}) are put into (\ref{defAPPeq7}), leading to
\vspace{-0.15cm}
\begin{align}
& 
  - \:  T \: \{\:
          ( \: s_{v} - s_{d} \: )
           \: - \:
      \overbrace{ [ \: (s_{v})_r - (s_{d})_r \: ]}^{c_{pd}\:\Lambda}
            \: \}
       \: \frac{d\,q_t}{dt}
 \nonumber \\
& 
\: - \: R_v \: T \, \left[
    \:\ln\left( \frac{e}{e_{sw}}\right) \frac{d_e\,q_l}{dt}
   + \ln\:\left( \frac{e}{e_{si}} \right) \frac{d_e\,q_i}{dt}
    \:\right]
 \nonumber \\
 & 
\: + \:\left( \: \mu_{v} -  \mu_{l} \: \right) \:  (\dot{q})_{eva} 
\: + \:\left( \: \mu_{v} -  \mu_{i} \: \right) \: (\dot{q})_{sub} 
\: + \:\left( \: \mu_{l} -  \mu_{i} \: \right) \: (\dot{q})_{fus}
  \: .\label{defAPPeq8} 
\end{align}

The last three terms forming the second line of (\ref{defAPPeq8}) exactly cancel out if the change of phases are reversible ones, i.e. if the chemical potentials are equal if one of the corresponding conversion rates $ (\dot{q})_{eva}$, $ (\dot{q})_{sub}$ or $(\dot{q})_{fus}$  exists.

The second line  of (\ref{defAPPeqS3}) doesn't exactly cancel out.
Nevertheless, it can be assumed that if some liquid water enters or leaves the parcel via the external diffusion fluxes (i.e. due to departures from the mean barycentric motion), the partial pressure $e$ for the water vapour will be equal to its saturating value $e_{sw}$, in order to deal with isentropic and reversible processes.
The same is true for isentropic and reversible changes in the ice water, for which it is assumed that ${e}=e_{si}$ if some $q_i$ enters or leaves the parcel.

Similarly, the  first line of (\ref{defAPPeq8}) doesn't cancel out, since $ s_{v} - s_{d}$  is not exactly equal to  $(s_{v})_r - (s_{d})_r$.
However, it is expected that the difference $ (s_{v} - s_{d}) - c_{pd}\:\Lambda$ must  be much smaller than $ (s_{v} - s_{d})$, leading to larger errors if the terms $c_{pd}\:\Lambda$  were omitted in (\ref{defAPPeq8}), as in the Betts formulation $\theta_l$.
If this term  was not included, a diffusion of $q_t$ into $q_d$ (or vice versa) would lead to an impact much more important than with $(\theta_s)_1$ defined by (\ref{defTHast4b1}) and leading to the first line of  (\ref{defAPPeq8}).

To assert these statements, let us write the difference $ (s_{v} - s_{d}) - [(s_{v})_r - (s_{d})_r]$ as
\vspace{-0.15cm}
\begin{align}
&( c_{pv} -  c_{pd} ) \ln\left( \frac{T}{T_r}\right)
   \: - \: R_{d}\:\ln\left( \frac{ p_d}{(p_d)_r}\right)  
   \: + \: R_{v} \:\ln\left( \frac{e}{e_r}\right)
  \: . \label{defAPPeq9}
\end{align}
As for the difference $ (s_{v} - s_{d}) $, it can be evaluated with $s_v^0$ and $ s_d^0$ as absolute reference values, leading to
\vspace{-0.15cm}
\begin{align}
& ( c_{pv} -  c_{pd} ) \ln\left( \frac{T}{T_0}\right)
   \: - \: R_{d}\:\ln\left( \frac{ p_d}{p_0}\right)
   + R_{v} \:\ln\left( \frac{e}{e_r}\right)
   + 
 \left[ \: R_{v} \:\ln\left( \frac{e_r}{p_0}\right)
          - (s_v^0 - s_d^0) \:
\right]
   .\label{defAPPeq9bis}
\end{align}
For the values  $T_r=T_0$ and $(p_d)_r \approx p_0$ retained in the present study, the difference of (\ref{defAPPeq9bis}) with (\ref{defAPPeq9}) is equal to the last bracketed terms of (\ref{defAPPeq9bis}).

For the values of the constants given in the Appendix-A, this difference can be evaluated to $-2352 - 3545 = - 5897$~J/K/kg. 
The other terms of (\ref{defAPPeq9}) are equal to zero for $T=T_r$, $p_d = (p_d)_r $ or $e=e_r$.
For the extreme tropospheric values $T=320$~K,  $p_d = 50$~hPa or  $e=0.1$~hPa, the three terms of  (\ref{defAPPeq9}) are  equal to $+134$, $+860$ and $-1896$~J/K/kg, respectively.
Therefore, the magnitudes of the first two terms depending on $\ln(T/T_r)$ and $\ln({ p_d}/{(p_d)_r})$ are indeed small in comparison of $5897$~J/K/kg.
The last term depending on $\ln(e/e_r)$ is less than one third of $5897$~J/K/kg for $e=0.1$~hPa (upper troposphere values).
For the FIRE-I region, $q_v$ varies between $2$ and $10$~g/kg for $p=850$ and $1000$~hPa, leading to values of $e$ varying between $3$ and $16$~hPa, with the last term $R_v\,\ln(e/e_r)$ varying between  $328$ and  $444$~J/K/kg.
It is thus less than one tenth of $5897$~J/K/kg.

As a consequence, the explanation on how the approximate entropy  equation (\ref{defAPPeqS3})  works with open systems and with variable values for $q_d$ and $q_t$ highlights the importance of the term $\Lambda\: q_t$  in the formulation of  $\theta_s$  or $(\theta_s)_1$, and in (\ref{defAPPeqS2}).

\vspace{4mm}
\noindent
{\bf Appendix D. The averaging operators.}
             \label{appendixAverage}
\renewcommand{\theequation}{D.\arabic{equation}}
  \renewcommand{\thefigure}{D.\arabic{figure}}
   \renewcommand{\thetable}{D.\arabic{table}}
      \setcounter{equation}{0}
        \setcounter{figure}{0}
         \setcounter{table}{0}
\vspace{1mm}

Conditionally linear averages can be applied to the specific contents $q_v$, $q_l$, $q_i$ or $q_t=1-q_q$.
However, they must not be applied to $\theta_l$ or $\theta_s$, because only the moist entropy $s$ verifies an additive property, with the moist entropy depending on $c_{pd}$ times the logarithm of $\theta_s$ and with $\overline{\ln(\theta)} \neq \ln(\overline{\theta})$.

Accordingly, the ``logarithmic mean value'' for $\theta_s$ will be denoted by $<\!\theta_s\!>$.
It is valid for either the clear-air, the in-cloud or the grid-cell averages of the entropy $\overline{s}$.
It is defined by averaging (\ref{defTHmSm1}) with $q_r$, $(s_d)_r$,  $(s_v)_r$, ${c}_{pd}$ and ${\theta}_{sr}$ constant, leading to
\vspace{-0.15cm}
\begin{align}
\overline{s}  
      &  \: = \; 
                    (1-q_r)\:(s_d)_r 
                  \: + \: q_r\:(s_v)_r
                  \: + \: {c}_{pd} \: 
               \ln\left(<\!{\theta}_{s}\!>\right)
               \: - \:  {c}_{pd} \: 
               \ln\left({\theta}_{sr}\right)
    \: , \label{defavTH1}
\end{align}
with
\vspace{-0.15cm}
\begin{align}
\ln\left( <\!{\theta}_{s} \!> \right)
     &  \: = \; 
\overline{\;
               \ln\left({\theta}_{s}  \right)
              \;}
    \: . \label{defavTH2}
\end{align}
Consequently, the logarithmic mean of  $< ({\theta}_{s})_1  >$ is defined by (\ref{defTHast4b0}) to (\ref{defTHast4b2}), leading to the result
\begin{align}
< ({\theta}_{s})_1  >
  & = 
\exp \left(
\overline{\;\ln\left[ \:({\theta}_{s})_1\: \right]\;}
  \right)
\; \;
\exp \left(
\Lambda \: \overline{\;q_t\:}
  \right)
\; \;
   \exp \left(  \: - \: 
          \frac{1}{{c}_{pd}} 
          \overline{\: \left[
          \frac{L_v}{T}\:q_l 
          \: + \:
          \frac{L_s}{T}\:q_i
                        \right]  }\:
  \right)
   . \label{defTHast4b1m}
\end{align}
The non-linearity concerns the logarithm term and the join variations of $T$ and $q_l$ or $q_i$ in the last exponential term of (\ref{defTHast4b1m}).

For the FIRE-I flights, the local values of $({\theta}_{s})_1$ mainly vary on the horizontal and they remain close to the mean value $\overline{({\theta}_{s})_1}$ with a discrepancy of a few percents.
In such a case, the departure term $({\theta}_{s})'_1  = ({\theta}_{s})_1 - \overline{({\theta}_{s})_1}$ is smaller than the average value $\overline{({\theta}_{s})_1}$ and the term $\overline{\;\ln[ \:({\theta}_{s})_1\:]\;} = \ln[ \:\overline{({\theta}_{s})_1}\:] + \overline{\;\ln[ \:1 + ({\theta}_{s})'_1 /\overline{({\theta}_{s})_1} \:]\;}$ can be approximated with $\ln(x)\approx x - x^2/2$ by $\ln[ \:\overline{({\theta}_{s})_1} ] \: + \: 0.5\: \overline{{({\theta}_{s})'_1}^{\,2}} / \overline{({\theta}_{s})_1}^{\,2} $, leading to
\begin{align}
< ({\theta}_{s})_1  >
  & \approx \: 
\overline{({\theta}_{s})_1}
\; \;
\exp \left(
\Lambda \: \overline{\;q_t\:}
  \right)
\; \;
\exp \left[
 \frac{\:
      \overline{ {({\theta}_{s})'_1}^{\,2} }
      \:}
      {
      2\:\overline{({\theta}_{s})_1 }^{\,2}
      }
  \right]
\; \;
   \exp \left(  \: - \: 
          \frac{1}{{c}_{pd}} 
          \overline{\: \left[
          \frac{L_v}{T}\:q_l 
          \: + \:
          \frac{L_s}{T}\:q_i
                        \right]  }\:
  \right)
   . \label{defTHast4b1m2}
\end{align}

For horizontal fluctuations  of  $({\theta}_{s})_1 $, the departure term $\overline{{({\theta}_{s})'_1}^{\,2}} / \overline{({\theta}_{s})_1}^{\,2} $ can be discarded because, for $|({\theta}_{s})'_1|$ less than $5$~K and for $\overline{({\theta}_{s})_1}$ equal to $300$~K, the departure term is about $3.10^{-4}$, leading to an impact of $0.05$~K on $< ({\theta}_{s})_1  >$.

As a consequence, the horizontal mean value for $({\theta}_{s})_1 $ is written
\vspace{-0.15cm}
\begin{align}
< ({\theta}_{s})_1  >
  & \approx \: 
\overline{({\theta}_{s})_1}
\: \: 
 \left( \frac{ { }^{ } }{ }
 1 \: + \: \Lambda \: \overline{\;q_t\:}
   \: - \: 
          \frac{1}{{c}_{pd}} 
           \left[ \: 
    \overline{\:
          \frac{L_v}{T}\:q_l 
             }
          \: + \:
    \overline{\:
          \frac{L_s}{T}\:q_i
             }
         \:  \right] \:
  \right)
   . \label{defTHast4b1m3}
\end{align}

The same analysis could not be retained for an application to a vertical mean of the moist entropy, with possible larger departure terms $|({\theta}_{s})'_1|$ (i.e. for an averaging of the PBL values and the free upper air regions).
In that case, the formulae (\ref{defTHast4b1m}) or (\ref{defTHast4b1m2}) must be retained.

For the specific contents $q_v$, $q_l$, $q_i$ or $q_d=1-q_t$, the standard deviation $\sigma_q$ is obtained from the linear mean value $m_q=\overline{\;q\:}$ and the corresponding variance $v^2_q = \overline{\;q^2\:}$, leading to the result 
\vspace{-0.15cm}
\begin{align}
\sigma_q  & \: = \: \sqrt{\; v^2_q -  (m_q)^2  \;}
   . \label{defqStd}
\end{align}
The method is different for a moist potential temperature like $<\!\theta_s\!>$ defined by (\ref{defTHast4b1m}).
If the mean and the variance of $\ln(\theta)$ are denoted by $m_{\ln(\theta)}=\overline{\;\ln(\theta)\:}$ and  $v^2_{\ln(\theta)} = \overline{\;\ln(\theta)^2\:}$, the standard deviation for $\ln(\theta)$ is given by (\ref{defqStlnTheta}) and  $\ln(\theta)$ can vary within $m_{\ln(\theta)} \pm \sigma_{\ln(\theta)}$.
 The standard deviation for $\theta$ can be set to half of the spread width $\exp[\:m_{\ln(\theta)} \pm \sigma_{\ln(\theta)}]$, leading to the results expressed by the product (\ref{defTHastStd}), valid for the potential temperature $\theta$.
\vspace{-0.15cm}
\begin{align}
\sigma_{\ln(\theta)}  & \: = \: \sqrt{\; v^2_{\ln(\theta)} -  [\:m_{\ln(\theta)}\:]^2  \;}
   , \label{defqStlnTheta} \\
\sigma_\theta  & = \exp{[\:m_{\ln(\theta)}\:]} \;
                  \sinh\!{\left[ \: \sigma_{\ln(\theta)} \:\right]\;}
   . \label{defTHastStd}
\end{align}




\vspace{5mm}
\noindent{\large\bf References}
\vspace{2mm}

 \noindent{$\bullet$ Bauer LA.} {1908}.
{The Relation between ``Potential Temperature'' and ``entropy''. 
{\it Phys. Rev., Series I}.
{\bf 26,} (2): pp.177--183.}

 \noindent{$\bullet$ Bauer LA.}, {1910}.
{Paper number XXII: The Relation between ``Potential Temperature'' and ``entropy''. 
{\it Smithsonian Miscellaneous collections. 
The mechanics of the Earth's Atmosphere.
A collection of Translations by Cleveland Abbe.
Third Collection}.
{\bf Vol. 51,} (4): pp.495--500.
Reprinted from the Physical Review paper (Bauer, 1908)}

 \noindent{$\bullet$ Betts AK.} {1973 (B73)}.
{Non-precipitating cumulus convection and its parameterization.
{\it Q. J. R. Meteorol. Soc.}
{\bf 99} (419):
178--196.}

 \noindent{$\bullet$ Betts AK., Dugan F.J.} {1973}.
{Empirical formula for saturation pseudoadiabats ans saturation equivalent potential temperature.
{\it J. Appl. Meteor.}
{\bf 12} (4):
731--732.}

 \noindent{$\bullet$ Bolton D.} {1980}.
{The computation of Equivalent Potential Temperature.
{\it Mon. Weather Rev.}
{\bf 108,} (7):
1046--1053.}

 \noindent{$\bullet$ Bougeault P, Lacarr\`ere P.} {1989}.
{Parameterization of orography-induced turbulence in a meso\-beta-scale model.
{\it Mon. Weather Rev.}
{\bf 117,} (8):
1872--1890.}

 \noindent{$\bullet$ Bretherton CS, Uttal T, Fairall CW, Yuter SE, Weller RA, Baumgardner D, Comstock K, Wood R, Raga GB.} {2004}.
{The EPIC 2001 Stratocumulus study.
{\it Bull. Amer. Meteor. Soc.}
{\bf 85,} (7):
967--977.}

 \noindent{$\bullet$ Brinkop S, Roeckner E.} {1995 (BR95)}.
{Sensitivity of a general circulation model to parameterizations of cloud-turbulence interactions in the atmospheric boundary layer.
{\it Tellus A.}
{\bf 47,} (2):
197--220.}

 \noindent{$\bullet$ Catry B, Geleyn JF, Tudor M, B\'enard P, Troj\'{a}kov\'{a} A.} {2007}.
{Flux-conservative thermodynamic equations in a mass-weighted
framework.
{\it Tellus A.}
{\bf 59,} (1):
71--79.}

 \noindent{$\bullet$ Cuijpers JWM, Bechtold P.} {1995}.
{A simple parameterization of cloud water related variables for use in boundary layer models.
{\it J. Atmos. Sci.}
{\bf 52} (13):
2486--2490.}

 \noindent{$\bullet$ Cuxart J, Bougeault P, Redelsperger JL.} {2000 (CBR00)}.
{A turbulent scheme allowing for mesoscale and large-eddy simulations.
{\it Q. J. R. Meteorol. Soc.}
{\bf 126} (562):
1--30.}

 \noindent{$\bullet$ Deardorff JW.} {1976}.
{Usefulness of Liquid-Water Potential Temperature in a Shallow-Cloud Model.
{\it J. Appl. Meteor.}
{\bf 15} (1):
98--102.}

 \noindent{$\bullet$ Deardorff JW.} {1980}.
{Cloud top entrainment instability.
{\it J. Atmos. Sci.}
{\bf 37} (1):
131--147.}

 \noindent{$\bullet$ Duynkerke PG, de Roode SR, van Zanten MC, Calvo J, Cuxart J, Cheinet S, Chlond A, Grenier G, Jonker PJ, K\"{o}hler M, Lenderink G, Lewellen D, Lappen C, Lock AP, Moeng C, M\"{u}ller F, Olmeda D, Piriou J, S\'anchez E, Sednev I.} {2004}.
{Observations and numerical simulations of the diurnal cycle of the EUROCS Stratocumulus case.
{\it Q. J. R. Meteorol. Soc.}
{\bf 130} (604):
3269--3696.} 

 \noindent{$\bullet$ Emanuel KA.} {1994}.
{Atmospheric convection.}
Pp.1--580.
{\it Oxford University Press: New York and Oxford.} 

 \noindent{$\bullet$ Gibbs,~J.~W.} {1873}.
{Graphical methods in the thermodynamics of fluids.
{\it Trans. Connecticut Acad.}
{\bf II}: p.309--342.
(Pp 1--32 in Vol. 1 of
{\it The collected works of J. W. Gibbs},
1928.
Longmans Green and Co.)} 

 \noindent{$\bullet$ Gibbs,~J.~W.} {1873}.
{A method of geometrical representation
of the thermodynamic properties of substance
by means of surfaces.
{\it Trans. Connecticut Acad.}
{\bf II}: p.382--404.
(Pp 33--54 in Vol. 1 of
{\it The collected works of J. W. Gibbs},
1928.
Longmans Green and Co.)} 

 \noindent{$\bullet$ Gibbs,~J.~W.} {1875-76-77-78}.
{On the equilibrium of heterogeneous substances.
{\it Trans. Connecticut Acad.}
{\bf III}: p.108--248, 1875-1876 and p.343-524, 1877-1878.
(Pp 55--353  in Vol. 1 of
{\it The collected works of J. W. Gibbs},
1928.
Longmans Green and Co.)} 

 \noindent{$\bullet$ Grenier H, Bretherton CS.} {2001 (GB01)}.
{A moist PBL parameterization for Large-Scale models and its
application to subtropical cloud-Topped marine boundary layers.
{\it Mon. Weather Rev.}
{\bf 129,} (3):
357--377.}

 \noindent{$\bullet$ De Groot SR, Mazur P.} {1962}.
{Non-equilibrium Thermodynamics. 
{\it North-Holland Publishing Company}.
Amsterdam}

 \noindent{$\bullet$ Hauf T, H\"{o}ller H.} {1987 (HH87)}.
{Entropy and potential temperature.
{\it J. Atmos. Sci.}
{\bf 44} (20):
2887--2901.}

 \noindent{$\bullet$ IPCC-AR4}{2007}.
{Summary for Policymakers.
In: Climate Change 2007. 
The physical science basis.
Contribution of working group I to 
the fourth assessment report of the
Intergovernmental Panel on Climate Change.
{\it Cambridge University Press, Cambridge, 
United Kingdom and New York, NY, USA.}
http://www.ipcc.ch/}

 \noindent{$\bullet$ Kuo H, Schubert WH.} {1988}.
{Stability of cloud-topped boundary layers.
{\it Q. J. R. Meteorol. Soc.}
{\bf 114} (482):
887--916.}

 \noindent{$\bullet$ Lilly DK.} {1968 (L68)}.
{Models of cloud-topped mixed layers under a strong inversion.
{\it Q. J. R. Meteorol. Soc.}
{\bf 94} (401):
292--309.} 

 \noindent{$\bullet$ Lilly DK.} {2002}.
{Entrainment into mixed layers. Part II: a new closure.
{\it J. Atmos. Sci.}
{\bf 59} (23):
3353--3361.}

 \noindent{$\bullet$ MacVean MK, Mason PJ.} {1990}.
{Cloud-top entrainment instability through small-scale mixing and its parameterization in numerical models.
{\it J. Atmos. Sci.}
{\bf 47} (8):
1012--1030.}

 \noindent{$\bullet$ Marquet P.} {1993 (M93)}.
{Exergy in meteorology: definition and properties
of moist available enthalpy.
{\it Q. J. R. Meteorol. Soc.}
{\bf 119} (511):
567--590.}

\noindent{$\bullet$ Marquet P.} {2011 (M11)}.
{Definition of a moist entropic potential temperature. Application to FIRE-I data flights.
{\it Q. J. R. Meteorol. Soc.}
{\bf 137} (656):
768--791.
\url{http://arxiv.org/abs/1401.1097}
{\tt arXiv:1401.1097 [ao-ph]}}

\noindent{$\bullet$ Marquet P, Geleyn J-F.} {2013}.
{On a general definition of the squared Brunt-V\"{a}is\"{a}l\"{a} Frequency associated with the specific moist entropy potential temperature.
{\it Q. J. R. Meteorol. Soc.}
{\bf 139} (670):
85--100.
\url{http://arxiv.org/abs/1401.2379}
{\tt arXiv:1401.2379 [ao-ph]}}

\noindent{$\bullet$ Geleyn J-F, Marquet P.} {2012}.
{Moist-entropic vertical adiabatic lapse rates: the standard cases and some lead towards inhomogeneous conditions.
{\it WGNE. Blue-Book}
(Addition to Marquet and Geleyn, 2013)
\url{http://arxiv.org/abs/1401.2383}
{\tt arXiv:1401.2383 [ao-ph]}}

\noindent{$\bullet$ Marquet P.} {2014}.
{On the definition of a moist-air potential vorticity.
{\it Q. J. R. Meteorol. Soc.}
Accepted in April 2013. Early view in January 2014. 
\url{http://arxiv.org/abs/1401.2006}
{\tt arXiv:1401.2006 [ao-ph]}}

\noindent{$\bullet$ Marquet P.} {2014}.
{On the definition of a moist-air specific thermal enthalpy.
{\it Q. J. R. Meteorol. Soc.}
Accepted in January 2014.
\url{http://arxiv.org/abs/1401.3125}
{\tt arXiv:1401.3125 [ao-ph]}}

 \noindent{$\bullet$ Neggers RAJ, Duynkerke PG, Rodts SMA.} {2003}.
{Shallow cumulus convection: A validation of 
large-eddy simulation against aircraft and 
Landsat observations
{\it Q. J. R. Meteorol. Soc.}
{\bf 129} (593):
2671--2696.} 

 \noindent{$\bullet$ Neggers RAJ, Siebesma AP, Jonker HJJ.} {2002}.
{A multiparcel model for shallow cumulus convection
{\it J. Atmos. Sci.}
{\bf 59} (10):
1655--1668.}

 \noindent{$\bullet$ Randall DA.} {1980}.
{Conditional instability of the first kind upside-down.
{\it J. Atmos. Sci.}
{\bf 37} (1):
125--130.}

 \noindent{$\bullet$ De Roode SR, Wang Q.} {2007 (RW07)}.
{Do Stratocumulus clouds detrain? FIRE I data revisited.
{\it Bound.-Layer Meteorol.}
{\bf 122,} (1):
479--491.}

 \noindent{$\bullet$ Stevens B, Lenschow DH, Faloona I, Moeng CH, Lilly DK, Blomquist B, Vali G, Bandy A, Campos T, Gerber H, Haimov S, Morley B, Thornton D.} {2003}.
{On entrainment rates in nocturnal marine Stratocumulus.
{\it Q. J. R. Meteorol. Soc.}
{\bf 129} (595):
3469--3493.} 

 \noindent{$\bullet$ Tripoli GJ, Cotton WR.} {1981 (TC81)}.
{The Use of lce-Liquid Water Potential Temperature as a Thermodynamic Variable In Deep Atmospheric Models.
{\it Mon. Weather Rev.}
{\bf 109,} (5):
1094--1102.}

 \noindent{$\bullet$ Von Bezold W.} {1891}.
{Paper number  XVI: On the Thermodynamics of the Atmosphere (second communication).
{\it Smithsonian Miscellaneous collections. 
The mechanics of the Earth's Atmosphere.
A collection of Translations by Cleveland Abbe.}
pp.243--256.
Translated from the paper ``Zur Thermodynamik der Atmosphaere'', in the
{\it Sitzungsberichte  der K\"onig. Akademie der Wissenschaften zu Berlin}, Vol. 46, p.1189-1206 (1888).
Available at: \\
http://openlibrary.org/b/OL6585650M/mechanics\_of\_the\_earth\_atmo\-sphere.}

 \noindent{$\bullet$ Von Helmholtz H.} {1891}.
{Paper number V: On Atmospheric Motions (First Paper).
{\it Smithsonian Miscellaneous collections. 
The mechanics of the Earth's Atmosphere.
A collection of Translations by Cleveland Abbe.}
pp.78--93.
Reprinted from the paper ``Ueber atmosphaerische Bewegungen'',
in the {\it Sitzungsberichte} of the Royal Prussian Academy of Science at Berlin, Vol. 46, p.647-663 (1888).
Available at: http://openlibrary.org/b/OL6585650M/mechanics\_of\_the\_earth\_atmo\-sphere.}

 \noindent{$\bullet$ Yamagushi T, Randall DA.} {2008}.
{Large-eddy simulation of evaporatively driven entrainment in cloud-topped mixed layers.
{\it J. Atmos. Sci.}
{\bf 65} (5):
1481--1504.}

 \noindent{$\bullet$ Zdunkowski W, Bott A.} {(2004)}.
{Thermodynamics of the Atmosphere.
A course in theoretical Meteorology 
{\it Cambridge University Press}.
}

 \noindent{$\bullet$ Zhu P, Bretherton CS, K\"{o}hler M, Cheng A, Chlond A, Geng Q, Austin P, Golaz JC, Lenderink G, Lock A, Stevens B.} {2005}.
{Intercomparison and interpretation of Single-Column Model simulations of a nocturnal Stratocumulus-topped marine boundary layer.
{\it Mon. Weather Rev.}
{\bf 133,} (9):
2741--2758.}


\end{document}